\documentstyle[psfig]{mn}

\def\etal{{\it et al.\ }}
\def\eg{{\it e.g.\ }}

\def\ie{{\it i.e.\ }}

\def\spose#1{\hbox to 0pt{#1\hss}}
\def\approxlt{\mathrel{\spose{\lower 3pt\hbox{$\sim$}}
	\raise 2.0pt\hbox{$<$}}}
\def\approxgt{\mathrel{\spose{\lower 3pt\hbox{$\sim$}}
	\raise 2.0pt\hbox{$>$}}}
\def\approxpropto{\mathrel{\spose{\lower 3pt\hbox{$\sim$}}
	\raise 2.0pt\hbox{$\propto$}}}
\mathchardef\twiddle="2218

\def\multleft#1{\hbox to size{\vbox {\halign {\lft{##}\cr #1}}\hfill}\par}
\def\multright#1{\hbox to size{\vbox {\halign {\rt{##}\cr #1}}\hfill}\par}

\def\today{\ifcase\month\or January\or February\or March\or April\or May\or
      June\or July\or August\or September\or October\or November\or December\fi
      \space\number\day, \number\year}
\def\<{\thinspace}
\def\apc{\rm atom cm$^{-2}$}
\def\cm{{\rm\thinspace cm}}
\def\erg{{\rm\thinspace erg}}

\def\keV{{\rm\thinspace keV}}

\def\km{{\rm\thinspace km}}

\def\Mpc{{\rm\thinspace Mpc}}
\def\Msun{\hbox{$\rm\thinspace M_{\odot}$}}

\def\s{{\rm\thinspace s}}
\def\yr{{\rm\thinspace yr}}

\def\ergpcmsqps{\hbox{$\erg\cm^{-2}\s^{-1}\,$}}

\def\ergps{\hbox{$\erg\s^{-1}\,$}}

\def\kmps{\hbox{$\km\s^{-1}\,$}}

\def\Msunpyr{\hbox{$\Msun\yr^{-1}\,$}}

\def\psqcm{\hbox{$\cm^{-2}\,$}}

\def\kmpspMpc{\hbox{$\kmps\Mpc^{-1}$}}

\def\apc{\rm atom cm$^{-2}$}
\title[ASCA and ROSAT observations of 
nearby cluster cooling flows]
{ASCA and ROSAT observations of nearby cluster cooling flows}
\author[S.W. Allen \etal]
{\parbox[]{6.in} {S.W. Allen$^1$, A.C. Fabian$^1$, R.M. Johnstone$^1$,
K.A. Arnaud$^2$ and P.E.J. Nulsen$^{3,4}$ \\
\footnotesize
1. Institute of Astronomy, Madingley Road, Cambridge CB3 0HA\\
2. Laboratory for High Energy Astrophysics, Goddard Space Flight Center, Greenbelt, MD 20771, USA \\
3. Department of Engineering Physics, University of Wollongong, Wollongong NSW 2522, Australia\\
4. Harvard-Smithsonian Center for Astrophysics, 60 Garden Street, Cambridge MA 02138, USA }}
\begin{document}
\date{Submitted to MNRAS 1999 September.}
\maketitle
\begin{abstract}
\noindent We present a detailed analysis of the X-ray properties of 
the cooling flows in a sample of nearby, X-ray bright clusters of 
galaxies using high-quality ASCA spectra and ROSAT X-ray 
images. We demonstrate the need for multiphase models to 
consistently explain the spectral and imaging X-ray data for 
the clusters. The mass deposition rates of the cooling flows,
independently determined from the ASCA spectra and ROSAT 
images, exhibit reasonable agreement. 
We confirm the presence of intrinsic 
X-ray absorption in the clusters using a variety of 
spectral models. We also report detections of extended 
$100\mu$m infrared emission, spatially coincident with the cooling 
flows, in several of the systems studied. The observed infrared 
fluxes and flux limits are in good agreement with the 
predicted values due to reprocessed X-ray emission from the 
cooling flows. We present precise 
measurements of the abundances of iron, magnesium, silicon and 
sulphur in the central regions of the Virgo and Centaurus clusters. 
Our results firmly favour models in which a high mass fraction ($70-80$ per 
cent) of the iron in the X-ray gas in these regions is due to 
Type Ia supernovae.  Finally, we present a series of methods 
which may be used to measure the ages of cooling flows from 
the X-ray data. The results for the present sample of clusters 
indicate ages of between $2.5$ and 7 Gyr. If the ages of cooling
flows are primarily set by subcluster merger events, then our results 
suggest that in the largest clusters, mergers with subclusters with masses of 
$\sim 30$ per cent of the final cluster mass are likely to disrupt 
cooling flows.   
\end{abstract}


\section{Introduction}

 In the central regions of most clusters of galaxies, the cooling time of
the hot intracluster medium is significantly shorter
than a Hubble time (Edge, Stewart \& Fabian 1992; White, Jones \& Forman 1997;
Peres \etal 1998). This is normally taken to
indicate the presence of a cooling flow, in which the hot gas cools and is
deposited throughout the cluster core 
(see Fabian 1994 for a review). Cooling flows are a phenomenon most easily
studied at X-ray wavelengths, where
the bulk of the cooling occurs. However, large Faraday rotation measures,
consistent with cooling-flow models, are also detected in radio observations
of cluster cores (Taylor, Barton \& Ge 1994; Taylor, Allen \& Fabian 1999) 
and strong optical emission-line 
nebulosities and UV/blue emission, associated with young, massive stars, 
are commonly observed in and around the dominant galaxies at the centres 
of cluster cooling-flows (\eg Johnstone, 
Fabian \& Nulsen 1987; McNamara \& O'Connell 1989; Allen 1995; Cardiel \etal 
1995, 1998; Crawford \etal 1999). The mass in young stars within the central 
$5-20$kpc regions of cooling-flow clusters 
typically accounts for $\approxlt 10$ per cent of the total mass inferred 
to be
deposited by the cooling flows within their cooling radii ($r_{\rm cool}\sim 
50-200$kpc) although can account for a significant fraction of the 
material deposited within the innermost part. At present,
however, the fate of the cooling gas at
larger radii remains uncertain (Daines, Fabian \& Thomas 1994; Fabian, 
Johnstone \& Daines 1994a; O'Dea \etal 1994; Voit \& Donahue 1995; 
O'Dea \& Baum 1996; 
Henkel \& Wiklind 1998; Allen 2000; Ferland, Fabian \& Johnstone 2000).

The X-ray surface brightness profiles of cooling-flow clusters 
require that the deposition of cooled matter is typically
distributed throughout the cooling radii with  ${\dot M}
\approxpropto r$.  This requires that cooling flows are multiphase (Nulsen 
1986; Thomas, Fabian \& Nulsen 1987). Spatially-resolved X-ray 
spectroscopy, carried out with the Position Sensitive 
Proportional Counter (PSPC) on ROSAT, has shown that the 
mean emission-weighted temperatures of cooling-flow clusters decrease 
towards their centres (\eg Allen \& Fabian 1994; Nulsen \& B\"ohringer 1995). 
More detailed multiphase spectral studies with ASCA have also revealed the 
presence of at least two gas phases in the central regions of cooling-flow
clusters, with a ratio of emission measures in good agreement
with the predictions from cooling-flow models (\eg Fukazawa \etal 1994; 
Fabian \etal 1994b; Allen 2000). 

 Early studies based on ASCA spectra for small samples of bright, 
nearby cooling flows (\eg Fabian \etal 1994b)  were complicated by 
uncertainties associated with the calculation of the Fe-L emission-line 
complex used in plasma emission codes. These calculations 
have since been significantly 
improved in the MEKAL code of Kaastra \& Mewe (1993; incorporating the Fe L calculations of 
Liedhal, Osterheld \& Goldstein 1995, although residual uncertainties in the 
calculations of other potentially important elements remain). Further 
complications in the modelling of the X-ray spectra arise from abundance
gradients (\eg Fukazawa \etal 1994; Matsumoto \etal 1996; 
Allen \& Fabian 1998), the effects 
of intrinsic absorption (\eg White \etal 1991; Allen \& Fabian 1997; Wise \& 
Sarazin 2000) and temperature variations within the flows, which depend 
upon the details of the local gravitational potentials.

 In this paper we study ASCA spectra and ROSAT X-ray imaging data 
for seven bright, nearby cooling flows; the Perseus Cluster (Abell 426), 
Abell 496, the Virgo Cluster, the Centaurus Cluster (Abell 3526), Abell 
1795, 2199 and 2597. We also include one non-cooling flow cluster, the Coma 
Cluster (Abell 1656), for comparison purposes. Our primary goal is to determine the 
properties of the cooling flows, using the improved MEKAL code in 
both our spectral and imaging analyses. We examine regions of the clusters 
large enough to encompass the entire cooling flows, thereby minimizing 
systematic problems associated with modelling the complex point spread 
functions of the ASCA mirrors and the underlying spatial distributions of the 
X-ray emitting and absorbing gas. 

Section 2 of this paper describes the observations and data reduction.
Sections 3 and 4 discuss the method and basic results from the 
spectral analysis. Section 5 examines the enrichment history of the 
ICM in the central regions of the Virgo and Centaurus clusters. 
Section 6 describes the deprojection analysis of the imaging data. 
Section 7 presents a comparison of the properties of the cooling flows 
determined from the independent spectral and imaging methods. The 
evidence for intrinsic X-ray absorption and reprocessed emission 
at infrared wavelengths are explored. Section 8 examines the constraints 
that may be placed on the ages of cooling flows from the X-ray data. 
Section 9 discusses the possible role of heating process in cooling flows
and the likely effects of subcluster merger events. 
The most important conclusions from our work are summarized in 
Section 10. Throughout this paper, we assume $H_0$=50 \kmpspMpc, $\Omega = 1$ 
and $\Lambda = 0$. For the Virgo Cluster, a distance of 18Mpc is assumed. 

\section{Observations and data reduction}

\subsection{The ASCA observations }

 The ASCA (Tanaka, Inoue \& Holt 1994) observations were made over a 
three-and-a-half year period between 1993 May and 1996 December. 
The ASCA X-ray telescope array (XRT) consists of four 
nested-foil telescopes, each focussed onto one of four detectors; two X-ray 
CCD cameras, the Solid-state Imaging Spectrometers (SIS; S0 and S1) and 
two Gas scintillation Imaging  Spectrometers (GIS; G2 and G3). The XRT 
provides a spatial resolution of $\sim 3$ arcmin (Half Power Diameter) 
in the energy range $0.3 - 12$ keV.  The SIS detectors provide excellent 
spectral resolution [$\Delta E/E = 0.02(E/5.9 {\rm keV})^{-0.5}$] over a 
$22 \times 22$ arcmin$^2$ field of view. The GIS detectors provide poorer 
energy resolution [$\Delta E/E = 0.08(E/5.9 {\rm keV})^{-0.5}$] but cover
a larger circular field of view of $\sim 50$ arcmin diameter. 

 For our analysis of the ASCA data we have used the screened event lists from 
the rev1 processing of the data sets available on the GSFC 
ASCA archive (for a detailed description of the rev1 processing 
see the GSFC ASCA Data Reduction Guide, published by GSFC.) 
The ASCA data were reduced using the FTOOLS software (version 3.6) issued 
by GSFC, from within the XSELECT environment (version 1.3). 
Further data-cleaning procedures as recommended in the ASCA Data Reduction 
Guide, including appropriate grade selection, gain  corrections and  manual
screening based on the individual instrument light curves, were followed. 
A summary of the ASCA observations, including the individual
instrument exposure times after all screening procedures were carried
out, is given in Table 1. 

 Spectra were extracted from all four ASCA detectors (except 
for Abell 1795 where the S1 data were lost due to saturation problems 
caused by flickering and `hot' pixels in the CCDs). The spectra were extracted 
from circular regions, centred on the peaks of the X-ray emission. 
For the SIS data, the radii of the regions used were selected to minimize 
the number of chip boundaries crossed (thereby minimizing the systematic 
uncertainties introduced by such crossings) whilst covering as large a 
region of the clusters as possible. 
Data from the regions between the chips were masked out and 
excluded. The final extraction radii for the SIS data are summarized in 
Table 2. The Table also notes the chip modes used for the observations 
(1,2 or 4 chip mode) and the number of chips from which the 
extracted data were drawn. For the GIS data, a fixed extraction radius of 
6 arcmin was adopted. 

 Background subtraction was carried out using the `blank sky' observations 
of high Galactic latitude fields compiled during the 
performance verification stage of the ASCA
mission. The background data were screened and grade selected in the same 
manner as the target observations and the background 
spectra were extracted from the same regions of the detectors as the 
cluster spectra. (We assume that the errors associated with the 
background subtraction are statistical in origin.)
For the SIS data, response matrices were generated 
using the FTOOLS SISRMG software. Where the spectra covered more than one 
chip, response matrices were created for each chip and combined 
to form a counts-weighted mean matrix. For the GIS analysis, the response 
matrices issued by GSFC on 1995 March 6 were used. For both the SIS and GIS data, 
auxiliary response files were 
generated with the ASCAARF software, with effective area calculations appropriate for 
extended sources.

\subsection{The ROSAT Observations}

 The ROSAT observations were carried out between 1991 February and 1994
August. For the more distant systems at redshifts $z \geq 0.03$, the 
observations were made with the High Resolution Imager (HRI), which provides a 
$\sim 5$ arcsec (FWHM) X-ray imaging facility covering a $\sim 40 \times 
40$ arcmin$^2$ field of view (David \etal 1996). For the clusters at lower 
redshifts, the PSPC was used, which has a lower angular resolution 
($\sim 25$ arcsec FWHM) but provides a larger 
($2\times2$ degree$^2$) field of view and enhanced sensitivity to extended, 
low-surface brightness X-ray emission. Using this combination of detectors 
both good spatial resolution and a field of view extending beyond the 
cooling radii in all clusters was obtained. 

 The reduction of the data was carried out using the Starlink 
ASTERIX package. For the HRI data, X-ray images were extracted on
a $2 \times 2$ arcsec$^2$ pixel scale. For the PSPC observations, only
data in the $0.4-2.0$ keV band were used, and the images were extracted 
with a pixel scale of $15 \times 15$ arcsec$^2$. Where more than one 
observation of a source was made, a mosaic was constructed from 
the individual observations. In each case the data were cleaned and
corrected for telescope vignetting and accurate centres were determined 
for the X-ray emission from the clusters. A summary of the ROSAT 
observations and the coordinates of the X-ray centroids is supplied in 
Table 3.

\section{Spectral Analysis of the ASCA data}

\subsection{The basic spectral models}

 The modelling of the X-ray spectra has been carried out using the XSPEC
spectral fitting package (version 9.0; Arnaud 1996). For the SIS data, 
only counts in pulse height analyser (PHA) channels
corresponding to energies between 0.6 and 10.0  \keV~  were included in the 
analysis (the energy range over which the calibration of the SIS 
instruments is best-understood). For the GIS data, only counts in the energy 
range $1.0  - 10.0$ \keV~were used. The spectra were grouped before fitting to 
ensure a minimum of 20 counts per PHA channel, allowing $\chi^2$ statistics 
to be used.

 The spectra have been modelled using the plasma codes of Kaastra \& Mewe
(1993; incorporating the Fe L calculations of Liedhal, Osterheld \&
Goldstein 1995) and the photoelectric absorption models of 
Balucinska-Church \& McCammon (1992). The data from
all four detectors were included in the
analysis, with the fit parameters linked to take the same values across 
the data sets. The exceptions to this were the emission measures of the 
ambient cluster gas in the four detectors which, due to the different 
extraction radii used and residual uncertainties in the flux calibration of 
the instruments, were maintained as independent fit parameters. 

 The spectra were first examined with a series of four, basic spectral 
models: model A consisted of an isothermal plasma 
in collisional equilibrium at the optically-determined
redshift for the cluster, and absorbed by
the nominal Galactic column density (Dickey \& Lockman 1990). 
The free parameters in this model were the temperature ($kT$) and 
metallicity ($Z$) of the plasma (measured relative to
the solar photospheric values of Anders \& Grevesse 1989, with the various 
elements assumed to be present in their solar ratios) 
and the emission measures in each of the four detectors. The second model, 
model B, was the same as model A but with the absorbing column density 
$(N_{\rm H})$ also included as a free parameter in the fits.  The third
model, model C, included an additional component explicitly accounting for 
the emission from the cooling flows in the clusters. The material in the 
cooling flows was assumed to cool at constant pressure from the ambient 
cluster temperature, following the prescription of Johnstone \etal (1992). The 
normalization of the cooling-flow component was parameterized in terms of a 
mass deposition rate (${\dot M_{\rm S}}$), which was a free parameter in the 
fits (linked to take the same value in all four detectors).  The 
metallicity of the cooling gas was assumed to be equal to that of the 
ambient cluster gas. The emission from the cooling flow was also assumed to 
be absorbed by an intrinsic column density, $\Delta N_{\rm H}$, of cold
gas, which was a further free parameter in the fits. The abundances of 
metals in the absorbing material were fixed to their solar values 
(Anders \& Grevesse 1989). 

 The fourth spectral model, model D, was similar to model C except 
that the constant-pressure cooling flow component
was replaced with a second isothermal emission component. Model D constitutes 
a more general model which should normally provide an adequate description 
of the more specific cooling-flow models at the spectral resolution and 
signal-to-noise ratios typical of ASCA observations (see discussion in 
Section 3.4). The temperature and normalization of the second, cooler 
emission component were included as free parameters in the fits 
(although the normalization 
of the second emission component was linked to take the same value in all four
detectors).  The second emission component was again  
assumed to be intrinsically absorbed by a column density, $\Delta N_{\rm H}$, 
of cold gas, which was a further free parameter in  the fits. 
The metallicities of the two emission components were assumed to be equal. 

 Fig. 1 shows the ASCA data and best-fitting spectral models 
for four of the clusters in the sample; Abell 426, the Virgo cluster, the
Centaurus cluster and Abell 2199. For clarity, only the results for the
S1 and G2 detectors are shown. Table 4 summarizes the fit results for 
the clusters, using the four basic spectral models. We note that the S0 
and G3 data for the Centaurus Cluster and the S0 and S1 data for Abell
2597 were found to exhibit small ($\sim 1$ per cent) gain
mis-calibrations, for which corrections were applied.

\subsection{Isothermal cooling flow models}

 The constant pressure cooling flow model (model C) provides
a crude approximation to the X-ray emission spectrum from a multiphase 
cooling flow. Recently, Nulsen (1998) has presented a more 
sophisticated, self-similar  
model for the emission from a cooling flow in which the mean gas 
temperature remains constant with radius {\it i.e.} an `isothermal' cooling 
flow. Although the isothermal cooling-flow model is
rather specific, the assumption of an approximately constant 
mass-weighted temperature over the bulk of the radius in a 
cluster cooling flow is consistent with current constraints on the 
distribution of gravitating matter in the core regions of 
massive clusters from X-ray and gravitational lensing studies (Allen 1998) and 
with previous, spatially-resolved spectroscopic X-ray studies of nearby 
clusters cooling flows with ASCA (\eg Fukazawa \etal 1994; Ohashi \etal 1997; 
Ikebe \etal 1999). The assumption 
of an approximately isothermal mean gas temperature in cooling flows 
also leads to a more plausible implied distribution of initial density inhomogeneities 
in the cluster gas than would be the case if the temperature profiles dropped 
significantly within the core regions (Thomas, Fabian \& Nulsen 1987). 

We have incorporated the Nulsen (1998) isothermal cooling flow model into the XSPEC
code and applied it to the ASCA data. As with the constant pressure cooling 
flow model (model C), the isothermal cooling-flow model introduces only 
two extra free parameters into the fits over and above those present in 
the single-phase spectral model (model B); the mass deposition rate
(${\dot M_{\rm S}}$) and the column density of intrinsic X-ray absorbing 
material ($\Delta N_{\rm H}$). Note, however, that whereas 
in the case of the constant pressure cooling flow model the tabulated
temperatures are the upper temperatures from which the gas cools, for the
isothermal cooling flow models the quoted temperatures are the mean 
temperatures (${\overline T}$ from the work of Nulsen 1998) in the
cooling flows, which are directly related to the gravitational potentials.
(For a comparison of the emission spectra from isothermal and constant 
pressure cooling flow models see Nulsen 1998).

For our initial analysis, we assumed that the slope of 
the mass deposition profile, $\eta =1$ (where ${\dot M} \propto
r^{\eta}$). The mean temperature of the gas in
the isothermal cooling flows was linked to be equal to 
that of the ambient gas at larger radii in the clusters (which, 
as with spectral models A--D, was a free fit parameter). 
The results determined with the isothermal cooling flow model with 
$\eta =1$ (hereafter spectral model E) are listed in Table 4. In most 
cases, the isothermal cooling flow 
model provides at least as good a fit to the ASCA spectra as the 
constant pressure model (model C), although for the Virgo cluster spectral 
model C provides a better fit. We note that for the Virgo 
cluster, the central $\sim 5$ arcmin radius (S0) aperture analysed 
corresponds to a spatial scale of only $\sim 25$ kpc. For this system, 
it seems plausible that the gravitational potential may flatten significantly 
in the region studied, and that the approximation of isothermality in the 
mass-weighted X-ray temperature profile may no longer hold. 

We have also investigated 
the effects on the spectral fits of using other values of $\eta$, in the range 
$0.75-2.5$. These results are shown in Table 5. As discussed in Section 8.3, 
a value of $\eta \sim 1.5$ provides a good match to the image 
deprojection analysis results for a number of the 
clusters studied, with statistically acceptable values spanning the range 
0.75 to 2.5. The effect of increasing the value of $\eta$ in the isothermal 
cooling flow models is always to increase the value of the mass deposition 
rate and to increase the intrinsic absorption required on that component. 
The Virgo cluster shows the largest increase 
in mass deposition rate; a factor of two when increasing $\eta$ from 0.75
to 2.5. Other clusters show more moderate increases of $20-70$ per cent. 
In general, the value of $\chi^2$ also decreases as $\eta$ is increased.
(The decrease in some cases is statistically significant when 
increasing $\eta$ from 0.75 to 2.5.)  Abell 426 is the only cluster
for which a minimum in $\chi^2$ is found. In most cases the reduction in 
$\chi^2$ continues to implausibly high values of $\eta$. We do not find a
significant improvement in $\chi^2$ when increasing $\eta$ from the value
$\eta = 1.0$ used in model E to the value of $\eta = 1.5$, the most likely 
value for several of the clusters determined from the imaging data.

In what follows, we will adopt the results determined with spectral model
E ($\eta = 1$) as our `best' spectral results on the cooling flows, except for 
the Virgo Cluster for which the results determined with the constant pressure 
model (model C) are preferred.

\subsection{The requirement for multiphase models and excess absorption}

The results determined from the basic spectral analysis, summarized in
Table 4, show that (with the exception of Abell 1795) the fits with the 
single-temperature 
models (A and B) are significantly improved 
(\ie a significant reduction in $\chi^2$ is obtained) when the 
absorbing column density acting along the line of sight is included as a free 
parameter in the fits. (For guidance, a reduction of $\Delta
\chi^2 \sim 7$, in a fit with $\nu \sim 1000$ degrees of freedom and a 
reduced chi-squared value $\chi^2_\nu \sim 1.00$, is significant at
approximately the 99 per cent confidence level.) A further highly 
significant reduction
in $\chi^2$ is obtained for all clusters when including either an absorbed 
cooling flow spectrum (models C, E) or an absorbed, cooler isothermal emission
component (model D) in the modelling. (For guidance, improvements in 
$\chi^2$ of $\sim 10$ and 15, with the introduction of 2 (models C, E) and 3 
(model D) extra fit parameters, are significant at approximately the 99 per 
cent confidence level in a fit with $\nu \sim 1000$ and $\chi^2_\nu \sim 
1.00$.) 

 The requirement for the introduction of cooler emission and absorption 
components is also illustrated in Fig. 2, where we show the residuals
(the data divided by the model predictions, in units of $\chi$) 
from fits to the individual SIS spectra at energies above 3~keV, using a 
simple, single temperature model with Galactic absorption (model A). The 
best-fitting models have then been extrapolated to cover the full $0.6-10.0$ 
keV band of the SIS detectors. In all cases,  except the non cooling-flow 
Coma cluster, a clear excess in the residuals is observed at energies 
between $\sim 0.8$ and 3.0 keV. This region of the spectrum is dominated by 
the Fe-L, Mg, Si and 
S line complexes, and the detection of positive residuals due to these 
features provides a clear indication of the presence of gas significantly 
cooler than the ambient cluster temperature.  In addition, the residuals
for a number of the clusters exhibit a clear deficit at low energies ($E
\approxlt 0.8$ keV) providing evidence for excess 
absorption, over and above that due to our own Galaxy, associated with the 
central regions of the clusters.

The temperatures and metallicities determined from the fits to the 
SIS data in the $3-10$ keV energy range are summarized in Table 6. 
The results are in good agreement with those determined from the 
more-sophisticated, multiphase analyses of the full ASCA data sets, 
using spectral models C, D and E.  

 The residual diagram for Abell 426, determined from the $3-10$ keV 
SIS data using the nominal (Dickey \& Lockman 1990) value for the 
Galactic column density ($N_{\rm H} = 1.49 \times 10^{21}$ \apc), shows 
a strong excess of emission below an energy, $E \sim 0.8$ keV (see 
the dotted points in Fig. 2). We suspect this excess to be due to an over-estimation of 
the Galactic column density in the region covered by the ASCA spectra, 
which is significantly smaller than the resolution of the HI
studies (Dickey \& Lockman 1990). In order for the residual diagram for
Abell 426 to appear like those for the other clusters included in this
study, a Galactic column density of $\sim 1.0 \times 10^{21}$\apc~
is required (solid
points in Fig. 2). The ROSAT PSPC X-ray colour ratio study 
of the Perseus cluster by Allen \& Fabian (1997) also suggests a Galactic 
column density in the region of the cluster core of 
$\sim 1.1\times 10^{21}\psqcm$. 
The 100$\mu$m IRAS map of the Perseus cluster is  complex (the system lies 
at a Galactic latitude of $b=-13$ degrees) and allows for a decrease in 
the Galactic column density towards the cluster centre. The result on the 
reduced Galactic column density for Abell 426 is also supported by the 
detailed modelling of the full ASCA data sets with spectral models C, D
and E, which determine values in the range $0.8-1.2 
\times 10^{21}$ \apc. The results on the Galactic column densities for the 
other clusters are in reasonable agreement with the values 
determined from the HI studies (Dickey \& Lockman 1990).

\subsection{A note on two-temperature models}

The results listed in Table 4 show that the two-temperature spectral
model (model D) generally provides a slightly better description of the 
ASCA spectra than the cooling-flow models (C, E). This has sometimes been 
taken to indicate that the cluster gas is distinctly two-phase, with the 
cooler phase being due to the central galaxy (\eg Makishima 1997; Ikebe
\etal 1999). However, the longevity of such a gas phase implied by the 
common occurrence of cooling flows would then require the 
presence of a heat source to carefully balance the radiative cooling. 
The very high cooling rates found in some distant, luminous clusters 
(\eg Allen 2000) are also difficult to explain with such models. 

The fact that the two-temperature models often provide a better
description of the ASCA data is not, a priori, surprising. Model D 
provides a more general fitting parameterization for multi-temperature 
plasmas observed with ASCA, with an extra degree of freedom over the 
cooling flow models. Simulated cooling-flow spectra, constructed at the 
spectral resolution and with count statistics comparable to those of typical 
ASCA observations, are well-described by two-temperature models 
(\eg Allen 2000; see also Buote, Canizares \& Fabian 1999). Where 
spectral model D provides a significantly better 
fit to the observations than models C and E, this need only indicate that 
the simple cooling flow models over-simplify the true emission spectra 
from the central regions of the clusters. In particular, given that the 
systems studied here are the nearest, X-ray brightest cluster cooling
flows, in which strong radial variations in metallicity and intrinsic 
absorption are known exist (Fukazawa \etal 1994, Matsumoto \etal 1996, 
Allen \& Fabian 1997) and in which clear deviations from solar abundance 
ratios are observed (Section 5), the fact that the basic cooling-flow 
models over-simplify the observed situation is to be expected. 

We note that for the very best ASCA data sets, such as the observation of 
the Centaurus Cluster presented here, the data are of sufficient 
quality to demonstrate the presence of more than two thermal components. 
Section 4 discusses the evidence for additional power-law components
in the fits with the two-temperature plasma models. For the Centaurus 
Cluster, the fit with the two-temperature plus power-law model
(also incorporating variable element abundance ratios; see Section 5) is 
further improved by the introduction of a third plasma component. Following 
Allen, Di Matteo \& Fabian (2000) and using only the data from the S1 and G2 
detectors for the fits incorporating the power-law components (which are the 
best calibrated detectors for that particular observation, the 
S0 and G3 data sets exhibiting noticeable gain offsets; Section 3.1), we 
find that the 
introduction of the third plasma component (which is assumed to be absorbed 
by the same intrinsic column density as the 
power-law and the cooler of the two initial plasma components) leads to a 
drop in $\chi^2$ of $\Delta \chi^2 = 37$ for the introduction of 2 additional 
degrees of freedom. The best fit temperatures in this case are 3.25, 
1.51 and 0.64 keV, respectively.

\subsection{Comparison with previous results}

A number of previous works have also presented results 
based on ASCA observations of the central regions of the 
clusters studied here. Fabian \etal (1994b) report results 
from an early analysis of ASCA data for the innermost 
regions of the Perseus (Abell 426), Centaurus and Coma 
clusters and Abell 1795. The regions studied by Fabian \etal (1994b) 
were smaller than those studied here and did not include the 
the entire cooling flows. For this reason the mass deposition 
rates measured by Fabian \etal (1994b) are approximately 
a factor two smaller than those reported here.

Fukazawa \etal (1994) present an analysis of a previous, 
shorter ASCA observation of central regions of the Centaurus 
Cluster (the same data studied by Fabian \etal 1994b). These authors 
reported the discovery of a strong metallicity gradient in the 
cluster, with a central metallicity and temperature in good 
agreement with the results presented here. Ikebe \etal (1999) 
present results from a detailed analysis of the longer, more recent 
observation of the Centaurus Cluster also 
studied in this paper. Their results from fits with 
two-temperature models incorporating variable element abundance ratios 
are in good general agreement with those reported here (Section 5). 

Matsumoto \etal (1996) present an analysis of ASCA observations of 
the central regions of the Virgo Cluster. Their results for the innermost 
regions of the cluster, determined using a two temperature model with 
variable element abundances, are in reasonable agreement with those
reported here, although small systematic differences exist. These
differences are presumably due to the different plasma codes 
used in the two studies. The MEKAL code (Kaastra \& Mewe 1993; Liedhal, 
Osterheld \& Goldstein 1995) employed in this paper contains significant 
improvements over the Raymond \& Smith (1977) code used in the earlier
Matsumoto \etal (1996) work. Our results on the temperature and 
abundance structure in the central regions of the Virgo and 
Centaurus clusters are in good agreement with the 
recent results of Buote \etal (1999). 

Xu \etal (1998) report results from a study of the central 
regions of Abell 1795 with ASCA. Their results, obtained using 
a simple, two-temperature model are in good agreement with 
those reported here from our analysis with spectral model D. 
The temperature in the central regions of the Coma cluster 
determined with spectral model B is in good agreement with the previous 
result of Honda \etal (1996). 

Markevitch \etal (1998, 1999) present detailed results on 
the temperature structure in Abell 496, 1795, 2199 and 2597.
These authors determine mean temperatures, corrected for the effects 
of cooling flows, for these clusters of $4.7 \pm 0.2$, $7.8 \pm 1.0$, 
$4.8 \pm 0.2$ and $4.4^{+0.4}_{-0.7}$ keV, respectively. In general, 
the mean cluster temperatures measured by Markevitch \etal (1998, 1999) 
are slightly hotter than the results for the central regions of the
clusters reported here, using spectral models C and E (the cooling-flow
models).

\section{The evidence for power-law components}

Allen \etal (2000) and Di Matteo \etal (2000) have previously reported the 
detection of hard, power-law emission components
in the ASCA data for the Virgo and Centaurus clusters.
The presence of these components 
is clearly evident in the residual diagrams shown in Fig. 2. 
We have extended this analysis to the other higher luminosity systems
reported here. Starting in each case with the best-fitting two temperature 
model, we have examined the improvements to the fits obtained by 
introducing an additional power-law emission component.\footnote{We use the 
two-temperature model (model D) as the starting point for our analysis of 
the power-law components since it
provides the most flexible fitting parameterization for the cluster 
emission, and therefore the most rigorous test for the presence of additional 
power-law components. The two-temperature model also allows for the inclusion 
of variable element abundance ratios (using the $vmekal$ model in XSPEC), 
which is firmly required by the data for the Virgo and Centaurus clusters 
(Section 5). We note that the inclusion of a power-law component into the 
fits with the cooling flow emission models (models C and E) also leads to 
significant improvements in $\chi^2$ and constraints on the power-law 
components consistent with those listed in Table 7 (although for 
Abell 496 the inferred $2-10$ keV flux is approximately twice as high and the 
$\chi^2$ value $+46$ worse). For the Virgo and Centaurus clusters, 
Abell 1795 and 2199, the mass deposition rates determined with the cooling 
flow-plus-power-law models are in good agreement with the values 
listed in Table 4, although for Abell 426, 496 and 2597, the rates are 
reduced by a factor $\geq 1.6$.}
The photon index and normalization of the power-law components were included 
as free fit parameters. (Note that for the Virgo and Centaurus clusters, 
the best 
fitting two-temperature models incorporating variable element abundances were 
used as the starting points for the analysis; see Section 5).
The power-law components were assumed to be absorbed by the same intrinsic 
column density as the cooler thermal emission components.\footnote{The 
ASCA spectra cannot easily constrain more complex spectral 
models in which the intrinsic column density acting on the power-law
component is included as a separate free parameter, although such 
models are probably more realistic. The intrinsic column densities 
associated with the power-law components measured with such models 
are, however, likely to be larger than the values listed in Table 7).} 
The results from the fits are summarized in Table 7. 

The results for Abell 426 and 2597 suggest the presence of relatively
luminous power-law components ($L_{\rm X,1-10} = 4.0^{+0.3}_{-0.4}\times
10^{44}$ and $2.0^{+1.1}_{-0.9}\times 10^{44}$ \ergps, respectively) with
photon indices ($\Gamma = 2.05^{+0.05}_{-0.05}$ and
$1.80^{+0.42}_{-0.66}$) consistent with the typical values determined 
for Seyfert nuclei (\eg Nandra \etal 1997). However, the flux associated with 
the power-law component in Abell 426 ($F_{\rm X,2-10} = 
1.80^{+0.12}_{-0.14}\times 10^{-10}$\ergpcmsqps) 
is approximately an order of magnitude larger than the 
nuclear point-source flux inferred from ROSAT HRI observations of the 
dominant cluster galaxy, NGC 1275, made in 1991 
February and 1994 August (using the same spectral model). This implies 
either the presence of significant intrinsic absorption associated with the 
nucleus of NGC 1275 (see also Section 7.3), that the power-law 
emission originates (at least in part) from some other source, 
or some other deficiency with the spectral model. 

The results for Abell 496, 1795 and 2199 are more analogous to those 
those previously reported for the Virgo and Centaurus clusters. For 
Abell 496,  we measure a photon index, $\Gamma =1.44^{+0.44}_{-0.38}$ and 
an intrinsic $1-10$ keV luminosity associated with the power-law component of 
$3.29^{+1.49}_{-0.84}\times 10^{43}$ \ergps, which is $\sim 80$ times larger 
than the value for M87. The results for Abell 1795 and 2199 indicate 
marginal improvements to the fits with the introduction of hard, power-law 
components, with intrinsic $1-10$ keV luminosities of $\sim 6$ and 
$3.5 \times 10^{43}$ \ergps, respectively. In all three cases the ASCA data 
allow for significant intrinsic absorption acting on power-law emission, 
which is consistent with the fact that a central point sources 
appear to contribute only weakly in the ROSAT band. 

Comparing the results in Table 7 with those in Table 4, we see that the
most noticeable effect on the best-fit parameter values measured with
spectral model D, on the introduction of the power-law component, is
to decrease the temperature of the hotter thermal component to a value in
better agreement with the results obtained using the cooling flow
models, C and E. For Abell 426, the mean metallicity is also increased
from $\sim 0.5 Z_{\odot}$ to $0.8Z_{\odot}$, which is consistent with the 
presence of a metallicity gradient in the cluster, albeit with the 
caveats given above. (We note that the relatively poor $\chi^2$ value for 
Abell 426 is primarily caused by residuals in the S0 data. 
If we repeat our analysis of Abell 426 incorporating the power-law 
component but excluding the S0 data, we obtain similar best-fit parameters 
and $\chi^2 = 1570$ for 1398 degrees of freedom.)

%In summary, the clusters studied exhibit a range of properties in 
%their power-law emission. (The results for the Coma cluster are presented 
%in Section 5.1 below). Unambiguous identification of the origin of these 
%components should be possible in the near future using observations made with 
%the Chandra Observatory and XMM-Newton.

\subsection{Power-law emission from the Coma Cluster}

The spectral fit to the Coma Cluster with spectral model D was not 
improved by the introduction of an additional power-law component. We
are able to place a 90 per cent confidence upper limit on the flux of 
any power-law component (with a canonical AGN photon index
$\Gamma=2.0$) incorporated into model D of $F_{\rm X,2-10} <2.2 
\times 10^{-12}$\ergpcmsqps.  However, the introduction of a power-law 
component was found to significantly improve the fit obtained with the 
more simple, single-temperature spectral model B 
($\Delta \chi^2 = 9.0$, giving $\chi^2 = 1265$ for 1225 degrees of freedom). 
The best-fit photon index and normalization for the power-law component,
determined using the single-temperature plus power-law model, are 
$\Gamma = 3.3^{+0.9}_{-1.6}$ and $A_1 = 3.1^{+4.0}_{-1.7} \times 10^{-3}$ 
photon keV$^{-1}$cm$^{-2}$s$^{-1}$. The mean cluster temperature measured
with this model is $8.5^{+0.6}_{-0.5}$ keV. The 90 per cent confidence
upper limits on the flux associated with the power-law component in the 
$2-10$ and $20-80$ keV bands are $1.9 \times 10^{-11}$ and 
$3.1 \times 10^{-11}$ \ergpcmsqps, respectively.

Fusco-Femiano \etal (1999) report the detection of hard, 
non-thermal X-ray emission from the the Coma Cluster using observations made 
with the Phoswhich Detection System (PDS) and High-Pressure Gas Scintillation 
Proportional Counter (HPGSPC) on the BeppoSAX
satellite. (The PDS and HPGSPC detectors have fields of view $\sim 1.3$ 
(hexagonal) and 1.0 degree across, respectively.) From a joint fit to the 
PDS and HPGSPC data, which together cover the $4.5-80$ keV range, using a 
single-temperature plus power-law model, these authors determine a mean 
temperature for the ICM of $8.5^{+0.6}_{-0.5}$ keV, and a best-fitting 
power-law photon index of $\Gamma = 1.7-3$. The $20-80$ keV flux associated 
with the power-law component (measured from the PDS data alone) is $\sim 2.2 
\times 10^{-11}$ \ergpcmsqps. These parameters are consistent 
with those determined from the ASCA data, using the same simple spectral 
model. 

Rephaeli, Gruber \& Blanco (1999) also report observations of the central
$\sim 1$ degree region of the Coma Cluster using the Proportional Counter 
Array (PCA) and High-Energy X-ray Timing 
Experiment (HEXTE) on the Rossi X-ray Timing Explorer (RXTE). Rephaeli 
\etal (1999) show that the RXTE data in the $3-100$ keV range can be 
equally well-modelled using either a two-temperature plasma model (with 
temperatures of $8.2^{+2.0}_{-1.2}$ and $4.7^{+1.6}_{-3.0}$ keV, 
respectively; with the cooler component accounting for $\sim 20$ per cent of 
the total flux) or a single-temperature plus power-law model. In this second 
case (favoured by the authors), the RXTE data measure a mean cluster 
temperature of $7.51 \pm 0.18$ keV and a photon index and normalization (at 
1 keV) associated with the power-law component of $\Gamma = 2.35 \pm 0.45$ 
and $A_1 = 1.9 \pm 0.4 \times 10^{-2}$ photon keV$^{-1}$cm$^{-2}$s$^{-1}$, 
respectively. This normalization is somewhat higher than that measured 
from the ASCA data, which may indicate that the source of the power-law
emission in this cluster is more extended than the regions included in 
the ASCA analysis.

\section{The abundances of elements in the Virgo and Centaurus clusters}

\subsection{The observed abundances ratios}

The metallicities and abundance ratios of the various elements in the
ICM place firm constraints on the enrichment history of the cluster
gas and the relative contributions from different types of supernovae.
Mushotzky \etal (1996) reported results from an ASCA study of the
outer regions of four nearby clusters (Abell 1060, AWM7 and
the same observations of Abell 496 and 2199 presented here) and
concluded, based on the mean metallicities of O, Ne, Si, S and Fe
determined from their work, that most of the enrichment in the outer
regions of clusters is likely to be due to Type II supernovae. However, 
these authors warned
that a significant contribution from Type Ia supernovae could not be
ruled out. Subsequent analyses of clusters and groups (Finoguenov \& 
Ponman 1999; Dupke \& White 2000a,b) have shown that their central regions
often exhibit enhanced contributions from Type Ia supernovae.

The high-quality ASCA spectra for the Centaurus and Virgo clusters 
analysed here provide an excellent opportunity to re-examine 
the issue of individual element abundances in the ICM in nearby clusters. 
(The SIS data for the Centaurus and Virgo clusters provide the most 
significant and reliable constraints of the eight systems examined here and 
we therefore concentrate our discussion on these two clusters). Using the S0 
and S1 data for the Virgo cluster, and the S1 data for the Centaurus Cluster 
(the data sets which are best-calibrated and most sensitive for the present 
task), and starting in each case with the best fitting two-temperature model 
(Model D; which has the abundances of all other elements 
linked to be equal to that of iron), we have systematically determined the
improvements to the fits obtained by allowing the metallicity 
of each element, in turn, to be included as a free parameter in the fit. 
Having identified the element giving the most significant improvement, 
the metallicity of that element was maintained as a free parameter, and the 
process repeated to determine the element providing the next most
significant improvement. This was continued until no further 
significant improvement (at $>95$ per cent confidence, as 
determined using an F-test) was obtained by allowing other elements to
fit freely.

For the Virgo cluster, the most significant improvement to the fit to the
joint S0 and S1 data with spectral model D was obtained by including the 
abundance of Mg as a free fit parameter ($\chi^2$ drops from 757.7 to
570.4). This was followed by Si (giving a further drop in $\chi^2$ to 511.5), 
Na ($\chi^2=488.8$) and S ($\chi^2=468.9$). At this point including the 
abundances of further elements as free parameters did not significantly 
improve the fit.  

For the Centaurus cluster, the most significant improvement in the fit to
the S1 data was also initially obtained by allowing the abundance of Mg 
to fit freely ($\chi^2$ drops from 478.4 to 409.5). This was followed, in 
turn, by O ($\chi^2= 369.3$), Si ($\chi^2=339.0$) and S ($\chi^2=322.3$). 

The results on the abundances of the individual elements in the
central regions of the Centaurus and Virgo clusters are summarized in 
Table 8. The two-dimensional confidence contours on the abundances of Mg, Si, 
S and Fe in the Virgo Cluster are shown in Fig. 3. The fits with the 
free-fitting element abundances provide highly significant improvements in 
the goodness-of-fit with respect to the values determined with the basic 
two-temperature model (model D). However, the 
best-fitting temperatures ($kT$ and $kT_2$), Galactic column densities 
($N_{\rm H}$) and intrinsic column densities ($\Delta N_{\rm H}$) are 
in good, general agreement with those listed in Table 4. The measured 
Galactic column densities are also in good agreement with the values
determined from HI studies (Dickey \& Lockman 1990). We note that the
results on the element abundances are not significantly modified by the 
inclusion of the power-law components discussed in Section 4.

Our results on the [Si/Fe] and [S/Fe] ratios in the central regions 
of the Virgo cluster are in good agreement with the previous 
measurements of Matsumoto \etal (1996; although the absolute element 
abundances differ, presumably due to the different plasma models used in 
the two studies; Section 3.5). Our results on the Si, S and Fe abundances in the 
central regions of the Centaurus cluster are similar to those reported 
by Ikebe \etal (1999). The measured Si, S and Fe abundances in both the 
Virgo and Centaurus clusters are in good agreement with the results 
of Buote \etal (1999).

We caution against over-interpretation of the result on the low 
oxygen abundance for the Centaurus Cluster. This result is primarily 
due to features at the lowest energies in the S1 spectrum, where systematic 
errors and complexities due to absorption in the cluster are at a maximum.
Similarly, the result on the high Na abundance in the Virgo cluster 
($11\pm3$ times the solar value) is due to a line-like emission feature at 
$E \sim 1.25$keV in the spectrum, where residual uncertainties in the modelling 
of the Fe-L emission complex remain. 

The element abundances most reliably determined from our data are
those of Fe, Mg, Si and S. For the Virgo cluster we determine
abundance ratios (using the solar photospheric abundance scale of
Anders \& Grevesse 1989) of [Mg/Fe]=-0.45 (with values in the range
-0.67 to -0.28 being consistent with the joint 90 per cent confidence
limits on the individual abundance measurements), [Si/Fe]=0.18
(0.12-0.24) and [S/Fe]=0.10 (0.03-0.17).  For the Centaurus cluster
we determine [Mg/Fe]=-0.49 (with a 90 per cent confidence range of
-0.82 to -0.25), [Si/Fe]=0.14 (0.07-0.22) and [S/Fe]=0.13
(0.05-0.22). Thus, we find the abundance ratios for the Virgo and
Centaurus clusters to be both well constrained and in excellent
agreement with each other. 

Our results on the abundance ratios for [Mg/Fe], [Si/Fe] and [S/Fe]
in the inner regions of the Centaurus and Virgo clusters differ from
those of Mushotzky \etal (1996) from their study of the outer regions
of other nearby clusters observed with ASCA. Mushotzky \etal (1996) 
determined mean abundance ratios for [Mg/Fe]=0.07, [Si/Fe]=0.31 and 
[S/Fe]=-0.11 from their sample, using the same solar photospheric 
abundance scale. The differences between these results are consistent
with the changes in the relative abundances with radius inferred to be present 
in other clusters (\eg Dupke \& White 2000b).

\subsection{Comparison with supernovae enrichment models}

Recently, a number of studies (Ishimaru \& Arimoto 1997; Gibson,
Lowenstein \& Mushotzky 1997; Nagataki \& Sato 1998; Dupke \& White
2000a) have presented detailed predictions for the $\alpha-$element
abundances in the ICM, as a function of the fraction of the iron
enrichment due to type Ia supernovae, using a variety of theoretical
supernovae models. Scaling the observed abundance ratios for the central 
regions of the Centaurus and Virgo clusters to the meteoric abundance scale 
of Anders \& Grevesse (1989), as used in the theoretical calculations, we 
determine mean, un-weighted abundance ratios of [Mg/Fe] $\sim -0.63$, 
[Si/Fe] $\sim 0.00$ and [S/Fe]$\sim 0.02$. Thus, the [Si/Fe] and 
[S/Fe] ratios in the central regions of the Centaurus and Virgo clusters 
are in good agreement with the solar meteoric values. In contrast, the 
observed [Mg/Fe] ratio indicates that magnesium is $\sim 4$ times
under-abundant with respect to the meteoric value. 

Comparing these results with the
predictions of Nagataki \& Sato (1998), our observed [Si/Fe] ratio
implies a mass fraction of the iron enrichment due to type Ia
supernovae, $M_{\rm Fe,SNIa}/M_{\rm Fe,total}$, in the range $M_{\rm
Fe,SNIa}/M_{\rm Fe,total} \sim 0.6-0.9$ (where the limits cover the
full range of supernovae models studied by those authors). A similar
comparison of the observed [Si/Fe] result with the range of models
discussed by Gibson \etal (1997) also requires $M_{\rm Fe,SNIa}/M_{\rm
Fe,total} \sim 0.6-0.8$.  The observed, mean [Mg/Fe] ratio 
favours a mass fraction due to type Ia supernovae of $M_{\rm
Fe,SNIa}/M_{\rm Fe,total} \approxgt 0.7$ (Gibson \etal 1997).

A comparison of the observed [S/Fe] ratio ([S/Fe]$\sim 0.02$) with the 
theoretical models of Nagataki \& Sato (1998) requires 
$M_{\rm Fe,SNIa}/M_{\rm Fe,total} =0.4-0.8$, for spherical type II supernovae 
explosions. However, the lower sulphur yields associated with aspherical 
type II explosion models imply somewhat lower $M_{\rm Fe,SNIa}/M_{\rm
Fe,total}$ values, depending on the degree of asphericity and the precise 
SNIa enrichment model used. In particular, the observed [S/Fe] ratio appears 
inconsistent with enrichment models incorporating both a high degree of 
asphericity in the SNeII explosions and simple deflagration of the type
Ia supernovae (Nagataki \& Sato 1998). Comparing the observed [S/Fe] value 
with Gibson \etal (1997) indicates $M_{\rm Fe,SNIa}/M_{\rm Fe,total}
\approxlt 0.7$. In contrast to previous results (Mushotzky \etal 1996; 
Gibson \etal 1997), 
our observations do not indicate a severe deficiency of sulphur with
respect to the other $\alpha$-elements, when compared with the theoretical 
predictions (at least for models incorporating relatively low degree of 
asphericity in the type-II supernovae). 

In summary, the observed $\alpha-$element abundance ratios in the
central regions of the Centaurus and Virgo clusters favour a high
mass fraction of the observed iron abundance being due to type Ia
supernovae, with a value $M_{\rm Fe,SNIa}/M_{\rm Fe,total} \sim
0.7-0.8$ providing consistent results.

\subsection{Type Ia supernovae and the formation of metallicity gradients}

Spatially resolved spectroscopy of the Centaurus and Virgo clusters
has revealed the presence of strong metallicity gradients in the
central regions of these systems (Fukazawa \etal 1994, Matsumoto \etal
1996).  Similar gradients have also been reported in a number of other
nearby cooling flows \eg Abell 496 (Hatsukade \etal 1997) and AWM7
(Ezawa \etal 1997).  Allen \& Fabian (1998) showed that the mean
emission weighted metallicity of luminous cooling-flow clusters is
approximately twice that of non cooling-flow systems, which is most
plausibly understood as being due to the presence of metallicity
gradients in the cooling flow systems.  In contrast, ASCA observations
of nearby, non-cooling flow clusters and clusters with relatively
small cooling flows \eg the Coma Cluster (Watanabe \etal 1997), Abell
1060 (Tamura \etal 1996) and the Ophiuchus Cluster (Matsuzawa \etal
1996), show little or no metallicity gradients.

The results on the individual element abundances in the central
regions the Centaurus and Virgo Clusters shows that, where the
metallicity gradients exist, the enrichment of the ICM is primarily
due to type-Ia supernovae. The mechanism by which metallicity
gradients form in cooling-flow clusters remains unclear, but may be due
to one or more of galaxy stripping, continuous enrichment by
supernovae within the central galaxy, or a Type Ia
supernova wind from the central galaxy (Dupke \& White 2000b).  The
absence of metallicity gradients in non-cooling flow systems appears
related to recent or ongoing subcluster merger events in these
systems, which will disrupt and mix the central high metallicity gas
with the surrounding lower-metallicity material (Allen \& Fabian
1998).  The presence of a metallicity gradient then appears a standard
property of regular, relaxed clusters which also, invariably, contain
cooling flows.

The presence of a cooling flow may act either to enhance or reduce 
any metallicity gradient present in a cluster, with the details depending on 
how well the metals are mixed with the cooling gas. If the mixing is
good then as the metals are injected or deposited into the cluster 
environment they will primarily enrich the least dense X-ray gas at any
radius, since this gas will have the largest volume filling factor. 
This less-dense material will flow furthest in towards the cluster center
in the cooling flow, whereas the denser, less-enriched material will be
deposited at larger radii.  Reisenegger, Miralda-Escud\'e \&  Waxman
(1996) discuss such a model for the production of the metallicity 
gradient observed in the Centaurus Cluster, wherein the observed enrichment is primarily due to type Ia 
supernovae within the central cluster galaxy. A prediction of their model is 
that metallicity gradients should be most obvious in relatively 
low X-ray luminosity clusters, with higher ratios for their stellar/X-ray gas mass 
contents, which appears consistent with the detections of strong 
gradients in the Centaurus and Virgo clusters, Abell 496 and AWM7. If the mixing of metals
with the ICM is poor, however, localized regions of high metallicity material 
may cool faster than the surrounding gas, leading to the gradual reduction
of any metallicity gradient apparent in the X-ray data, and to a decrease in 
the mean, emission-weighted metallicity of the cluster.  Further 
discrimination on the origin of metallicity gradients in cooling
flow clusters will be possible using observations made with the Chandra 
Observatory and XMM-Newton.

\section{Deprojection analysis of the ROSAT images}

The analysis of the ROSAT images has been carried out using an 
extensively updated version of the deprojection code of Fabian 
\etal (1981; see also White \etal 1997). Azimuthally-averaged 
X-ray surface brightness profiles 
were determined for each cluster from the ROSAT images. 
These profiles were background-subtracted, 
corrected for telescope vignetting and re-binned 
to provide sufficient counts in each annulus for the deprojection
analysis to be extended beyond the cooling radii.

 With the X-ray surface brightness profiles as the primary input, 
and under assumptions of spherical symmetry and hydrostatic equilibrium, 
the deprojection technique can be used 
to study the basic properties of the intracluster gas 
(temperature, density, pressure, cooling rate) as a 
function of radius. The deprojection code uses a monte-carlo method to 
determine the statistical uncertainties on the results and 
incorporates the appropriate HRI and PSPC spectral response matrices issued 
by GSFC. The cluster metallicities were fixed at the values 
determined from the spectral analysis and the 
absorbing column densities were initially set to their respective Galactic 
values. 

The deprojection code requires the total 
mass profiles for the clusters, which define the
pressure profiles, to be specified. We have iteratively determined 
the mass profiles that result in deprojected
temperature profiles (which approximate the mass-weighted
temperature profiles in the clusters) that are isothermal within the regions 
probed by the ROSAT data and which are consistent with the best-fit 
temperatures determined from the ASCA spectra with spectral model E (or 
spectral models C and D for the Virgo and Coma clusters, respectively;  
Section 3). The validity of the assumption of 
isothermal mass-weighted temperature profiles in the 
cluster cores is discussed by Allen (1998). The use of approximately 
isothermal deprojected temperature profiles also allows for a direct 
comparison between the mass deposition rates from the cooling flows 
determined from image deprojection and spectral analyses (see Section 7.1).

 The dominant components of the cluster mass profiles have been 
parameterized using the model for an isothermal sphere given in
Equation 4-125 of Binney \& Tremaine 1987) with 
adjustable core radii, $r_{\rm c}$, and velocity dispersions, $\sigma$. 
The core radii were adjusted until the deviations from isothermality in
the deprojected temperature profiles were minimized. The velocity 
dispersions were then adjusted  until the temperatures determined from the 
deprojection code came into agreement with the values measured from the
spectral analyses. An initial estimate for the pressure in the outermost 
radial bin used in the analysis is also required by the deprojection code, 
although the  uncertainties in these values do not significantly affect 
the results. 

For several of the clusters, these simple single-component mass models 
could not adequately satisfy the requirement for isothermal 
deprojected temperature profiles, particularly in the central regions of the 
clusters. However, for these systems it was also found that   
the introduction of a second `linear' mass component (providing a constant 
contribution to the total mass per kpc radius {\it cf} Thomas \etal 1987) 
significantly improved the fits. These linear components (which may 
be intuitively associated with the central dominant galaxies in the clusters) 
were truncated at outer radii of 20 kpc. Their normalizations  
were adjusted iteratively (together with the $r_{\rm c}$ and $\sigma$ values) 
in order to obtain the best match to the required isothermal temperature profiles.

 The mass distributions determined from the deprojection analysis 
are summarized in columns $3-5$ of Table 9.  The errors on the velocity 
dispersions are the range of values that result in isothermal 
deprojected temperature profiles 
that are consistent, at the 90 per cent confidence limit, with the 
spectrally-determined temperatures. These results may be used to provide 
direct comparisons with independent mass constraints from dynamical studies. 
We note that the mean core radius for the mass distributions in the cooling 
flow clusters of $48\pm15$ kpc is in good agreement with the value of 
$\sim 50$ kpc determined from the combined X-ray and gravitational lensing 
study of more distant, luminous lensing clusters by Allen (1998).

\section{The properties of the cooling flows}

In the following Section, we examine the basic properties of 
the cooling flows in the clusters. We compare the mass deposition 
rates determined from the spectral and image 
deprojection methods. We examine the evidence for intrinsic 
X-ray absorption in the clusters and compare our results with 
those from previous studies. Finally, we report detections of 
extended infrared emission, spatially coincident with the cooling 
flows, in several of the systems studied. The observed infrared 
fluxes and flux limits are shown to be in good agreement with the 
predicted values due to reprocessed X-ray emission from the 
cooling flows. 

\subsection{The mass deposition rates }

The spectral and image deprojection analyses presented 
in Sections 3 and 6 provide essentially independent estimates of the mass 
deposition rates in the clusters. A comparison of the results obtained from 
these analyses therefore provides a firm test of the validity of the 
cooling flow model. 

The deprojection method describes the X-ray emission from 
a cluster as arising from a series of concentric spherical shells. 
The luminosity in a particular shell, $j$, may be written as the sum of 
four components (Arnaud 1988).

\begin{equation}
L_j = \Delta{\dot M}_jH_j + \Delta{\dot M}_j \Delta \Phi_j +
\left[\sum_{i=1}^{j-1} {\Delta{\dot M}_i (\Delta \Phi_j + \Delta H_j
})\right],
\label{swa_eq1}
\end{equation}

\noindent where $\Delta{\dot M}_j$ is the mass deposited in shell $j$,
$H_j$ is the enthalpy of the gas in shell $j$, and $\Delta \Phi_j$ is
the gravitational energy released in crossing that shell. $\sum_{i=1}^{j-1}
{\Delta{\dot M}_i }$ is the mass flow rate through shell $j$, and $\Delta
H_j$ the change of enthalpy of the gas as it moves through that shell.
The first term in equation \ref{swa_eq1} thus accounts for the enthalpy of
the gas deposited in shell $j$. The second term is the gravitational work
done on the gas deposited in shell $j$. The third and fourth terms
respectively account for the gravitational work done on material flowing 
through shell $j$ to interior radii, and the enthalpy released by that 
material as it passes  through the shell.

In any particular shell, the densest material in the cooling flow is assumed 
to cool out and be deposited. Since the cooling time of this 
material will be short compared to the flow time, the cooling
can be assumed to take place at a fixed radius. Thus, the luminosity 
contributed by the first term in equation 1 should have a spectrum 
appropriate for gas cooling at constant pressure from the ambient 
cluster temperature \ie the same spectrum as the cooling flow component 
in spectral model C (Section 3).
For the bulk of the material continuing to flow inwards towards the
cluster centre, the cooling via X-ray emission is assumed to be offset by 
the gravitational work done on the gas as it moves inwards. The emission 
accounted for in the second and third terms of equation 1 should therefore 
have a spectrum that can be approximated by an isothermal plasma at the 
appropriate ambient temperature for the cluster \ie the spectrum of 
the isothermal emission component also included in model C.  Since the 
mass-weighted 
temperature profiles in the clusters are assumed to remain approximately 
isothermal with radius, the luminosity contributed by the fourth term 
of equation \ref{swa_eq1} should be relatively small. Overall, the integrated 
emission from the cooling flow can be expected to have a spectrum 
similar to that described by spectral model E. (In general, the results on 
the mass deposition rates determined with spectral models C and E show good 
agreement; Table 4.)

 The mass deposition rates determined from the deprojection analysis
(${\dot M_{\rm I}}$; Table 9) are the mass inflow rates measured at the 
cooling radii in the clusters {\it i.e.} the radii at which the cooling time 
of the cluster gas first exceeds a Hubble time ($1.3 \times 10^{10}$ yr). 
Thus, if the cooling flow model is correct, and if the cooling flows have 
existed for a significant fraction of a Hubble time, the mass deposition 
rates determined from the deprojection analysis should be comparable to 
the values measured independently from the spectral data.

 Fig. 4(a) shows the mass deposition rates determined from 
the spectral analysis (${\dot M_{\rm S}}$) versus the results 
from the deprojection study (${\dot M_{\rm I}}$). In all 
cases, except the Virgo Cluster, we have used the 
spectral results determined with model E as the ordinate, since this is the 
more physical of the two cooling flow models and typically provides at least 
as good a fit to the data as the constant pressure approximation. 

 The results plotted in Fig. 4(a)
exhibit an approximately linear correlation. A fit to the data
with a power-law model of the form ${\dot M_{\rm S}} = P {\dot 
M_{\rm I}^Q}$, using the Akritas \& Bershady (1996) bisector
modification of the ordinary least-squares statistic (which accounts 
for the possibility of intrinsic scatter in the data)  gives a 
best-fitting slope, $Q=1.15\pm 0.33$ and a normalization, 
$P= 0.36\pm 0.65$ (where the errors are the standard deviations 
determined by bootstrap re-sampling). 

The spectral analysis presented in Section 3 also indicates that the 
cooling gas is intrinsically absorbed by equivalent hydrogen column 
densities of, typically, a few $10^{21}$ \apc. In principle, therefore, 
the mass deposition rates determined from the deprojection analysis should 
also be corrected for the effects of this absorbing material. 
Such corrections, however, depend upon both the spatial 
distribution and nature of the absorbing material, which at present 
remain uncertain. Nevertheless, under the specific 
assumption that the intrinsic absorption is due to cold gas which lies in a 
uniform screen in front of the cooling flows, the effects of 
intrinsic absorption on the deprojection results may be estimated.

The corrections for intrinsic absorption due to a uniform foreground 
screen of cold gas have been carried out by re-running the
deprojection code with the absorbing column densities set to the total 
values determined with spectral model E (model C for the Virgo Cluster). 
Since the intrinsic column densities measured with model E 
are redshifted quantities, we have set the total column densities
used in the revised deprojection analysis, which assumes zero redshift 
for the absorber, to values $N_{\rm H} + \Delta N_{\rm H}/(1+z)^3$, where
$N_{\rm H}$ is the Galactic column density.) 
The absorption-corrected mass deposition rates (${\dot M_{\rm C}}$) 
so determined are summarized in Table 10. Fig. 4(b) compares the 
results on the mass deposition rates from the spectral analysis with the 
absorption-corrected deprojection values. 
A fit to the data with a power-law model of the form ${\dot M_{\rm S}} = 
P {\dot M_{\rm C}^Q}$ gives a slope, $Q=1.09\pm 0.32$ and a
normalization, $P=0.18\pm0.37$ (where the errors are the
standard deviations determined by bootstrap re-sampling).

We conclude that the results on the mass deposition rates 
determined from the spectral and imaging methods 
exhibit an approximately linear correlation and marginal agreement in 
normalization in the case where no corrections for 
intrinsic absorption on the imaging results are made. 
The application of simple correction factors, appropriate for 
absorption by a uniform screen of cold gas, lead to the 
imaging results exceeding the spectral values by 
typical factors of $\sim 2-4$, although the details of these corrections 
remain highly uncertain at present. As we shall discuss in 
Section 8, an improved agreement between the imaging and 
spectral results on the mass deposition rates is 
obtained once the different ages of the cooling flows 
are also accounted for in the analysis.

\subsection{Intrinsic X-ray absorption}

 The negative residuals at low energies ($E<0.8$keV) in Fig. 2 
provide a clear indication of the presence of 
excess absorption, over and above the nominal Galactic values 
(Dickey \& Lockman 1990) along the lines of sight to the clusters. The 
analyses of the full ASCA data sets with spectral models B, C, D and E also 
indicate the presence of intrinsic X-ray absorbing material in the 
clusters. However, the precise column density measurements, summarized in 
Table 11, are sensitive to the spectral model used. 

 Using the simple isothermal model with free-fitting absorption (model B),
we measure a mean excess column density for the seven cooling-flow 
clusters of $2.4 \pm 2.8 \times 10^{20}$ \apc. We note that the value 
measured for the Perseus Cluster (Abell 426) is negative, supporting the 
conclusion drawn in Section 3.3 that the nominal Galactic column density 
determined from HI studies ($N_{\rm H} = 1.49 \times 10^{21}$ \apc; 
Dickey \& Lockman 1990) overestimates the true column density along the line 
of sight to the cluster core. Interestingly, the excess column density 
measured for the Coma cluster ($1.5 \pm 0.5 \times 10^{20}$ \apc) 
is consistent with the mean value determined for the cooling-flow 
clusters, using the same spectral model. Our results on the excess 
column densities measured with spectral model B may be compared to 
the 90 per cent confidence upper limits on the size of any systematic 
excess column density present in the S0 and S1 detectors due to calibration 
errors of $<8 \times 10^{19}$ and $<2 
\times 10^{20}$ \apc, respectively (Orr \etal 1988; these values are valid 
for observations made before mid 1996. See also Dotani \etal 1996).

The results on the intrinsic column densities for the cooling-flow clusters 
are quite different, however, when the more-sophisticated multiphase spectral
models (C, D and E) are used. As discussed in Section 3, models C, D and E
generally provide a significantly better fit to the ASCA data for the 
cooling-flow clusters than the simple single-phase models (A and B). Using 
spectral models C and E, which are our preferred models in that they provide a 
consistent description for the spectral and imaging X-ray data for the 
clusters (Sections 7.1, 8.3), we measure a mean intrinsic column density 
acting on the cooling-flow components of $3-4 \times 10^{21}$ \apc. Using the 
two-temperature model (model D) we determine a similar mean intrinsic column 
density. Thus, the typical intrinsic column 
densities measured with the multiphase spectral models are 
approximately an order of magnitude larger than those determined using the 
single-phase model B. These results are in good agreement with those presented 
by Allen (2000) from an analysis of larger sample of more luminous, distant 
cooling flow clusters. 

 We have also measured the intrinsic column densities in the 
clusters using one further spectral model, which we refer to as 
model C'. Model C' is identical to model C except that it assumes that
the intrinsic absorption acts on the entire cluster spectrum, rather than
just the cooling gas, and that the absorbing material lies at zero redshift. 
Model C' has been used in a number of previous studies (\eg White \etal
1991; Allen \& Fabian 1997) and is included here for comparison purposes. 
The results obtained for the present sample of clusters using model C'
are also listed in Table 11. The mean excess column density for the
cooling-flow clusters determined with this model is $4.9 \pm 4.5 
\times 10^{20}$ \apc.

 Allen \& Fabian (1997) presented results from an X-ray colour 
deprojection study of 18 clusters observed with the ROSAT PSPC, from which
they determined intrinsic column densities across the central 30 
arcsec (radius) regions  of Abell 426, 496, 1795, 2199 and 2597 
(using spectral model C') of $0.50\pm0.55$, $2.57\pm0.62$, 
$2.20 \pm 0.20$, $2.67\pm0.25$  and $2.36 \pm 0.58 \times 10^{20}$ \apc, 
respectively. The results for Abell 496, 1795 and 2199 are in reasonable 
agreement with those presented here, although our value for Abell 2597 is 
$\sim 3$ times larger than the Allen \& Fabian (1997) result. We note, 
however, that the ASCA observations of Abell 2597 were carried out in 1996 
December and are likely to have been affected by radiation damage to the 
SIS detectors, which is known to lead to overestimates of the measured
column densities in observations made after (approximately) late 1995 
(Hwang \etal 1999). The Allen \& Fabian (1997) result for Abell 426 was 
determined using a lower value for the Galactic column density of 
$1.37\times10^{21}$\apc 
(Stark \etal 1992). If we also adopt this lower value, our ASCA measurement for
the intrinsic column density in this cluster, using model C', rises to
$0.10^{+0.50}_{-0.40} \times 10^{20}$, consistent with the previous 
PSPC result. 

 Six of the cooling-flow clusters included in the present work were also 
studied by White \etal (1991), using Einstein Observatory SSS data. 
The analysis of the SSS observations (which covered central, three arcmin 
radius circular apertures in the clusters) was also carried out using 
spectral model C'. White \etal (1991) presented measurements of the 
intrinsic column densities in Abell 426, 496, 1795, 2199 and the Virgo cluster 
of $13\pm2$, $20^{+4}_{-2}$, $8\pm3$, $14\pm2$ and $15\pm2 \times 10^{20}$ 
\apc, respectively. (The SSS observation of the Centaurus cluster was offset 
from the centroid listed in Table 3 by $\sim 4$ arcmin, prohibiting any 
direct comparison with the present work.) The intrinsic column densities 
measured by White \etal (1991) are significantly larger (by factors ranging 
from two to $\approxgt$ an order of magnitude) than the values determined 
from the ASCA and ROSAT data. The ASCA result for the non-cooling flow Coma 
cluster, measured with spectral model B, of $\Delta N_{\rm H} = 1.5 \pm 0.5 
\times 10^{20}$ \apc~is consistent with the White \etal (1991)
limit of $\Delta N_{\rm H} < 5 \times 10^{20}$ \apc. 

 Finally, we have examined the constraints that may be placed on the 
covering fraction, $f$, of the intrinsic X-ray absorbing material using
the ASCA data. Using spectral model E, we measure a typical best-fit 
covering fraction of unity and can rule out covering 
fractions significantly less than 90 per cent. The exception to this is the 
result for the Virgo cluster, for which we measure a best-fit covering 
fraction with spectral model E, $f = 0.66^{+0.15}_{-0.16}$ (for spectral 
model C we find $f = 0.86^{+0.07}_{-0.07}$). The results on the covering 
fractions, and the agreement of the ASCA and PSPC results, suggest that the 
White \etal (1991) results are likely to have 
systematically over-estimated the intrinsic column densities in clusters, 
although the qualitative result on the detection of significant intrinsic 
X-ray absorption in cooling-flow clusters is confirmed.

\subsection{The luminosity reprocessed in other wavebands}

 The luminosities absorbed at X-ray wavelengths must eventually be
reprocessed in other wavebands. If the absorbing material is dusty 
(as is likely to be the case in the central regions of 
the clusters \eg Voit \& Donahue 1995; Fabian \etal 1994a) then 
the bulk of this reprocessed 
emission is likely to emerge in the far infrared. Table 12 summarizes 
the reprocessed luminosities ({\it i.e.} the bolometric luminosities 
absorbed within the clusters) measured with spectral model E (or 
model C for the Virgo Cluster). The reprocessed 
luminosities range from  $\sim 4 \times 10^{42}$ \ergps~for the Virgo 
Cluster to $\sim 5 \times 10^{44}$ \ergps for Abell 2597. 

 Table 12 also lists the observed 60 and $100\mu$m fluxes within a 
four arcmin (radius) aperture centred on the X-ray centroids for the 
clusters (Table 3). The observed fluxes were measured using the IPAC 
SCANPI software and co-added IRAS scans. (The median of the 
co-added scans was used.) The root-mean-square deviations in the residuals 
external to the source extraction regions after baseline subtraction, and
the in-scan separations (in arcmin) between the peaks of the $100\mu$m 
emission and the X-ray centres, are also included in the table.
 
Three of the clusters listed in Table 12, Abell 426, the Centaurus
Cluster and Abell 2199, provide convincing detections of $100\mu$m 
flux associated with the peaks of their X-ray emission. 
For the Centaurus cluster and Abell 2199 (and the slightly offset source
associated with Abell 496), the infrared emission appears spatially 
extended. For Abell 426, the detected infrared flux is point-like and 
presumably associated with the active nucleus of the central galaxy, 
NGC1275. Cox, Bregman \& Schombert (1995) have previously reported 
detections of infrared emission from the central regions of the 
Centaurus Cluster and Abell 2199, from an analysis of the same 
IRAS scan data. These authors did not detect significant 
($\geq 98$ per cent confidence) infrared emission from within 
1 arcmin (radius) of the dominant galaxies in 
Abell 496, 1795 or 2597, which is consistent with our results. 
The presence of a strong, point-like infrared source in Abell 426 was 
previously reported by Wise \etal (1993) who measured 60 and $100\mu$m fluxes 
associated with this source of 6.48 and 8.73 Jy, respectively. 

Following Helou \etal (1988) and Wise \etal (1993), the total 
infrared luminosities in the $1-1000\mu$m band may be estimated 
from the observed IRAS fluxes using the relation 

\begin{equation}
L_{\rm 1-1000\mu m} \sim 2.8 \times 10^{44} (\frac{z}{0.05})^2(2.58S_{60}+S_{100}) \ergps,
\end{equation}

\noindent where $S_{60}$ and $S_{100}$ are the 60 and $100\mu$m IRAS
fluxes in units of Jy. This relation assumes a dust temperature of $\sim 30$K and an 
emissivity index $n$ in the range $0-2$, where the emissivity is proportional to 
the frequency, $\nu^n$. We associate a systematic uncertainty of $\pm 30$ per cent with
the estimated $1-1000\mu$m luminosities, which is combined in quadrature
with the random errors. (In most cases, the systematic error 
exceeds the random errors. Note that the systematic errors associated with 
the measured IRAS fluxes may be $\gg 30$ per cent in some cases \eg see the
note on the Virgo Cluster below.) The $1-1000\mu$m luminosities 
calculated from this relation are summarized in Table 12. Fig. 5 
shows the results plotted as a function of the predicted reprocessed
luminosities. 

For the Centaurus Cluster and Abell 2199, the estimated $1-1000\mu$m
luminosities are in good agreement with the reprocessed
luminosities predicted from the X-ray models. The observed
$S_{60}/S_{100}$ flux ratios for these clusters indicate dust
temperatures in the range $20-50$K. For Abell 426, the 
$1-1000\mu$m luminosity exceeds the predicted reprocessed X-ray
luminosity by a factor $\sim 4$, consistent with the presence of
a strong, intrinsically absorbed active nucleus in this system 
({\it cf} Section 4). For Abell 496, the observed flux also exceeds the
predicted flux by a factor of $\sim 4$, although the $\sim 3$ arcmin 
spatial offset between the 100$\mu$m and X-ray centroids suggests that the 
infrared emission is likely to originate, at least in part, from some
source other than the cooling flow. (This is also evidenced by the
implausibly low $S_{60}/S_{100}$ flux ratio measured for the cluster). 
For the Virgo Cluster, the predicted reprocessed X-ray luminosity exceeds 
the $1-1000\mu$m luminosity determined from the IRAS data by a factor of 
three. However, this is unsurprising since the SCANPI software is not well suited to 
measuring fluxes from highly extended objects (the cooling flow in the
Virgo Cluster has the largest angular extent of any of the systems studied
in this paper, although this may also affect the IRAS results for the 
Perseus Cluster.)  Finally, we note that it if the bulk 
of the absorbing material in the clusters is very cold ($T < 10$K), as 
suggested by some models (\eg Johnstone, Fabian \& Taylor 1998 and
references therein), then a significant fraction of reprocessed emission 
may emerge in the sub-mm band.

\section{Measuring the ages of cooling flows}

The natural state for a regular, relaxed
cluster of galaxies appears to include the presence of a cooling flow 
in its core (\eg Edge \etal 1992; Peres \etal 1998). Simulations suggest 
that cooling flows are only likely to 
be disrupted to the extent that they are `turned off' when a cluster 
merges with a subcluster of comparable size (\eg McGlynn \& Fabian 1984).
In this section, we discuss three methods which 
may be used to measure the ages of cooling flows from the X-ray data and
apply them to the ASCA and ROSAT observations of the nearby cluster cooling
flows. 

\subsection{Method 1: X-ray colour deprojection}

 The first constraints on the ages of cooling flows in clusters 
were presented by Allen \& Fabian (1997) from their X-ray colour
deprojection study of 18 nearby systems observed with the
ROSAT PSPC. These authors used their X-ray colour deprojection technique 
to measure the spatial distributions of cool(ing) gas in the clusters and
compared their results with the predictions from simple, time-dependent 
cooling flow models. Essentially, Allen \& Fabian (1997) identified the 
ages of the cooling flows with the cooling time of the ICM at the 
outermost radii at which significant cooling is observed. These authors 
provide cooling-flow age measurements for seven clusters, four of 
which are also included in the present work. Their results for 
Abell 496, 1795, 2199 and 2597 are $3.3-6.8$, $3.6-6.7$, $5.3-7.7$ and
$2.5-5.1$ Gyr, respectively.

\subsection{Method 2: Comparison of the spectral and imaging mass
deposition rates}

A second estimate of the ages of cooling flows may be obtained by 
comparing the essentially independent results on the mass 
deposition rates determined from the spectral and image 
deprojection studies. The deprojection analysis, discussed in Section 6,
measures the mass deposition profiles in the clusters {\it i.e.} the mass
deposition rates as a function of radius. The integrated mass 
deposition rates inferred from the deprojection analysis are the
mass deposition rates integrated out to some critical radius ($r_{\rm
cool}$) at which the mean cooling time of the X-ray gas ($t_{\rm cool}$) 
becomes equal to an assumed age for the system. (The results presented in 
Table 9 are for an assumed age of $1.3 \times 10^{10}$ yr). The spectral data, 
in contrast, provide a measure of the current, total mass deposition rates 
in the clusters. Thus, the ages of the cooling flows may simply be associated 
with the cooling time of the X-ray gas at the radii where the integrated 
mass deposition 
rates determined from the deprojection analysis become equal to the
values measured directly from the spectral data. 

The primary uncertainty with this method lies in the corrections for 
the effects on intrinsic absorption, especially on the deprojection 
results. The larger the correction for intrinsic absorption at a
particular radius, the larger the mass deposition rate inferred 
at that radius and, therefore, the smaller the age of the cooling flow. 
The cooling-flow ages determined with this method, both with and without 
corrections for intrinsic absorption due to a uniform foreground 
screen of cold gas on the deprojection analysis, are summarized in 
columns 3 and 4 of 
Table 13.

\subsection{Method 3: Breaks in the deprojected mass deposition profiles} 

 The X-ray colour deprojection method (method 1) identifies the 
age of a cooling flow with the mean cooling time of the cluster gas 
at the outermost radius where the spectral signature of cooling gas is
observed. Since the presence of cooling gas will tend to enhance the 
X-ray luminosity of a cluster, the outermost radius at which cooling 
occurs may also be expected to be associated with a `break' in the 
X-ray surface brightness profile and, more evidently, the mass 
deposition profile determined from the deprojection analysis. 

 We have searched for the presence of breaks in the mass deposition
profiles determined from the deprojection analysis by fitting these 
profiles with broken power-law models. Only those data from radii 
interior to the 90 percentile upper limits to the cooling radii were 
included in the modelling, with the exception of the Virgo cluster where, 
due to systematic uncertainties associated with the effects of the 
detector support structure, the analysis was limited to the central 
78 kpc (15 arcmin) radius. An unweighted, ordinary least-squares statistic
was used in the fits. (A reduced $\chi^2$ value of 1.0 was assumed in 
estimating the $\Delta \chi^2 =2.71$ errors.) The results from the fits 
are summarized in Table 14. 

Several of the clusters studied (\eg Abell 426, 496) 
exhibit clear breaks in their mass deposition profiles at radii 
$r<r_{\rm cool}$. In other systems, the identification of a 
break radius is less clear, although still formally significant (\eg Abell 2199). 
Following method 1, we identify the ages of the 
cooling flows with the mean cooling time of the cluster gas at 
the break radii. In calculating the ages, we assume that the 
X-ray emission at the break radii is unaffected by intrinsic 
absorption, which is reasonable if the absorbing material in the clusters 
is gradually accumulated by the cooling flows. 

The only cluster for which we do not detect a significant break in the
mass deposition profile is the Centaurus Cluster which, interestingly,
also has the largest age measured with method 2. (The age measurement
for the Centaurus Cluster with method 2 is consistent with 
the cooling flow having survived intact since the cluster was formed. In
this case no break in the mass deposition profile is expected). Fig. 6 
shows the mass deposition profiles for a representative subsample of the 
clusters studied, with the best-fitting broken power-law models overlaid. The 
mean cooling time of the cluster gas as a function of radius is shown 
in the lower panels.

We have measured the slopes of the mass deposition profiles interior 
and external to the break radii in the clusters. (These slopes may be
used to constrain the range of plausible isothermal cooling flow models
discussed in Section 3.2). The mean slope interior to the break radii 
({\it i.e.} in the regions where mass deposition is expected 
to occur) is $1.37\pm0.27$. Beyond this point, the mean slope flattens to a value of 
$0.59\pm0.21$.  

Finally, in Table 14, we also list the integrated mass deposition 
rates interior to the break radii, both corrected and uncorrected for 
absorption due to a uniform foreground screen of cold gas (as described 
in Section 7.1). Fig. 7 shows these values plotted 
as a function of the mass deposition rates determined from the spectral
data. (The Centaurus Cluster does not exhibit a clear break within the
cooling radius and we assume ${\dot M_{\rm Brk}} = {\dot M_{\rm C}}$ or
${\dot M_{\rm Brk}} = {\dot M_{\rm I}}$ from Table 10, as appropriate.)
In general, the agreement between the results is good, 
particularly for the uncorrected data. (Recall that significant 
uncertainties remain in the corrections for intrinsic absorption,  
due both to the unknown geometry and physical nature of the absorbing 
matter. See also Section 9.1). A fit to the results 
with a power-law model of the form ${\dot M_{\rm S}} = 
P {\dot M_{\rm Brk}^Q}$, using the Akritas \&
Bershady (1996) bisector modification of the ordinary least-squares 
statistic, gives a slope $Q=1.08\pm 0.14$ and a normalization, 
$P=0.33\pm0.25$ for the absorption-corrected data, and a slope
$Q=1.11\pm 0.19$ and a normalization, $P=0.70\pm0.61$ for the 
uncorrected results (where the errors are the
standard deviations determined by bootstrap re-sampling).

\subsection{Comparison of results}

Table 13 summarizes the ages for the cooling flows determined 
using the three different methods. These results are also shown 
in graphical form in Fig. 8, where we have averaged the 
absorption-corrected and uncorrected results determined with method 2. 
In general, the results on the ages show good agreement, suggesting 
that the individual methods may indeed be used to reliably measure the 
ages of the cooling flows. 

One source of uncertainty with the age measurements is the relation between 
the cooling time ({\it i.e.} the time taken for the gas to cool to zero 
degrees at constant pressure) of the ICM at the outermost radii 
where cooling gas is observed and the ages of the cooling flows. 
In detail, the outermost radius at which significant mass deposition 
occurs will depend upon the spectrum of density inhomogeneities in the 
cluster gas, which at present remains unknown. 

As discussed above, the effects of intrinsic absorption provide an 
additional source of 
uncertainty. In particular, the ages
determined with method 2 (the comparison of the spectral and deprojected 
mass deposition rates) are sensitive to such effects, although the results 
based on the break radii (method 3) are insensitive to any uniform change 
in column density across the cooling flows, and those from the X-ray colour 
deprojection analysis (method 1) are largely independent 
of column density uncertainties (other than in the determination 
of the cooling time at the outermost edge of the cooling flows,
where we assume zero intrinsic absorption). Future observations with 
the Chandra Observatory and XMM-Newton will directly address such 
uncertainties, permitting
spatially-resolved spectroscopic determinations of both the mass
deposition and intrinsic absorption profiles in clusters on spatial
scales $\approxgt$ a few arcsec. 

Taking into account the uncertainties associated with the different 
methods, we identify method 1 as the most robust, followed by 
method 3, with method 2 being the most sensitive to systematic effects. 
Method 1 requires the best data, with good spatial and moderate 
spectral resolution and a high signal-to-noise ratio. (To date this
method has has only been applicable to PSPC observations of a few of the
nearest, brightest clusters of galaxies). In contrast, method 3 should be
applicable to a large number of ROSAT HRI observations. We note that method 
2 is, at present, also complicated by the fact 
that it combines results from different detectors, which introduces additional
systematic uncertainties.

Averaging the absorption-corrected and uncorrected results determined
with method 2 (which provides a reasonable estimate of our uncertainty 
in these quantities) 
and taking an unweighted mean of the results
determined with the three separate methods, we arrive at the following 
ages for the cooling flows in the clusters: Abell 426 (2.5 Gyr),
Abell 496 (5.4 Gyr), the Virgo Cluster (2.6 Gyr), the Centaurus Cluster
(7.1 Gyr), Abell 1795 (5.4 Gyr), Abell 2199 (5.9 Gyr) and Abell 2597 (3.2 
Gyr).

\section{Discussion}

\subsection{The possible role of heating processes}

The age-corrected mass deposition rates determined 
from the image deprojection analysis, with no corrections for the effects 
of intrinsic absorption applied, are generally in good 
agreement with the spectrally-determined values. 
However, the absorption-corrected, age-corrected deprojection 
results are typically about a 
factor of two larger than the spectral measurements. Although this 
discrepancy may simply be due to incorrect assumptions about the 
spatial distribution and physical nature of the absorbing medium 
(Voit \& Donahue 1995, Arnaud \& Mushotzky 1998 and Allen 2000 
discuss the possibility that the intrinsic absorption in clusters 
may be primarily due to dust grains) it is sensible to also 
consider other possibilities.

The location of the break radii in the clusters is crucial 
in measuring the integrated mass deposition rates from the 
imaging data. One possibility for reducing the imaging values 
is to associate the ages of the cooling flows with the cooling time
of the cluster gas at radii of about one half of the observed 
break radii. This would, however, lead to significantly shorter ages for 
the cooling flows, which would be puzzling for a phenomenon
which appears to be so common. Moreover the X-ray colour deprojection
technique (Method 1) clearly shows that 
cool(ing) gas extends out to about the break radii. 
The assumption of constant metallicity within the cooling flows 
(where significant metallicity gradients are observed; Section 5.3) 
will also introduce systematic errors, although these
are unlikely to result in discrepancies as large as a factor of two. 
The use of a larger value for $\eta$ in the isothermal cooling flow 
spectral models (Table 5) can also slightly reduce the discrepancy between 
the spectral and absorption-corrected imaging mass deposition rates, but 
not fully account for it.  
It remains possible that a combination of some or all of these factors, 
together with issues such as the correctness of the plasma code, details 
depending on the assumed element abundances (the current cooling flow 
models do not account for variable element abundance ratios) 
and/or the geometry and effects of the magnetic field, are responsible. 

A further plausible alternative is that there may be some heating as well as
cooling in the central region. Many authors have proposed various heating
models (see Fabian 1994 for references) but none has been found to be
compelling and essential in the face of the clear spectral evidence
for cool, and probably cooling, gas components. Nevertheless the
presence of radio sources at the centres of most cooling flows
demonstrates that at least some heating occurs. An important
point with any viable heating process, given the X-ray observations, is 
that it must allow some of the gas to continue cooling, 
which means that the heating rate must be say one third to one half of 
the total cooling rate. Such a close balance suggests feedback {\it i.e.} 
the heat source is fed by the cooling gas.

One obvious solution is that the central engines in the radio sources are 
fuelled by the flows (Bailey 1980; Nulsen, Stewart \& Fabian 1984;
Fabian \& Crawford 1990). This has the useful properties that the
radio sources are spatially extended and can therefore supply heat 
(probably via low energy cosmic rays, depending on the magnetic 
topology and connection to the various gas phases, or by shocks) 
over an extended region, and that the cooling gas fueling the source 
will, in a multiphase flow, be the hottest phase in which the 
effects of cosmic ray and/or shock heating are greatest 
(in general, the hottest phase will have the longest cooling time and 
shock or cosmic ray heating will increase its cooling time by the 
greatest absolute amount. Presumably heating will also be 
strongest near the centre). This hottest phase is then likely to be the 
phase through which the feedback operates. If there is temporarily too much 
heating, this may stop the cooling and reduce the central fuel supply to 
the radio source. Conversely, if there is too little heating 
the accretion rate onto the nucleus will rise (for a discussion in the 
context of single-phase gas models see \eg Tucker \& David 1997).

Pedlar \etal al (1992) and Owen \& Eilek (1998) have emphasised the
importance of heating by the central radio sources in the 
Perseus Cluster and Abell 2199, respectively. Our results provide an 
indication of the possible strength of this heating in such systems. 
The heating may occur intermittently, but overall it 
appears that it may cause up to a factor of two reduction in the total mass 
deposition rates deduced from the image deprojection method, 
once absorption corrections have been made. We note that these rates
remain close to the values obtained when no absorption correction is 
made and are therefore similar to the pre-1990 published values, which 
have always been regarded as uncertain by about a factor of two (\eg 
Arnaud 1988; Fabian 1994) due to uncertainties in the gravitational 
potentials of the clusters.

The tentative identification of a heating signature discussed here 
does not affect the question of what becomes of the cooled material 
deposited by cooling flows. The spectral evidence from ASCA and other X-ray 
data suggest that tens to hundreds of solar masses of gas cool out of 
cooling flows per year. Over the billion year age of a flow this amounts to 
$5\times 10^{10}$ to greater than $10^{12}\Msun$ of cooled material, 
of which only a small fraction is directly observed at other wavelengths.

\subsection{The masses of merging subclusters that 
disrupt cooling flows}

Observations and theoretical models suggest that cooling flows 
are likely to be disrupted when a cluster merges with a subcluster 
of comparable size (\eg McGlynn \& Fabian 1984; Edge, Stewart \& Fabian
1992; Allen 1998). When the density in the core of the merging subcluster 
is sufficiently high, the gas in the central regions of the dominant
cluster is likely to be shocked and disrupted as the systems merge, 
leading to complex X-ray morphologies, temperature inhomogeneities and an 
inflation of the X-ray core radii (Navarro, Frenk \& White 
1995; Schindler 1996; Roettiger, Burns \& Loken 1996). 
The time taken for the cooling flow to reform after the dark matter 
distributions merge and hydrostatic equilibrium is restored will depend upon 
the degree to which cool gas originally present in the central regions of 
the clusters is heated and mixed with the surrounding ICM. However, the fact 
that $\sim 70$ per cent of bright, nearby clusters contain cooling flows 
(Peres \etal 1998) suggests that cooling flows are `turned off' 
on average for $\sim 30$ per cent of the time. In this case, we may 
expect the typical time taken for a cooling flow to reestablish itself 
following a major subcluster merger event to be of the order of a few Gyr. 
Given the results on the ages of the cooling flows presented in Section 8, 
we may therefore expect the present sample of clusters to have experienced 
their last major subcluster merger events during the last $5-10$ Gyr (with
the Coma, Perseus and Virgo clusters having experienced the most recent 
merger events and the cooling flow in the Centaurus Cluster having been 
undisturbed for the longest period). 

Assuming that subcluster mergers are the primary cause of the disruption of
cooling flows, we can estimate the masses of the merging subclusters
which cause significant disruption by integrating the merger rates 
predicted by hierarchical formation models, using parameters appropriate for 
rich clusters. Such rates are given by the extended Press-Schechter formalism 
of Lacey \& Cole (1993). Examples of their application to rich clusters are 
given by Cavaliere, Menci \& Tozzi (1999).

The results for a cluster mass of $10^{15}\Msun$ are shown in Fig. 9,
where we plot the fraction of clusters that have undergone a merger 
with a subcluster of mass greater than $10-80$ per cent of the final 
cluster mass, within a given interval. (We assume $\Omega=1.0, \Lambda = 
0, b=1$ and a cosmic fluctuation index $n=-1.5$.) It is clear from the 
figure that for there to have been a good chance ($\approxgt 50$ per cent) 
of a merger within the last 10 Gyr, subclusters with masses of 
$\approxlt 30 - 40$ per cent of the total final mass are required. Major 
mergers involving subclusters of similar mass ({\it i.e.} half of the final 
mass) have a probability of $\sim 20$ per cent within that time. The 
results are similar for an open cosmology with $\Omega=0.3$ (other 
cosmological parameters the same) for which subclusters with 
masses of $\approxlt 20 - 30$ per cent of the total, final mass have an 
$\approxgt 50$ per cent chance of merging within the last 10 Gyr and the 
probability of a merger with a subcluster of similar mass is $\sim 10$ per 
cent. Thus, if subcluster mergers are the primary cause of the disruption of 
cooling flows in rich clusters, it seems likely that subclusters with 
masses of $\sim 30$ per cent of the final cluster mass can disrupt cooling 
flows. 

The results for a cluster of mass $10^{14}\Msun$ (a mass similar to 
that of the lowest luminosity clusters in our sample) show 
a much higher rate of significant mergers, with a $\sim 50$ per cent chance of 
a major merger within the last 5 Gyr. It is therefore not clear why the
ages of the cooling flows in the present sample of objects appear 
roughly independent of their luminosity (and therefore mass). A possible 
explanation is that mergers may not be the only cause of cooling flow 
disruption. Activity in the nuclei of the central, dominant galaxies 
may also be important. If the ages of the cooling flows (taking into 
account the time required for the flows to settle down) are around 8 Gyr, 
then the 
disruption is likely to have occurred at redshift of $z= 1 - 2$ and 
so could be related to the peak of quasar/radio activity. 

The application of similar techniques to future observations made with 
the Chandra Observatory and XMM-Newton should permit precise 
measurements of the ages of cooling flows for large samples 
of clusters within redshifts $z \approxlt 1$.

\section{Conclusions}

 The main conclusions that may be drawn from this paper 
may be summarized as follows:
\vskip 0.2cm

  (i) We have demonstrated the need for multiphase models to 
consistently explain the spectral and imaging X-ray data for 
the nearest, brightest cooling-flow clusters. In general, the 
diffuse, thermal emission from these systems can be modelled 
using either two temperature or cooling flow 
models, with excess absorption acting on the cooler emission 
components. 

 (ii) The mass deposition rates for the cooling flows, independently 
inferred from the ASCA spectra and ROSAT images, exhibit reasonable
agreement, providing support for the basic cooling 
flow picture. We have discussed the systematic uncertainties associated 
with the results, which are present at the factor $\sim 2$ level, 
and commented on the possible effects of intermittent heating in 
the cooling flows. 

 (iii) The ASCA spectra for the central regions of the Virgo 
and Centaurus clusters place firm constraints on the abundances 
of iron, magnesium, silicon and sulphur in the ICM. Our results 
firmly favour enrichment models in which a high mass fraction 
($70-80$ per cent) of the iron in the ICM in the central regions of 
these clusters (in which strong metallicity gradients are observed) is due 
to Type Ia supernovae. 

 (iv) We have confirmed the presence of intrinsic 
X-ray absorption in the ASCA spectra for the 
clusters using a variety of spectral models. The measured 
column densities are sensitive to the spectral models 
used and range (on average) from a few $10^{20}$ \apc~using a simple 
single-temperature model to a few $10^{21}$ \apc~using our preferred 
cooling-flow models. 

  (v) We have reported detections of 60 and $100\mu$m infrared 
emission, coincident with the X-ray centroids, from a number of the 
systems studied (confirming the original detections 
by Wise \etal 1993 and Cox \etal 1995). For the Centaurus 
Cluster and Abell 2199, the coadded IRAS
scan data show that the infrared emission is spatially extended. The 
infrared fluxes and flux limits are in good agreement with the predicted 
values due to reprocessed X-ray emission from the intrinsic X-ray 
absorbing material in the cooling flows. 

 (vi) We have discussed three methods which may be used to 
measure the ages of cooling flows from the X-ray data. The 
application of these techniques to the present sample of objects indicates 
cooling-flow ages of between $2.5$ and 7 Gyr. If the ages of cooling
flows in the most massive clusters clusters are primarily set by subcluster 
merger events, 
then our results suggest that mergers with subclusters with masses of 
$\sim 30$ per cent of the final cluster mass are sufficient to disrupt 
cooling flows.   

\section*{Acknowledgements}
ACF and SWA thank the Royal Society for support. PEJN gratefully acknowledges 
the hospitality of the Harvard-Smithsonian Center for Astrophysics.
This work was funded in part by NASA grants NAG8-1881, NAG5-3064 and
NAG5-2588.

\clearpage

 %%%%%%%%%%%%%%%%%%%%%%%%%% TABLES %%%%%%%%%%%%%%%%%%%%%%%%%%%%%

\begin{table*}
\vskip 0.2truein
\begin{center}
\caption{Summary of the ASCA Observations. Column 2 lists the 
optically-determined redshifts for the
clusters (for the Virgo Cluster we assume a distance of 18 Mpc).
Column 3 summarizes the dates of the observations. Columns $4-7$ list the net 
exposure times (in seconds) in each of the four detectors 
after all screening and cleaning procedures were carried out.}
\vskip 0.2truein
\begin{tabular}{ c c c c c c c c c }
%\multicolumn{1}{c}{} &
%\multicolumn{1}{c}{} &
%\multicolumn{1}{c}{} &
%\multicolumn{1}{c}{} &
%\multicolumn{1}{c}{} &
%\multicolumn{4}{c}{} \\
 \hline                                                                               
Cluster         & ~ &  z      &  Date     & ~ &    S0  &   S1  &  G2   &  G3  \\  
 \hline                                                                               
&&&&&&&& \\                                                                         
Abell 426       & ~ & 0.0183  &  1993 Aug 06  & ~ &  13910 & 6814  & 15618 & 14916 \\       
Abell 496       & ~ & 0.0320  &  1993 Sep 21  & ~ &  29685 & 21515 & 40050 & 39986 \\    
%Hydra A        & ~ & 0.0522  &  1993 May 28  & ~ &  15127 & 13119 & 19602 & 19274 \\      
Virgo           & ~ & 18Mpc   &  1993 Jun 07  & ~ &  13633 & 14668 & 16874 & 16874 \\      
Centaurus       & ~ & 0.0104  &  1995 May 19  & ~ &  68820 & 67650 & 70981 & 70999 \\       
Coma            & ~ & 0.0232  &  1993 Jun 14  & ~ &  8732  & 6637  & 9312  & 9312  \\      
Abell 1795      & ~ & 0.0634  &  1993 Jun 16  & ~ &  31284 & ----- & 37649 & 37641 \\    
Abell 2199      & ~ & 0.0309  &  1993 Jul 25  & ~ &  19153 & 13932 & 29758 & 29758 \\       
Abell 2597      & ~ & 0.0852  &  1996 Dec 06  & ~ &  33933 & 34697 & 39249 & 39239 \\       
&&&&&&&& \\                                                                         
\hline 
&&&&&&&& \\                                                                         
\end{tabular}
\end{center}
\parbox {7in}
{}
\vskip 1cm
\end{table*}

\begin{table*}
\vskip 0.2truein
\begin{center}
\caption{The radii of the circular extraction regions used in the analysis of 
the ASCA SIS data (in
arcmin and kpc) and the chip modes used in the observations (either 1,2 or 4-CCD mode). 
The numbers in parentheses indicate the number of 
chips contributing to the extracted spectra. For the GIS data a fixed
extraction radius of 6 arcmin was used.}
\vskip 0.2truein
\begin{tabular}{ c c c c c }
\hline                                                                               
Cluster         & ~ & S0   & S1     & SIS Chip Mode \\  
                & ~ & (amin/kpc) & (amin/kpc)   &           \\
\hline                                                         
&&&& \\                                                   
Abell 426       & ~ & 4.9/152  & 4.7/145    & 4(1)    \\       
Abell 496       & ~ & 3.9/206  & 3.2/169    & 4(1)    \\        
Virgo           & ~ & 4.9/25.5 & 4.1/21.3    & 4(2)    \\      
Centaurus       & ~ & ---      & 3.7/66.0    & 1(1)    \\      
Coma            & ~ & 10.0/389 & 9.5/370    & 4(4)    \\      
Abell 1795      & ~ & 4.5/447  & ---    & 4(2)    \\       
Abell 2199      & ~ & 5.3/271  & 4.2/215    & 4(2)    \\    
Abell 2597      & ~ & 4.5/580  & 3.7/477    & 2(2)    \\       
&&&& \\                                                                     
\hline 
&&&& \\                                                                     
\end{tabular}
\end{center}
\parbox {7in}
{}
\end{table*}

\clearpage

\begin{table*}
\vskip 0.2truein
\begin{center}
\caption{Summary of the ROSAT Observations. Columns 2 and 3 
list the dates of the observations and the instrument
used. Column 4 lists the exposure times (in seconds).
Where more than a single observation of a cluster was made, details 
for each observation are given. 
Columns 5 and 6 list the coordinates for the centroids of the X-ray
emission from the clusters.}
\vskip 0.2truein
\begin{tabular}{ c c c c c c c c }
%\multicolumn{1}{c}{} &
%\multicolumn{1}{c}{} &
%\multicolumn{1}{c}{} &
%\multicolumn{1}{c}{} &
%\multicolumn{1}{c}{} &
%\multicolumn{1}{c}{} &
%\multicolumn{2}{c}{} \\                            
\hline                                                                                                                               
 Cluster     & ~ &  Date    & Instrument  & Exposure    & ~ &     R.A.  (J2000). &    Dec. (J2000.)                 \\  
\hline                                                                                                                               
&&&&&&& \\                                                                                                                         
Abell 426          & ~ &  1992 Feb 02  & PSPC & 4787    & ~ &   $03^{\rm h}19^{\rm m}48.5^{\rm s}$ & $ 41^{\circ}30'27''$  \\  
Abell 496          & ~ &  1992 Sep 12  & HRI  & 14488   & ~ &   $04^{\rm h}33^{\rm m}38.1^{\rm s}$ & $-13^{\circ}15'42''$  \\  
Virgo              & ~ &  1992 Dec 17  & PSPC & 9961    & ~ &   $12^{\rm h}30^{\rm m}49.8^{\rm s}$ & $12^{\circ}23'32''$  \\  
Centaurus          & ~ &  1994 Jul 05  & PSPC & 3192    & ~ &   $12^{\rm h}48^{\rm m}48.7^{\rm s}$ & $-41^{\circ}18'44''$  \\  
Coma               & ~ &  1991 Jun 17  & PSPC & 22108   & ~ &   $12^{\rm h}59^{\rm m}46.1^{\rm s}$ & $27^{\circ}56'21''$  \\  
Abell 1795 \#1     & ~ &  1992 Jun 25  & HRI  & 2768    & ~ &   $13^{\rm h}48^{\rm m}52.7^{\rm s}$ & $26^{\circ}35'27''$  \\  
Abell 1795 \#2     & ~ &  1993 Jan 21  & HRI   & 11088   & ~ &   --- & --- \\
Abell 1795 \#3     & ~ &  1994 Jun 23  & HRI   & 11080   & ~ &   --- & --- \\  
Abell 2199 \#1     & ~ &  1991 Feb 10  & HRI  & 5308    & ~ &   $16^{\rm h}28^{\rm m}38.5^{\rm s}$ & $39^{\circ}33'03''$  \\  
Abell 2199 \#2     & ~ &  1994 Feb 03  & HRI   & 26528   & ~ &   --- & --- \\  
Abell 2199 \#3     & ~ &  1994 Aug 31  & HRI   & 20976   & ~ &   --- & --- \\  
Abell 2597         & ~ &  1992 Jun 06  & HRI  & 16228   & ~ &   $23^{\rm h}25^{\rm m}19.7^{\rm s}$ &  $-12^{\circ}07'27''$  \\  
&&&&&&& \\                                                                                                                         
\hline
&&&&&&& \\                                                                                                                         
\end{tabular}
\end{center}
\parbox {7in}
{}
\end{table*}

\clearpage

\begin{table*}
\vskip 0.2truein
\begin{center}
\caption{The best-fit parameter values and 90 per cent  ($\Delta \chi^2 =
2.71$) confidence limits from the basic spectral analysis of the ASCA data. 
Temperatures ($kT$), metallicities ($Z$), column densities ($N_{\rm H}$), 
intrinsic column densities ($\Delta N_{\rm H}$), mass deposition rates
(${\dot M_{\rm S}}$) and the normalization of the intrinsically absorbed
(cooler) emission components in spectral model D, were linked to take the same
values in all four detectors. Only the normalization of the hotter isothermal 
emission component was allowed to vary independently for each detector. 
Temperatures are quoted in keV and  metallicities as a fraction of the solar
photospheric value (Anders \& Grevesse 1989). Column densities and
intrinsic column densities are in units of $10^{21}$ atom cm$^{-2}$. Mass 
deposition rates are in \Msunpyr.} 
\vskip 0.2truein
\begin{tabular}{ c c c c c c c c c c c c c}
%&&&&&&&&&&&  \\                                                                                                                                               
\hline                                                                                                                                                         
            & ~ &   Parameters   & ~~~ &     Model A            & ~~~ & Model B                & ~~~ &     Model C            & ~~~ &  Model D                & ~~~ &  Model E                \\
\hline                                                                                                                                                                                        
%all pt source arfs except a426,cen,virgo
&&&&&&&&&&&  \\			                                
% NULSEN MODELS FOR CP AND ISO FLOWS
% EXTENDED ARF ON ALL
            & ~ &   $kT$         & ~~~ & $4.22^{+0.03}_{-0.03}$ & ~~~ & $4.31^{+0.04}_{-0.04}$ & ~~~ & $4.70^{+0.18}_{-0.13}$ & ~~~ & $6.00^{+0.27}_{-0.29}$ & ~~~ & $4.57^{+0.09}_{-0.08}$ \\    
            & ~ &   $Z$          & ~~~ & $0.47^{+0.02}_{-0.02}$ & ~~~ & $0.47^{+0.01}_{-0.02}$ & ~~~ & $0.46^{+0.02}_{-0.02}$ & ~~~ & $0.45^{+0.02}_{-0.02}$ & ~~~ & $0.46^{+0.02}_{-0.02}$ \\     
Abell 426   & ~ &   $N_{\rm H}$  & ~~~ & $1.49$                 & ~~~ & $1.37^{+0.03}_{-0.03}$ & ~~~ & $1.14^{+0.10}_{-0.15}$ & ~~~ & $0.83^{+0.12}_{-0.10}$ & ~~~ & $0.93^{+0.16}_{-0.19}$ \\ 
            & ~ &   ${\dot M}$   & ~~~ &  ---                   & ~~~ &  ---                   & ~~~ & $258^{+89}_{-89}$      & ~~~ &  ---                   & ~~~ & $303^{+75}_{-74}$      \\       
            & ~ &   $kT_2$       & ~~~ &  ---                   & ~~~ &  ---                   & ~~~ &  ---                   & ~~~ & $2.17^{+0.14}_{-0.16}$ & ~~~ & ---                    \\ 
       & ~ & $\Delta N_{\rm H}$  & ~~~ &  ---                   & ~~~ &  ---                   & ~~~ & $3.6^{+0.4}_{-0.5}$    & ~~~ & $2.5^{+0.5}_{-0.4}$    & ~~~ & $3.4^{+0.4}_{-0.4}$    \\ 
            & ~ &   $\chi^2$/DOF & ~~~ & 2487/1659              & ~~~ & 2443/1658              & ~~~ & 2405/1656              & ~~~ & 2229/1655              & ~~~ & 2389/1656              \\                 
\hline                                                                                                                                                                                      
&&&&&&&&&&&  \\			         														             	                          
%PT SOURCE ARF 																		             	                          
            & ~ &   $kT$         & ~~~ & $3.78^{+0.04}_{-0.04}$ & ~~~ & $3.54^{+0.05}_{-0.05}$ & ~~~ & $3.86^{+0.15}_{-0.15}$ & ~~~ & $3.85^{+0.10}_{-0.08}$ & ~~~ & $3.67^{+0.08}_{-0.07}$ \\    
            & ~ &   $Z$          & ~~~ & $0.50^{+0.03}_{-0.03}$ & ~~~ & $0.51^{+0.03}_{-0.03}$ & ~~~ & $0.50^{+0.04}_{-0.03}$ & ~~~ & $0.50^{+0.03}_{-0.03}$ & ~~~ & $0.50^{+0.03}_{-0.03}$ \\     
Abell 496   & ~ &   $N_{\rm H}$  & ~~~ & $0.46$                 & ~~~ & $0.83^{+0.05}_{-0.05}$ & ~~~ & $0.49^{+0.21}_{-0.21}$ & ~~~ & $0.74^{+0.08}_{-0.07}$ & ~~~ & $0.44^{+0.22}_{-0.24}$ \\ 
            & ~ &   ${\dot M}$   & ~~~ &  ---                   & ~~~ &  ---                   & ~~~ & $136^{+46}_{-55}$      & ~~~ &  ---                   & ~~~ & $110^{+39}_{-42}$      \\       
            & ~ &   $kT_2$       & ~~~ &  ---                   & ~~~ &  ---                   & ~~~ & ---                    & ~~~ & $0.47^{+0.08}_{-0.08}$ & ~~~ & ---                    \\ 
       & ~ & $\Delta N_{\rm H}$  & ~~~ &  ---                   & ~~~ &  ---                   & ~~~ & $4.3^{+0.6}_{-0.9}$    & ~~~ & $11.9^{+2.0}_{-1.5}$   & ~~~ & $3.6^{+0.6}_{-0.7}$    \\ 
            & ~ &   $\chi^2$/DOF & ~~~ & 1546/1283              & ~~~ & 1400/1282              & ~~~ & 1379/1280              & ~~~ & 1319/1279 & ~~~ & 1380/1280              \\                 
\hline                                                                                                                                                                                      
&&&&&&&&&&&  \\			         														             	                          
%EXTENDED ARF %%																	             	                          
            & ~ &   $kT$         & ~~~ & $2.05^{+0.01}_{-0.02}$ & ~~~ & $2.02^{+0.01}_{-0.02}$ & ~~~ & $2.23^{+0.03}_{-0.03}$ & ~~~ & $2.19^{+0.02}_{-0.03}$ & ~~~ & $2.08^{+0.02}_{-0.02}$ \\    
            & ~ &   $Z$          & ~~~ & $0.67^{+0.02}_{-0.02}$ & ~~~ & $0.65^{+0.02}_{-0.02}$ & ~~~ & $0.80^{+0.03}_{-0.03}$ & ~~~ & $0.79^{+0.04}_{-0.03}$ & ~~~ & $0.75^{+0.02}_{-0.03}$ \\     
Virgo       & ~ &   $N_{\rm H}$  & ~~~ & $0.25$                 & ~~~ & $0.35^{+0.04}_{-0.03}$ & ~~~ & $0.05^{+0.15}_{-0.05}$ & ~~~ & $0.31^{+0.07}_{-0.07}$ & ~~~ & $0.37^{+0.19}_{-0.18}$ \\ 
            & ~ &   ${\dot M}$   & ~~~ &  ---                   & ~~~ &  ---                   & ~~~ & $10.7^{+1.0}_{-1.4}$   & ~~~ &  ---                   & ~~~ & $5.8^{+1.0}_{-0.9}$    \\       
            & ~ &   $kT_2$       & ~~~ &  ---                   & ~~~ &  ---                   & ~~~ & ---                    & ~~~ & $0.84^{+0.04}_{-0.03}$ & ~~~ & ---                    \\ 
       & ~ & $\Delta N_{\rm H}$  & ~~~ &  ---                   & ~~~ &  ---                   & ~~~ & $3.76^{+0.28}_{-0.48}$ & ~~~ & $5.0^{+0.6}_{-0.8}$    & ~~~ & $1.88^{+0.53}_{-0.65}$ \\ 
            & ~ &   $\chi^2$/DOF & ~~~ & 2527/1153              & ~~~ & 2502/1152              & ~~~ & 1975/1150              & ~~~ & 1950/1149              & ~~~ & 2018/1150              \\                 
\hline                                                                                                                                                                                      
&&&&&&&&&&&  \\                                                                                                                                                                                
%EXTENDED ARF%																		             	                          
            & ~ &   $kT$         & ~~~ & $2.77^{+0.01}_{-0.02}$ & ~~~ & $2.70^{+0.01}_{-0.02}$ & ~~~ & $3.20^{+0.07}_{-0.07}$ & ~~~ & $3.24^{+0.08}_{-0.08}$ & ~~~ & $2.92^{+0.03}_{-0.03}$ \\    
            & ~ &   $Z$          & ~~~ & $1.18^{+0.03}_{-0.03}$ & ~~~ & $1.16^{+0.02}_{-0.03}$ & ~~~ & $1.24^{+0.03}_{-0.03}$ & ~~~ & $0.93^{+0.04}_{-0.03}$ & ~~~ & $1.23^{+0.03}_{-0.03}$ \\     
Centaurus   & ~ &   $N_{\rm H}$  & ~~~ & $0.81$                 & ~~~ & $1.01^{+0.03}_{-0.03}$ & ~~~ & $1.00^{+0.20}_{-0.20}$ & ~~~ & $1.28^{+0.08}_{-0.10}$ & ~~~ & $1.01^{+0.21}_{-0.16}$ \\ 
            & ~ &   ${\dot M}$   & ~~~ &  ---                   & ~~~ &  ---                   & ~~~ & $42.0^{+3.7}_{-3.7}$   & ~~~ & ---                    & ~~~ & $36.3^{+2.4}_{-2.8}$   \\       
            & ~ &   $kT_2$       & ~~~ &  ---                   & ~~~ &  ---                   & ~~~ & ---                    & ~~~ & $1.34^{+0.04}_{-0.04}$ & ~~~ & ---                    \\ 
       & ~ & $\Delta N_{\rm H}$  & ~~~ &  ---                   & ~~~ &  ---                   & ~~~ & $2.46^{+0.36}_{-0.38}$ & ~~~ & $0.23^{+0.35}_{-0.23}$ & ~~~ & $1.92^{+0.25}_{-0.33}$ \\ 
            & ~ &   $\chi^2$/DOF & ~~~ & 4928/1553              & ~~~ & 4782/1552              & ~~~ & 2703/1549              & ~~~ & 2584/1549              & ~~~ & 2696/1549              \\                 
%mdot is for gis2 
\hline
&&&&&&&&&&  \\                                                                                                                                               
\end{tabular}
\end{center}
 
\parbox {7in}
{}
\end{table*}

\clearpage

\addtocounter{table}{-1}
\begin{table*}
\vskip 0.2truein
\begin{center}
\caption{Spectral Results - continued}
\vskip 0.2truein
\begin{tabular}{ c c c c c c c c c c c c c }
%&&&&&&&&&&  \\                                                                                                                                               
\hline                                                                                                                                                         
            & ~ &   Parameters   & ~~~ &     Model A            & ~~~ & Model B                & ~~~ &     Model C            & ~~~ &  Model D               & ~~~ &  Model E                 \\
\hline                                                                                                                                                                                        
&&&&&&&&&&  \\                                                                                                                                                                              
            & ~ &   $kT$         & ~~~ & $6.44^{+0.58}_{-0.52}$ & ~~~ & $8.08^{+0.26}_{-0.25}$ & ~~~ &  ---                   & ~~~ & $9.07^{+0.62}_{-0.89}$ & ~~~ &  ---                   \\           
            & ~ &   $Z$          & ~~~ & $0.20^{+0.07}_{-0.08}$ & ~~~ & $0.19^{+0.04}_{-0.02}$ & ~~~ &  ---                   & ~~~ & $0.21^{+0.04}_{-0.04}$ & ~~~ &  ---                   \\           
Coma        & ~ &   $N_{\rm H}$  & ~~~ & $0.09 $                & ~~~ & $0.24^{+0.05}_{-0.05}$ & ~~~ &  ---                   & ~~~ & $0.11^{+0.07}_{-0.07}$ & ~~~ &  ---                   \\           
            & ~ &   ${\dot M}$   & ~~~ &  ---                   & ~~~ &        ---             & ~~~ &  ---                   & ~~~ &  ---                   & ~~~ &  ---                   \\           
            & ~ &   $kT_2$       & ~~~ &  ---                   & ~~~ &  ---                   & ~~~ &  ---                   & ~~~ & $<0.87$                & ~~~ &  ---                   \\           
       & ~ & $\Delta N_{\rm H}$  & ~~~ &  ---                   & ~~~ &  ---                   & ~~~ &  ---                   & ~~~ & $>10.7$                & ~~~ &  ---                   \\           
            & ~ &   $\chi^2$/DOF & ~~~ & 1696/1228              & ~~~ & 1274/1227              & ~~~ &  ---                   & ~~~ & 1254/1224              & ~~~ &  ---                   \\           
\hline				         														     	       	                          
&&&&&&&&&&  \\			         														     	       	                          
%PT SOURCE ARF%																		     	       	                          
            & ~ &   $kT$         & ~~~ & $5.40^{+0.08}_{-0.09}$ & ~~~ & $5.33^{+0.10}_{-0.11}$ & ~~~ & $5.87^{+0.25}_{-0.19}$ & ~~~ & $6.21^{+0.25}_{-0.22}$ & ~~~ & $5.61^{+0.17}_{-0.15}$ \\    
            & ~ &   $Z$          & ~~~ & $0.36^{+0.03}_{-0.02}$ & ~~~ & $0.36^{+0.03}_{-0.02}$ & ~~~ & $0.37^{+0.02}_{-0.03}$ & ~~~ & $0.37^{+0.03}_{-0.03}$ & ~~~ & $0.37^{+0.02}_{-0.03}$ \\     
Abell 1795  & ~ &   $N_{\rm H}$  & ~~~ & $0.12$                 & ~~~ & $0.17^{+0.05}_{-0.05}$ & ~~~ & $0.00^{+0.06}_{-0.00}$ & ~~~ & $0.00^{+0.03}_{-0.00}$ & ~~~ & $0.00^{+0.06}_{-0.00}$ \\
            & ~ &   ${\dot M}$   & ~~~ &  ---                   & ~~~ &  ---                   & ~~~ & $301^{+76}_{-76}$      & ~~~ &  ---                   & ~~~ & $246^{+71}_{-70}$      \\       
            & ~ &   $kT_2$       & ~~~ &  ---                   & ~~~ &  ---                   & ~~~ & ---                    & ~~~ & $1.48^{+0.54}_{-0.29}$ & ~~~ & ---                    \\ 
       & ~ & $\Delta N_{\rm H}$  & ~~~ &  ---                   & ~~~ &  ---                   & ~~~ & $3.1^{+0.7}_{-0.5}$    & ~~~ & $5.2^{+2.8}_{-2.9}$    & ~~~ & $2.6^{+0.7}_{-0.5}$    \\ 
            & ~ &   $\chi^2$/DOF & ~~~ & 1442/1236              & ~~~ & 1440/1235              & ~~~ & 1419/1233              & ~~~ & 1387/1232              & ~~~ & 1422/1233              \\                 
\hline				         														     	       	                          
&&&&&&&&&&  \\                                                                                                                                                                                
%PT SOURCE ARF% scale factor 1.097																		     	       	                          
            & ~ &   $kT$         & ~~~ & $4.16^{+0.05}_{-0.05}$ & ~~~ & $3.91^{+0.06}_{-0.05}$ & ~~~ & $4.38^{+0.12}_{-0.22}$ & ~~~ & $4.97^{+0.35}_{-0.25}$ & ~~~ & $4.09^{+0.08}_{-0.09}$ \\    
            & ~ &   $Z$          & ~~~ & $0.42^{+0.02}_{-0.02}$ & ~~~ & $0.43^{+0.02}_{-0.03}$ & ~~~ & $0.43^{+0.03}_{-0.03}$ & ~~~ & $0.40^{+0.03}_{-0.03}$ & ~~~ & $0.42^{+0.03}_{-0.03}$ \\     
Abell 2199  & ~ &   $N_{\rm H}$  & ~~~ & $0.09$                 & ~~~ & $0.40^{+0.05}_{-0.05}$ & ~~~ & $0.00^{+0.25}_{-0.00}$ & ~~~ & $0.00^{+0.10}_{-0.00}$ & ~~~ & $0.00^{+0.19}_{-0.00}$ \\ 
            & ~ &   ${\dot M}$   & ~~~ &  ---                   & ~~~ &  ---                   & ~~~ & $180^{+22}_{-76}$      & ~~~ &  ---                   & ~~~ & $147^{+20}_{-46}$      \\       
            & ~ &   $kT_2$       & ~~~ &  ---                   & ~~~ &  ---                   & ~~~ & ---                    & ~~~ & $1.84^{+0.36}_{-0.28}$ & ~~~ & ---                    \\ 
       & ~ & $\Delta N_{\rm H}$  & ~~~ &  ---                   & ~~~ &  ---                   & ~~~ & $4.1^{+0.5}_{-0.9}$    & ~~~ & $2.8^{+1.6}_{-1.1}$    & ~~~ & $3.5^{+0.5}_{-0.7}$    \\ 
            & ~ &   $\chi^2$/DOF & ~~~ & 1693/1295              & ~~~ & 1571/1294              & ~~~ & 1550/1292              & ~~~ & 1516/1291              & ~~~ & 1549/1292              \\                 
\hline                                                                                                                                                                                      
&&&&&&&&&&  \\			         														     	       	                          
%PT SOURCE ARF% scale factor 1.073*0.9579 distance factor=1.028																		     	       	                          
            & ~ &   $kT$         & ~~~ & $3.90^{+0.07}_{-0.06}$ & ~~~ & $3.38^{+0.06}_{-0.07}$ & ~~~ & $3.88^{+0.29}_{-0.30}$ & ~~~ & $2.31^{+0.23}_{-0.28}$ & ~~~ & $3.45^{+0.11}_{-0.10}$ \\    
            & ~ &   $Z$          & ~~~ & $0.32^{+0.03}_{-0.03}$ & ~~~ & $0.35^{+0.03}_{-0.04}$ & ~~~ & $0.36^{+0.05}_{-0.04}$ & ~~~ & $0.29^{+0.04}_{-0.04}$ & ~~~ & $0.35^{+0.04}_{-0.04}$ \\     
Abell 2597  & ~ &   $N_{\rm H}$  & ~~~ & $0.25$                 & ~~~ & $0.99^{+0.07}_{-0.08}$ & ~~~ & $0.21^{+0.44}_{-0.21}$ & ~~~ & $0.65^{+0.55}_{-0.28}$ & ~~~ & $0.00^{+0.38}_{-0.00}$ \\ 
            & ~ &   ${\dot M}$   & ~~~ &  ---                   & ~~~ &  ---                   & ~~~ & $774^{+169}_{-332}$    & ~~~ &  --- & ~~~ &  $717^{+66}_{-208}$      \\       
            & ~ &   $kT_2$       & ~~~ &  ---                   & ~~~ &  ---                   & ~~~ &  ---                   & ~~~ & $6.22^{+1.86}_{-1.23}$ & ~~~ & ---                    \\ 
       & ~ & $\Delta N_{\rm H}$  & ~~~ &  ---                   & ~~~ &  ---                   & ~~~ & $5.4^{+0.9}_{-0.9}$    & ~~~ & $2.4^{+1.7}_{-2.4}$    & ~~~ &  $5.0^{+0.7}_{-0.7}$    \\ 
            & ~ &   $\chi^2$/DOF & ~~~ & 1412/984               & ~~~ & 1127/983               & ~~~ &  1114/981              & ~~~ & 1083/980 & ~~~ & 1110/981               \\                 
\hline                                                                                                                                                   
\end{tabular}
\end{center}
 
\parbox {7in}
{}
\end{table*}

\clearpage

\begin{table*}
\vskip 0.2truein
\begin{center} 
\caption{The mass deposition rates (in \Msunpyr) and intrinsic column densities 
(in units of $10^{21}$ atom cm$^{-2}$) determined from the fits to the 
ASCA spectra with the isothermal cooling flow models for various  
values of the slope parameter $\eta$.}
 \vskip 0.2truein
\begin{tabular}{ c c c c c c c c c c c c c}
%&&&&&&&&&&&  \\                                                                                                                                               
\hline
            & ~ &   Parameters        & ~~~ & $\eta=0.75$            & ~~~ & $\eta=1.0$            & ~~~ & $\eta=1.5$           & ~~~ & $\eta=2.0$           & ~~~ & $\eta=2.5$            \\
\hline                    
&&&&&&&&&&&  \\			                                
            & ~ & ${\dot M}$          & ~~~ & $283^{+68}_{-74}$      & ~~~ & $303^{+75}_{-74}$    & ~~~ & $329^{+81}_{-83}$   & ~~~ & $340^{+89}_{-83}$   & ~~~ & $347^{+95}_{-83}$    \\
Abell 426   & ~ & $\Delta N_{\rm H}$  & ~~~ & $3.2^{+0.3}_{-0.4}$    & ~~~ & $3.4^{+0.4}_{-0.4}$  & ~~~ & $3.7^{+0.3}_{-0.4}$ & ~~~ & $3.8^{+0.4}_{-0.4}$ & ~~~ & $3.9^{+0.4}_{-0.4}$  \\
            & ~ & $\chi^2$/DOF        & ~~~ & 2391.1/1656            & ~~~ & 2388.9/1656          & ~~~ & 2389.8/1656         & ~~~ & 2390.5/1655         & ~~~ & 2390.7/1656          \\
\hline                    
&&&&&&&&&&&  \\			                                
            & ~ & ${\dot M}$          & ~~~ & $93.3^{+37.6}_{-37.2}$ & ~~~ & $110^{+39}_{-42}$    & ~~~ & $136^{+46}_{-50}$   & ~~~ & $152^{+50}_{-54}$   & ~~~ & $162^{+52}_{-57}$    \\
Abell 496   & ~ & $\Delta N_{\rm H}$  & ~~~ & $3.2^{+0.6}_{-0.7}$    & ~~~ & $3.6^{+0.6}_{-0.7}$  & ~~~ & $4.0^{+0.6}_{-0.7}$ & ~~~ & $4.2^{+0.6}_{-0.7}$ & ~~~ & $4.4^{+0.6}_{-0.7}$  \\
            & ~ & $\chi^2$/DOF        & ~~~ & 1382.7/1280            & ~~~ & 1379.9/1280          & ~~~ & 1377.5/1280         & ~~~ & 1376.2/1280         & ~~~ & 1375.4/1280          \\
\hline                    
&&&&&&&&&&&  \\			                                
            & ~ & ${\dot M}$          & ~~~ & $5.0^{+0.8}_{-0.9}$    & ~~~ & $5.8^{+1.0}_{-0.9}$  & ~~~ & $7.9^{+1.1}_{-1.2}$ & ~~~ & $9.2^{+1.3}_{-1.4}$ & ~~~ & $10.1^{+1.4}_{-1.5}$ \\
Virgo       & ~ & $\Delta N_{\rm H}$  & ~~~ & $1.4^{+0.5}_{-0.7}$    & ~~~ & $1.9^{+0.5}_{-0.7}$  & ~~~ & $2.7^{+0.5}_{-0.6}$ & ~~~ & $3.1^{+0.4}_{-0.5}$ & ~~~ & $3.3^{+0.4}_{-0.5}$  \\
            & ~ & $\chi^2$/DOF        & ~~~ & 2022.9/1150            & ~~~ & 2017.9/1150          & ~~~ & 2004.0/1150         & ~~~ & 1996.3/1150         & ~~~ & 1991.5/1150          \\
\hline                    
&&&&&&&&&&&  \\			                                
            & ~ & ${\dot M}$          & ~~~ & $33.3^{+2.6}_{-2.0}$   & ~~~ & $36.3^{+2.4}_{-2.8}$ & ~~~ & $41.7^{+3.2}_{-3.2}$& ~~~ & $45.3^{+3.6}_{-3.6}$& ~~~ & $47.7^{+3.9}_{-3.8}$ \\
Centaurus   & ~ & $\Delta N_{\rm H}$  & ~~~ & $1.7^{+0.2}_{-0.3}$    & ~~~ & $1.9^{+0.3}_{-0.3}$  & ~~~ & $2.3^{+0.3}_{-0.3}$ & ~~~ & $2.6^{+0.3}_{-0.3}$ & ~~~ & $2.7^{+0.3}_{-0.3}$  \\
            & ~ & $\chi^2$/DOF        & ~~~ & 2708.4/1549            & ~~~ & 2696.3/1549          & ~~~ & 2681.6/1549         & ~~~ & 2678.2/1549         & ~~~ & 2678.0/1549          \\
\hline                    
&&&&&&&&&&&  \\			                                
            & ~ & ${\dot M}$          & ~~~ & $222^{+69}_{-67}$      & ~~~ & $246^{+71}_{-70}$    & ~~~ & $284^{+77}_{-77}$  & ~~~ & $306^{+186}_{-80}$  & ~~~ & $321^{+83}_{-83}$   \\
Abell 1795  & ~ & $\Delta N_{\rm H}$  & ~~~ & $2.3^{+0.7}_{-0.4}$    & ~~~ & $2.6^{+0.7}_{-0.5}$  & ~~~ & $2.9^{+0.7}_{-0.5}$ & ~~~ & $3.0^{+0.8}_{-0.5}$ & ~~~ & $3.2^{+0.7}_{-0.5}$  \\
            & ~ & $\chi^2$/DOF        & ~~~ & 1422.7/1233            & ~~~ & 1421.5/1233          & ~~~ & 1420.3/1233         & ~~~ & 1419.6/1233         & ~~~ & 1419.2/1233          \\
\hline                    
&&&&&&&&&&&  \\			                                
            & ~ & ${\dot M}$          & ~~~ & $125^{+17}_{-40}$      & ~~~ & $147^{+20}_{-46}$    & ~~~ & $168^{+22}_{-54}$   & ~~~ & $186^{+23}_{-59}$   & ~~~ & $197^{+24}_{-64}$    \\
Abell 2199  & ~ & $\Delta N_{\rm H}$  & ~~~ & $3.2^{+0.5}_{-0.7}$    & ~~~ & $3.5^{+0.5}_{-0.7}$  & ~~~ & $3.8^{+0.5}_{-0.6}$ & ~~~ & $4.0^{+0.5}_{-0.7}$ & ~~~ & $4.2^{+0.5}_{-0.7}$  \\
            & ~ & $\chi^2$/DOF        & ~~~ & 1550.1/1292            & ~~~ & 1548.7/1292          & ~~~ & 1547.6/1292         & ~~~ & 1547.1/1292         & ~~~ & 1546.8/1292          \\
\hline                    
&&&&&&&&&&&  \\			                                
            & ~ & ${\dot M}$          & ~~~ & $613^{+59}_{-186}$     & ~~~ & $717^{+66}_{-208}$    & ~~~ & $867^{+77}_{-247}$   & ~~~ & $963^{+85}_{-277}$   & ~~~ &  $1029^{+91}_{-299}$ \\
Abell 2597  & ~ & $\Delta N_{\rm H}$  & ~~~ & $4.7^{+0.7}_{-0.6}$    & ~~~ & $5.0^{+0.7}_{-0.7}$  & ~~~ & $5.4^{+0.7}_{-0.7}$ & ~~~ & $5.6^{+0.7}_{-0.7}$ & ~~~ & $5.7^{+0.7}_{-0.6}$  \\
            & ~ & $\chi^2$/DOF        & ~~~ & 1112.2/981             & ~~~ & 1110.2/981           & ~~~ & 1108.4/981          & ~~~ & 1107.5/981          & ~~~ & 1106.9/981 \\
&&&&&&&&&&&  \\			                                
\hline                    
&&&&&&&&&&&  \\			                                
\end{tabular}
\end{center}
\parbox {7in}
{}
\end{table*}

\clearpage

\begin{table*}
\vskip 0.2truein
\begin{center}
\caption{The results from the fits to individual SIS data sets (instrument 
specified in parentheses) over the $3.0-10.0$ keV energy range, using 
spectral model A.}
\vskip 0.2truein
\begin{tabular}{ c c c c c }
\hline                                             
Cluster         & ~ &  $kT$                   &  $Z$           & $\chi^2$/DOF \\
                & ~ &  (keV)                  &  ($Z_{\odot}$) &              \\
\hline       
&&&& \\       
Abell 426 (S1)  & ~ & $5.77^{+0.46}_{-0.41}$ & $0.46^{+0.07}_{-0.06}$  & 148.1/143 \\  
Abell 496 (S0)  & ~ & $4.14^{+0.38}_{-0.34}$ & $0.35^{+0.07}_{-0.07}$  & 117.7/118 \\  
Virgo (S0)      & ~ & $2.17^{+0.11}_{-0.11}$ & $0.51^{+0.10}_{-0.09}$  & 108.5/108 \\  
Centaurus (S1)  & ~ & $3.18^{+0.18}_{-0.16}$ & $0.97^{+0.13}_{-0.11}$  & 146.5/142 \\  
Coma (S0)       & ~ & $8.48^{+1.04}_{-0.85}$ & $0.17^{+0.06}_{-0.06}$  & 146.7/149 \\  
Abell 1795 (S0) & ~ & $6.10^{+0.49}_{-0.43}$ & $0.38^{+0.06}_{-0.05}$  & 165.0/146 \\  
Abell 2199 (S0) & ~ & $4.80^{+0.42}_{-0.37}$ & $0.40^{+0.07}_{-0.07}$  & 153.3/126 \\ 
Abell 2597 (S1) & ~ & $4.05^{+0.66}_{-0.53}$ & $0.34^{+0.13}_{-0.11}$  & 67.6/78 \\  
&&&& \\       
\hline 
&&&& \\       
\end{tabular}
\end{center}
\parbox {7in}
{} 
\end{table*}

\begin{table*}
\vskip 0.2truein
\begin{center}
\caption{The best-fit parameter values and 90 per cent  ($\Delta \chi^2 =
2.71$) confidence limits from the spectral analysis with the two 
temperature model incorporating the
power-law components. The normalizations of the power-law components 
($A_1$) are quoted at an energy of 1keV in units of $10^{-5}$ photon
keV$^{-1}$cm$^{-2}$s$^{-1}$. The $2-10$ keV fluxes associated with the 
power-law components  ($F_{\rm X, 2-10}$) are in units of \ergpcmsqps~and 
are not corrected for absorption. The intrinsic luminosities in the 
$1-10$ keV band ($L_{\rm X, 1-10}$) are corrected for absorption and
are in units of \ergps. $\Delta \chi^2$ values are the improvements in 
$\chi^2$ obtained with the introduction of the power-law component into 
the best-fitting two-temperature model (model D; or in the case of the
Coma Cluster the best fitting single temperature model B). The
final $\chi^2$ values and number of degrees of freedom in the fits are also
listed ($\chi^2$/DOF). The results for the Virgo and Centaurus clusters 
are from Allen \etal (2000). For these two systems the two temperature models 
incorporating variable element abundances (Section 4) were used in the 
analyses.}
\vskip 0.2truein
\begin{tabular}{ c c c c c c c c c }
%&&&&&&&& \\   
\hline                                                                                                                                                         
        Parameters   & ~~~ &     Abell 426          & ~~~ & Abell 496              & ~~~ &    Virgo                & ~~~ & Centaurus                 \\
\hline                                                                                                                                                                                        
&&&&&&&& \\   
        $kT$         & ~~~ & $4.43^{+0.35}_{-0.27}$ & ~~~ & $3.42^{+0.12}_{-0.12}$ & ~~~ & $2.01^{+0.04}_{-0.03}$ & ~~~ & $3.27^{+0.26}_{-0.12}$ \\                
        $Z$          & ~~~ & $0.78^{+0.08}_{-0.07}$ & ~~~ & $0.53^{+0.06}_{-0.03}$ & ~~~ & $0.72^{+0.04}_{-0.04}$ & ~~~ & $0.86^{+0.03}_{-0.07}$ \\                
        $N_{\rm H}$  & ~~~ & $0.43^{+0.19}_{-0.23}$ & ~~~ & $0.64^{+0.10}_{-0.16}$ & ~~~ & $0.30^{+0.08}_{-0.08}$ & ~~~ & $0.96^{+0.22}_{-0.26}$ \\                
        ${\dot M}$   & ~~~ &  ---                   & ~~~ & ---                    & ~~~ & ---                    & ~~~ & ---                    \\                
        $kT_2$       & ~~~ & $2.14^{+0.45}_{-0.42}$ & ~~~ & $0.50^{+0.12}_{-0.10}$ & ~~~ & $0.79^{+0.05}_{-0.05}$ & ~~~ & $1.46^{+0.07}_{-0.06}$ \\                
 $\Delta N_{\rm H}$  & ~~~ & $3.3^{+0.5}_{-0.4}$    & ~~~ & $12.3^{+2.8}_{-1.7}$   & ~~~ & $4.88^{+0.72}_{-0.86}$ & ~~~ & $0.59^{+1.01}_{-0.54}$ \\                
        $\Gamma$     & ~~~ & $2.05^{+0.05}_{-0.05}$ & ~~~ & $1.44^{+0.44}_{-0.38}$ & ~~~ & $1.40^{+0.37}_{-0.46}$ & ~~~ & $0.76^{+0.59}_{-0.65}$ \\      
        $A_1$        & ~~~ & $7878^{+1007}_{-1061}$ & ~~~ & $100^{+150}_{-56}$     & ~~~ & $137^{+159}_{-55}$     & ~~~ & $22.0^{+62.1}_{-16.4}$ \\   
%&&&&&&&& \\   
  $F_{\rm X, 2-10}$  & ~~~ & $1.80^{+0.12}_{-0.14}\times 10^{-10}$ & ~~~ & $5.58^{+1.42}_{-1.00}\times 10^{-12}$ & ~~~ & $8.69^{+1.70}_{-1.56} \times 10^{-12}$ & ~~~ & $4.20^{+1.29}_{-0.99}  \times 10^{-12}$  \\ 
  $L_{\rm X, 1-10}$  & ~~~ & $3.96^{+0.34}_{-0.36}\times 10^{44}$  & ~~~ & $3.29^{+1.49}_{-0.84}\times 10^{43}$  & ~~~ & $4.22^{+1.37}_{-1.06} \times 10^{41}$  & ~~~ & $2.14^{+0.99}_{-0.60} \times 10^{42}$  \\ 
     $\Delta \chi^2$ & ~~~ & 171                    & ~~~ & 49.7                   & ~~~ & 90.0                   & ~~~ & 38.8                      \\           
        $\chi^2$/DOF & ~~~ & 2058/1653              & ~~~ & 1269/1277              & ~~~ & 1468/1143              & ~~~ & 954.7/746                 \\           
&&&&&&&& \\   
&&&&&&&& \\   
                     & ~~~ &   Abell 1795           & ~~~ &     Abell 2199         & ~~~ &  Abell  2597  & ~~~ &     Coma \\
\hline                                                                                                                                                                                        
        $kT$         & ~~~ & $5.50^{+0.51}_{-0.88}$ & ~~~ & $3.64^{+1.11}_{-0.19}$ & ~~~ & $3.06^{+0.25}_{-0.22}$ & ~~~ & $8.46^{+0.63}_{-0.48}$ \\           
        $Z$          & ~~~ & $0.36^{+0.04}_{-0.03}$ & ~~~ & $0.44^{+0.04}_{-0.06}$ & ~~~ & $0.45^{+0.09}_{-0.09}$ & ~~~ & $0.20^{+0.06}_{-0.04}$ \\           
        $N_{\rm H}$  & ~~~ & $0.00^{+0.09}_{-0.00}$ & ~~~ & $0.27^{+0.15}_{-0.27}$ & ~~~ & $0.36^{+0.35}_{-0.36}$ & ~~~ & $0.54^{+0.28}_{-0.24}$ \\           
        ${\dot M}$   & ~~~ & ---                    & ~~~ &  ---                   & ~~~ & ---                    & ~~~ & ---                    \\           
        $kT_2$       & ~~~ & $1.38^{+0.46}_{-0.45}$ & ~~~ & $0.91^{+0.39}_{-0.46}$ & ~~~ & $0.86^{+0.16}_{-0.11}$ & ~~~ & ---                 \\           
 $\Delta N_{\rm H}$  & ~~~ & $7.2^{+5.7}_{-3.3}$    & ~~~ & $10.8^{+6.7}_{-3.8}$   & ~~~ & $12.4^{+4.1}_{-1.7}$   & ~~~ & ---                    \\           
        $\Gamma$     & ~~~ & $0.12^{+0.98}_{-2.05}$ & ~~~ & $1.01^{+0.78}_{-1.61}$ & ~~~ & $1.80^{+0.42}_{-0.66}$ & ~~~ & $3.34^{+0.82}_{-1.63}$   \\ 
        $A_1$        & ~~~ & $5.96^{+80.0}_{-5.92}$ & ~~~ & $62.3^{+239}_{-61.6}$  & ~~~ & $133^{+187}_{-102}$    & ~~~ & $310^{+191}_{-170}$   \\ 
%&&&&&&&& \\   
  $F_{\rm X, 2-10}$  & ~~~ & $3.53^{+5.51}_{-2.20}\times 10^{-12}$ & ~~~ & $7.24^{+3.02}_{-5.65}\times 10^{-12}$ & ~~~ & $4.21^{+1.25}_{-1.40}\times 10^{-12}$ & ~~~ & $1.28^{+17.5}_{-0.80}\times 10^{-12}$ \\ 
  $L_{\rm X, 1-10}$  & ~~~ & $5.96^{+12.45}_{-4.10}\times 10^{43}$ & ~~~ & $3.52^{+2.74}_{-1.00}\times 10^{43}$  & ~~~ & $1.99^{+1.11}_{-0.92}\times 10^{44}$  & ~~~ & $8.49^{+48.6}_{-3.78}\times 10^{42}$ \\ 
     $\Delta \chi^2$ & ~~~ & 10.8                   & ~~~ & 7.3                    & ~~~ & 32.2                    & ~~~ & 9.4                    \\           
        $\chi^2$/DOF & ~~~ & 1376/1230              & ~~~ & 1509/1289              & ~~~ & 1051/978                & ~~~ & 1265/1225               \\           
&&&&&&&& \\   
\hline                                                                                                                                                   
&&&&&&&& \\   
\end{tabular}
\end{center}
\parbox {7in}
{}
\end{table*}

\clearpage

\begin{table*}
\vskip 0.2truein
\begin{center}
\caption{The results on the individual element abundances in the central regions of 
the Virgo and Centaurus clusters 
using the two-temperature model with free-fitting element abundances 
(see Section 5). Error bars are the 90 per cent confidence limits 
($\Delta \chi^2=2.71$) 
on a single interesting parameter. The abundances of elements not listed
in the table were linked to have the same ratios relative to their solar values 
as that of Fe.}
\vskip 0.2truein
\begin{tabular}{ c c c c c }
%&&&& \\                                                                                                                                               
\hline                                                                                                                                                         
Parameter          & ~ &      Virgo             & ~~~ & Centaurus \\
\hline                                                                                           
&&&& \\
$kT$               & ~ & $2.13^{+0.04}_{-0.04}$ & ~~~ & $4.09^{+0.61}_{-0.59}$ \\    
$N_{\rm H}$        & ~ & $0.33^{+0.08}_{-0.08}$ & ~~~ & $0.85^{+0.24}_{-0.22}$ \\     
$\Delta N_{\rm H}$ & ~ & $3.41^{+0.96}_{-1.23}$ & ~~~ & $0.62^{+0.74}_{-0.62}$ \\ 
$kT_2$             & ~ & $0.86^{+0.06}_{-0.05}$ & ~~~ & $1.55^{+0.09}_{-0.11}$  \\ 
Fe                 & ~ & $0.79^{+0.05}_{-0.04}$ & ~~~ & $0.93^{+0.05}_{-0.04}$  \\ 
Mg                 & ~ & $0.28^{+0.11}_{-0.10}$ & ~~~ & $0.30^{+0.20}_{-0.15}$  \\ 
Si                 & ~ & $1.20^{+0.09}_{-0.09}$ & ~~~ & $1.29^{+0.19}_{-0.14}$  \\ 
Na                 & ~ & $11.0^{+3.4}_{-3.2}$   & ~~~ & ---  \\ 
S                  & ~ & $1.00^{+0.10}_{-0.09}$ & ~~~ & $1.25^{+0.24}_{-0.16}$  \\ 
O                  & ~ & ---                    & ~~~ & $0.00^{+0.06}_{-0.00}$  \\ 
$\chi^2$/DOF       & ~ & 468.9/357              & ~~~ & 322.3/216              \\                 
&&&& \\                                                                                                                                               
\hline                                                                                                                                                   
&&&& \\                                                                                                                                               
\end{tabular}
\end{center}
 
\parbox {7in}
{} 
\end{table*}

\begin{table*}
\vskip 0.2truein
\begin{center}
\caption{Results from the deprojection analysis of the ROSAT images. 
Column 2 lists the binsize (in arcsec and kpc) used in the
analysis. Columns $3-5$ summarize the velocity dispersions, core
radii and galaxy linear mass (GLM) components (where required) used to 
parameterize the cluster mass profiles. Columns 6-8 summarize the results 
on the cooling time ($t_{\rm cool}$; the time 
for the gas in the central bin to cool from the ambient cluster
temperature at constant pressure), the cooling radius ($r_{\rm cool}$; the radius where the cooling 
time of the cluster gas first exceeds a Hubble time) and the mass deposition 
rates (${\dot M_{\rm I}}$; the integrated mass deposition rates within 
the cooling radii). Errors on the velocity dispersions are 
90 per cent confidence limits. Errors on the cooling times are the 10 and 90 percentile values from 100 
Monte Carlo simulations. The 
upper and lower confidence limits on the cooling radii are the
points where the 10 and 90 percentiles exceed and become less than the Hubble
time, respectively. Errors on the mass deposition rates are the 90 and 10
percentile values at the upper and lower limits for the cooling radius.
We note that the errors on the deprojection results do not account 
for uncertainties in the total mass profiles. }
\vskip 0.2truein
\begin{tabular}{ c c c c c c c c c c}
\hline                                                                                                                               
\multicolumn{1}{c}{} &
\multicolumn{1}{c}{} &
\multicolumn{1}{c}{} &
\multicolumn{3}{c}{MASS PROFILES} &
 \multicolumn{1}{c}{} &
\multicolumn{3}{c}{DEPROJECTION RESULTS} \\
&&&&&&&&& \\                                                                                                                                                                   
Cluster     &   binsize       & ~ &  $\sigma$      & $r_{\rm c}$ & GLM                 & ~ &   $t_{\rm cool}$          &   $r_{\rm cool}$  &   ${\dot M_{\rm I}}$             \\
            &   (arcsec/kpc)  & ~ &  (\kmps)        &    (kpc)   & (\Msun kpc$^{-1}$)  & ~ &  ($10^{9}$ yr)            &     (kpc)         &   (\Msunpyr) \\  
\hline                                                                                                                                                         
&&&&&&&&& \\                                                                                                                                                                         
Abell 426   &  30/15.5        & ~ &  $670^{+10}_{-10}$      & 40   &  ---                & ~ &   $0.74^{+0.04}_{-0.04}$  & $188^{+36}_{-10}$  & $587^{+94}_{-43}$ \\  
Abell 496   &  12/10.6        & ~ &  $550^{+10}_{-10}$      & 40   & $8.0\times 10^{10}$ & ~ &   $0.55^{+0.15}_{-0.11}$  & $126^{+81}_{-17}$ & $111^{+194}_{-44}$   \\  
Virgo       &  30/2.60        & ~ &  $450^{+10}_{-10}$      & 50   & $8.0\times 10^{10}$ & ~ &   $0.10^{+0.01}_{-0.01}$  & $106^{+28}_{-19}$ & $23.2^{+9.1}_{-5.6}$   \\  
%model E Virgo       &  30/2.60        & ~ &  $420^{+10}_{-10}$      & 50   & $8.0\times 10^{10}$ & ~ &   $0.10^{+0.01}_{-0.01}$  & $107^{+24}_{-17}$ & $24.0^{+10.6}_{-5.7}$   \\  
Centaurus   &  30/8.91        & ~ &  $430^{+10}_{-10}$      & 35   & $1.2\times 10^{11}$ & ~ &   $0.44^{+0.03}_{-0.03}$  & $124^{+41}_{-12}$ & $39.3^{+25.5}_{-7.8}$ \\  
Coma        &  30/19.4        & ~ &  $760^{+30}_{-20}$      & 300  & $5.0\times 10^{10}$ & ~ &   $14.2^{+8.3}_{-3.8}$    &  $<29$             & $<1$                    \\  
Abell 1795  &  16/26.5        & ~ &  $700^{+20}_{-20}$      & 50   & $1.0\times 10^{11}$ & ~ &   $1.37^{+0.09}_{-0.08}$     & $191^{+34}_{-19}$  & $462^{+108}_{-56}$     \\  
Abell 2199  &  16/13.6        & ~ &  $570^{+10}_{-10}$      & 80   & $1.5\times 10^{11}$ & ~ &   $1.01^{+0.05}_{-0.04}$  & $173^{+11}_{-30}$ & $197^{+20}_{-41}$ \\  
Abell 2597  &  12/25.8        & ~ &  $610^{+10}_{-10}$      & 40   & ---                 & ~ &   $0.86^{+0.10}_{-0.08}$  & $162^{+82}_{-21}$  & $423^{+91}_{-99}$   \\  
&&&&&&&& \\                                                                                                                                                           
\hline                                                                                                                              
&&&&&&&& \\                                                                                                                                                           
\end{tabular}			    
\end{center}
\parbox {7in}
{}
\end{table*}

\clearpage

\begin{table*}
\vskip 0.2truein
\begin{center}
\caption{A comparison of the integrated mass deposition rates determined from 
the deprojection analysis, both before correction (${\dot M_{\rm I}}$) and
after correction for the effects of intrinsic absorption due to a uniform 
screen of cold gas (${\dot M_{\rm C}}$; Section 7.1), with the values 
determined from the 
spectral analysis (${\dot M_{\rm S}}$) using spectral model E (model C for the
Virgo Cluster). Errors on the corrected deprojection results
account for the statistical uncertainties in the deprojected quantities but 
not the errors in the the intrinsic column densities.}
\vskip 0.2truein
\begin{tabular}{ c c c c c c }
\hline                                                                                                                               
\multicolumn{1}{c}{} &
\multicolumn{1}{c}{} &
\multicolumn{1}{c}{INTRINSIC} &
\multicolumn{1}{c}{ORIGINAL} &
\multicolumn{1}{c}{CORRECTED} &
\multicolumn{1}{c}{SPECTRAL} \\
\multicolumn{1}{c}{} &
\multicolumn{1}{c}{} &
\multicolumn{1}{c}{ABSORPTION} &
\multicolumn{1}{c}{DEPROJ.} &
\multicolumn{1}{c}{DEPROJ.} &
\multicolumn{1}{c}{ANALYSIS} \\
&&&& \\       
Cluster         & ~ & $\Delta N_{\rm H}$       &  ${\dot M_{\rm I}}$ & ${\dot M_{\rm C}}$ & ${\dot M_{\rm S}}$         \\
                & ~ & ($10^{21}$ \apc)         &  (\Msunpyr) & (\Msunpyr)  & (\Msunpyr)   \\
\hline                                                                                                                    
&&&& \\       
Abell 426       & ~ &  $3.4^{+0.4}_{-0.4}$    &  $587^{+94}_{-43}$     & $1175^{+319}_{-112}$   &  $303^{+75}_{-74}$      \\   
Abell 496       & ~ &  $3.6^{+0.6}_{-0.7}$    &  $111^{+194}_{-44}$    & $289^{+470}_{-208}$    &  $110^{+39}_{-42}$ \\  
Virgo           & ~ &  $3.76^{+0.28}_{-0.48}$ &  $23.2^{+9.1}_{-5.6}$  & $47.5^{+29.6}_{-15.7}$ &  $10.7^{+1.0}_{-1.4}$   \\  
%model E Virgo           & ~ &  $1.88^{+0.53}_{-0.65}$ &  $24.0^{+10.6}_{-5.7}$ & $39.8^{+12.4}_{-11.8}$ &  $5.8^{+1.0}_{-0.9}$    \\  
Centaurus       & ~ &  $1.92^{+0.25}_{-0.33}$ &  $39.3^{+25.5}_{-7.8}$ & $71.9^{+22.6}_{-24.6}$ &  $36.3^{+2.4}_{-2.8}$   \\  
Abell 1795      & ~ &  $2.6^{+0.7}_{-0.5}$    &  $462^{+108}_{-56}$    & $1212^{+164}_{-208}$   &  $246^{+71}_{-70}$      \\ 
Abell 2199      & ~ &  $3.5^{+0.5}_{-0.7}$    &  $197^{+20}_{-41}$     & $599^{+79}_{-77}$      &  $147^{+20}_{-46}$      \\   
Abell 2597      & ~ &  $5.0^{+0.7}_{-0.7}$    &  $423^{+91}_{-99}$     & $1159^{+268}_{-654}$   &  $717^{+66}_{-208}$     \\  
&&&& \\       
\hline                                                                                                                                                   
&&&& \\       
\end{tabular}
\end{center}
\parbox {7in}
{}
\vskip 1cm
\end{table*}

\begin{table*}
\vskip 0.2truein
\begin{center}
\caption{A summary of the results on intrinsic absorption 
from the ASCA data. Columns $2-6$ list the differences between the 
spectrally-determined column densities and the nominal Galactic values (in 
units of $10^{21}$\apc) for spectral models B, C, C', D and E. 
Also listed are the (90 per cent
confidence) constraints on the covering fraction of the intrinsic
absorber, $f$, determined using spectral model E. All results assume solar 
metallicity in the absorbing material. The errors on the mean values are the standard
deviations for the sample of cooling-flow clusters (the Coma cluster 
is excluded).}
\vskip 0.2truein
\begin{tabular}{ c c c c c c c }

%\multicolumn{1}{c}{} &
%\multicolumn{1}{c}{} &
%\multicolumn{1}{c}{} &
%\multicolumn{1}{c}{} &
%\multicolumn{1}{c}{} &
%\multicolumn{1}{c}{} &
%\multicolumn{1}{c}{} \\
\hline
Cluster           &        B-A               &     C-A              &          C'-A          &       D-A              &     E-A                 & $f$                    \\
\hline                                                                                                                                          	                                     	   
&&&&&& \\	  						                                                                                	                                     	   
Abell 426         & $-0.12^{+0.03}_{-0.03}$  & $3.6^{+0.4}_{-0.5}$  & $-0.11^{+0.05}_{-0.04}$& $2.5^{+0.5}_{-0.4}$    &  $3.4^{+0.4}_{-0.4}$    & $>0.99$                \\
Abell 496         & $0.37^{+0.05}_{-0.05}$   & $4.3^{+0.6}_{-0.9}$  & $0.38^{+0.10}_{-0.06}$ & $11.9^{+2.0}_{-1.5}$   &  $3.6^{+0.6}_{-0.7}$    & $>0.97$                \\ 
Virgo             & $0.10^{+0.04}_{-0.03}$   & $3.8^{+0.3}_{-0.5}$  & $0.65^{+0.06}_{-0.06}$ & $5.0^{+0.6}_{-0.8}$    &  $1.88^{+0.53}_{-0.65}$ & $0.66^{+0.15}_{-0.16}$ \\ 
Centaurus         & $0.20^{+0.03}_{-0.03}$   & $2.5^{+0.4}_{-0.4}$  & $1.25^{+0.06}_{-0.06}$ & $0.23^{+0.35}_{-0.23}$ &  $1.92^{+0.25}_{-0.33}$ & $>0.98$                \\
Coma              & $0.15^{+0.05}_{-0.05}$   &       ---            &           ---          &       ---              &       ---               &  ---                   \\ 
Abell 1795        & $0.05^{+0.05}_{-0.05}$   & $3.1^{+0.7}_{-0.5}$  & $0.10^{+0.11}_{-0.09}$ & $5.2^{+2.8}_{-2.9}$    &  $2.6^{+0.7}_{-0.5}$    & $>0.85$                \\ 
Abell 2199        & $0.31^{+0.05}_{-0.05}$   & $4.1^{+0.5}_{-0.9}$  & $0.39^{+0.09}_{-0.09}$ & $2.8^{+1.6}_{-1.1}$    &  $3.5^{+0.5}_{-0.7}$    & $>0.96$                \\ 
Abell 2597        & $0.74^{+0.07}_{-0.08}$   & $5.4^{+0.9}_{-0.9}$  & $0.74^{+0.07}_{-0.08}$ & $2.4^{+1.7}_{-2.4}$    &  $5.0^{+0.7}_{-0.7}$    & $>0.98$                \\ 
&&&&&& \\	  						            					                                
\hline                                                           						                                
MEAN              & $0.24 \pm 0.28$          & $3.8 \pm 0.9$        & $0.49 \pm 0.45$      &    $4.3 \pm 3.8$       & $3.1 \pm 1.1$           &                        \\
\hline                                                                                       
&&&&&& \\
\end{tabular}
\end{center}
\parbox {7in}
{}
\end{table*}

\clearpage

\begin{table*}
\vskip 0.2truein
\begin{center}
\caption{Column 2 lists the X-ray luminosities expected to be reprocessed into 
(predominantly) far infrared emission by the intrinsic absorbing material in the 
clusters. The best-fit parameters determined with spectral model E have been 
used in the calculations (except for the Virgo cluster, for which model C
is preferred). Columns 3 and 4 lists the observed $60$ and 
$100\mu$m IRAS fluxes within a 4 arcmin (radius) aperture 
centered on the X-ray centres (Table 3) measured using the IPAC SCANPI 
software and 
co-added IRAS scans. Error bars are the root-mean-square deviations in the residuals, 
external to the source extraction regions, after baseline subtraction.  
Where no detection was made an upper limit equal to three times the r.m.s.
deviation in the residuals is given. Column 5 lists the in-scan separation 
$\Delta R$ (in arcmin) between the peak of the $100\mu$m emission and the X-ray centre. 
Column 6 lists the total $1-1000\mu$m luminosities calculated using equation 2.
For Abell 2597, the predicted reprocessed luminosity may be overestimated due 
to the effects of radiation damage to the SIS detectors (see 
Section 7.2.)}
\vskip 0.2truein
\begin{tabular}{ c c c c c c }
\hline                                                               
\multicolumn{1}{c}{} &
\multicolumn{1}{c}{$L_{\rm repro.}$} &
\multicolumn{1}{c}{$S_{\rm 60}$} &
\multicolumn{1}{c}{$S_{\rm 100}$} &
\multicolumn{1}{c}{$\Delta R$} & 
\multicolumn{1}{c}{$L_{\rm 1-1000\mu m}$} \\
Cluster      &  ($10^{44}$ \ergps)  & (mJy) & (mJy) & (arcmin)   &  ($10^{44}$ \ergps)  \\
\hline                                             	      
&&&&& \\						      
Abell 426    & $2.4 \pm 0.7$          & $7020 \pm 66$ & $6310 \pm 440$ & 0.40   & $9.2 \pm 2.8$ \\
Abell 496    & $0.74 \pm 0.30$        & $60 \pm 38$   & $2210 \pm 130$ & -3.02  & $2.7 \pm 0.8$  \\
Virgo        & $4.3 \pm 0.7 \times 10^{-2}$   & $520 \pm 44$  & $80 \pm 99$ & 1.25   & $1.4 \pm 0.5  \times 10^{-2}$\\
Centaurus    & $0.20 \pm 0.02$        & $420 \pm 37$  & $890 \pm 182$  & 0.68   & $0.25 \pm 0.08$ \\
Coma         & ---    & $<141$        & $<408$         &  ---   & $<0.47$ \\
Abell 1795   & $1.9 \pm 0.5$          & $<144$        & $<354$         &  ---   & $<3.3$  \\
Abell 2199   & $0.99^{+0.15}_{-0.40}$ & $150 \pm 26$  & $510 \pm 82$   & -0.13  & $0.95 \pm 0.31$ \\ 
Abell 2597   & $5.1^{+0.4}_{-1.5}$    & $80 \pm 26$   & $130 \pm 134$  & 0.63   & $2.7 \pm 1.5$  \\
&&&&& \\	    					  
\hline              					  
&&&&& \\
\end{tabular}
\end{center}
\parbox {7in}
{}
\vskip 1cm
\end{table*}

\begin{table*}
\vskip 0.2truein
\begin{center}
\caption{The ages of the cooling flows, in units of Gyr, determined with the
three methods described in Section 8. Column 2 lists the measurements of 
Allen \& Fabian (1997) from their X-ray colour deprojection study (method 1).
Columns 3 and 4 lists the results obtained from the comparisons of the
mass deposition rates determined from the spectral and image deprojection
analyses, both excluding (no $\Delta N_{\rm H}$) and including ($\Delta
N_{\rm H}$) corrections for 
the effects of intrinsic absorption by a uniform screen of cold gas 
on the deprojection results (method 2). 
Column 5 summarizes the ages inferred from the breaks in the 
deprojected mass deposition profiles (method 3).}
\vskip 0.1truein
\begin{tabular}{ c c c c c }
\hline                                             
\multicolumn{1}{c}{Cluster} &
%\multicolumn{1}{c}{} &
\multicolumn{1}{c}{METHOD 1} &
\multicolumn{2}{c}{METHOD 2} &
\multicolumn{1}{c}{METHOD 3} \\
              &            &  (no $\Delta N_{\rm H}$) &   ($\Delta N_{\rm H}$) &                \\
\hline      	                     			                           
&&&& \\     	                     			                           
Abell 426     & ---        &  $2.16^{+0.40}_{-0.17}$  &  $1.69^{+0.23}_{-0.24}$  & $3.10^{+0.54}_{-0.72}$  \\  
Abell 496     & $3.3-6.8$  &  $9.44^{+14.0}_{-5.08}$ &  $2.94^{+1.53}_{-0.87}$  & $4.85^{+0.69}_{-0.80}$  \\  
Virgo         &  ---       &  $2.83^{+0.18}_{-0.29}$  &  $1.02^{+0.20}_{-0.16}$  & $3.32^{+0.71}_{-0.63}$  \\  
Centaurus     &  ---       &  $9.95^{+0.88}_{-0.97}$  &  $4.31^{+1.52}_{-0.74}$  &    ---                  \\  
Abell 1795    & $3.6-6.7$  &  $5.41^{+0.95}_{-1.66}$  &  $2.87^{+0.36}_{-0.35}$  & $6.76^{+1.29}_{-2.52}$  \\  
Abell 2199    & $5.3-7.7$  &  $8.98^{+2.34}_{-3.19}$  &  $3.91^{+0.50}_{-0.72}$  & $4.82^{+1.00}_{-0.84}$  \\ 
%Abell 2597   & $2.5-5.1$  &  $5.46^{+1.17}_{-1.26}$  &  $1.95^{+0.22}_{-0.25}$  & $3.30^{+2.28}_{-1.43}$  \\  
Abell 2597    & $2.5-5.1$  &  $> 5.2$                 &  $2.51^{+0.59}_{-0.80}$  & $3.30^{+2.28}_{-1.43}$  \\

&&&& \\       
\hline                                                                                                                                                   
&&&& \\       
\end{tabular}
\end{center}
\parbox {7in}
{}
\end{table*}

\clearpage

\begin{table*}
\vskip 0.2truein
\begin{center}
\caption{The results from fits to the mass deposition profiles determined 
from the deprojection analysis with the broken power-law models. Only those 
data from radii interior to the 90 percentile upper limit to the cooling radii 
were included in the modelling, with the exception of the Virgo cluster where 
(due to uncertainties associated with the effects of the PSPC rib support
structure) the analysis was limited to the central 78 kpc (15 arcmin) radius. 
Columns 2 and 3 list the slopes, $\eta_1$ and $\eta_2$, of the mass
deposition profiles interior to and external to the break radii ($r_{\rm
break}$; Column 4), respectively. Column 5 lists the mean cooling time of the 
cluster gas at the break radius (the errors on the cooling time are the 
differences relative to the values measured at the upper and lower limits to 
the break radii). Columns 6 and 7 list the integrated mass deposition rates 
within the break radii, with (${\dot M_{\rm C}} (r<r_{\rm break})$) and 
without (${\dot M_{\rm I}} (r<r_{\rm break})$) corrections for intrinsic 
absorption due to a uniform screen of cold gas. Error bars are pseudo-90 per 
cent confidence limits determined from the unweighted least-squares fits,
assuming a reduced $\chi^2$ value $\chi^2_\nu = 1.0$.}
\vskip 0.2truein
\begin{tabular}{ c c c c c c c c }
\hline                                             
Cluster         & ~ & $\eta_1$              & $\eta_2$                  & $r_{\rm break}$    &  $t_{\rm cool} (r_{\rm break})$    & ${\dot M_{\rm I}} (r<r_{\rm break})$   & ${\dot M_{\rm C}} (r<r_{\rm break})$   \\
                & ~ & $(r<r_{\rm break})$   & $(r>r_{\rm break})$       & (kpc)              &  ($10^9$ yr)   & (\Msunpyr)  & (\Msunpyr)  \\
\hline       	                                                          
&&&&&&& \\       	                                                          
Abell 426       & ~ & $1.58^{+0.47}_{-0.26}$  & $0.36^{+0.08}_{-0.08}$  & $82^{+9}_{-12}$    & $3.10^{+0.54}_{-0.72}$  & $417^{+29}_{-66}$    & $785^{+110}_{-173}$  \\  
Abell 496       & ~ & $1.19^{+0.27}_{-0.17}$  & $0.55^{+0.08}_{-0.08}$  & $77^{+11}_{-13}$   & $4.85^{+0.69}_{-0.80}$  & $87^{+10}_{-16}$     & $188^{+56}_{-54}$    \\  
Virgo           & ~ & $0.94^{+0.09}_{-0.07}$  & $0.65^{+0.06}_{-0.07}$  & $38^{+6}_{-5}$     & $3.32^{+0.71}_{-0.63}$  & $12.8^{+1.1}_{-1.9}$ & $29.6^{+3.8}_{-6.6}$ \\  
%model EVirgo   & ~ & $0.95^{+0.09}_{-0.07}$  & $0.66^{+0.06}_{-0.07}$  & $38^{+6}_{-5}$     & $3.08^{+0.52}_{-0.42}$  & $****^{+5.2}_{-5.5}$ & $20.9^{+5.2}_{-5.5}$ \\  
Abell 1795      & ~ & $1.35^{+0.18}_{-0.16}$  & $0.67^{+0.12}_{-0.11}$  & $119^{+19}_{-37}$  & $6.76^{+1.29}_{-2.52}$  & $305^{+60}_{-95}$    & $632^{+215}_{-243}$  \\  
Abell 2199      & ~ & $1.69^{+0.25}_{-0.22}$  & $0.92^{+0.06}_{-0.05}$  & $74^{+12}_{-8}$    & $4.82^{+1.00}_{-0.84}$  & $86^{+15}_{-12}$     & $227^{+61}_{-57}$    \\ 
Abell 2597      & ~ & $1.45^{+1.63}_{-0.73}$  & $0.37^{+0.31}_{-0.37}$  & $91^{+72}_{-45}$   & $3.30^{+2.28}_{-1.43}$  & $309^{+160}_{-165}$  & $895^{+562}_{-547}$  \\  
&&&&&&& \\       
\hline                                                                                                                                                   
&&&&&&& \\       
\end{tabular}
\end{center}
\parbox {7in}
{}

\end{table*}

\clearpage

%%%%%%%%%%%%%%%%% FIGURES %%%%%%%%%%%%%%%%%%%%%%%%%%%%%%%%

\begin{figure*}
\hbox{
\hspace{0cm}\psfig{figure=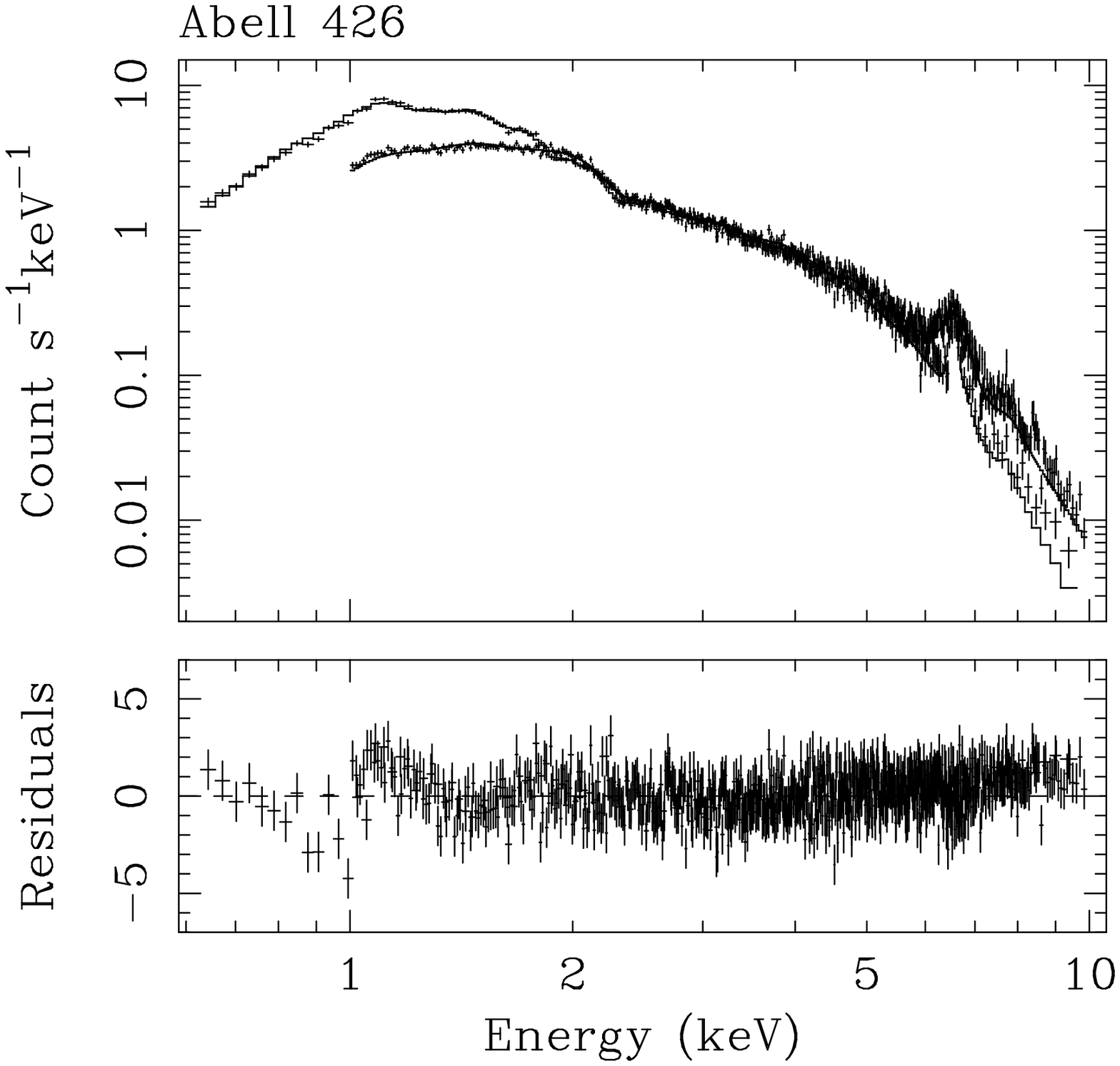,width=0.65\textwidth,angle=270}
\hspace{-2.5cm}\psfig{figure=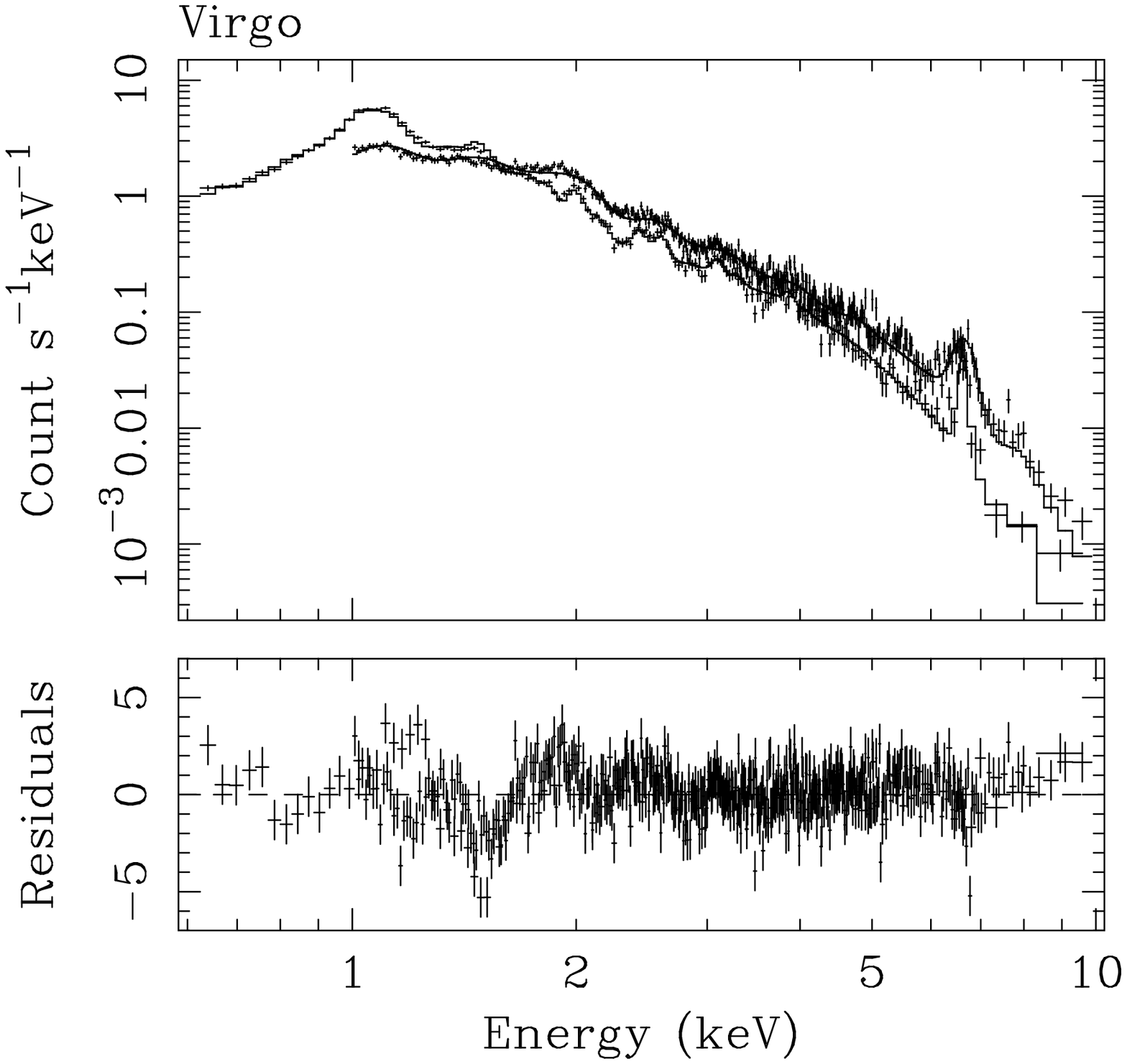,width=0.65\textwidth,angle=270}
}
\vspace{-0.2cm}
\hbox{
\hspace{0cm}\psfig{figure=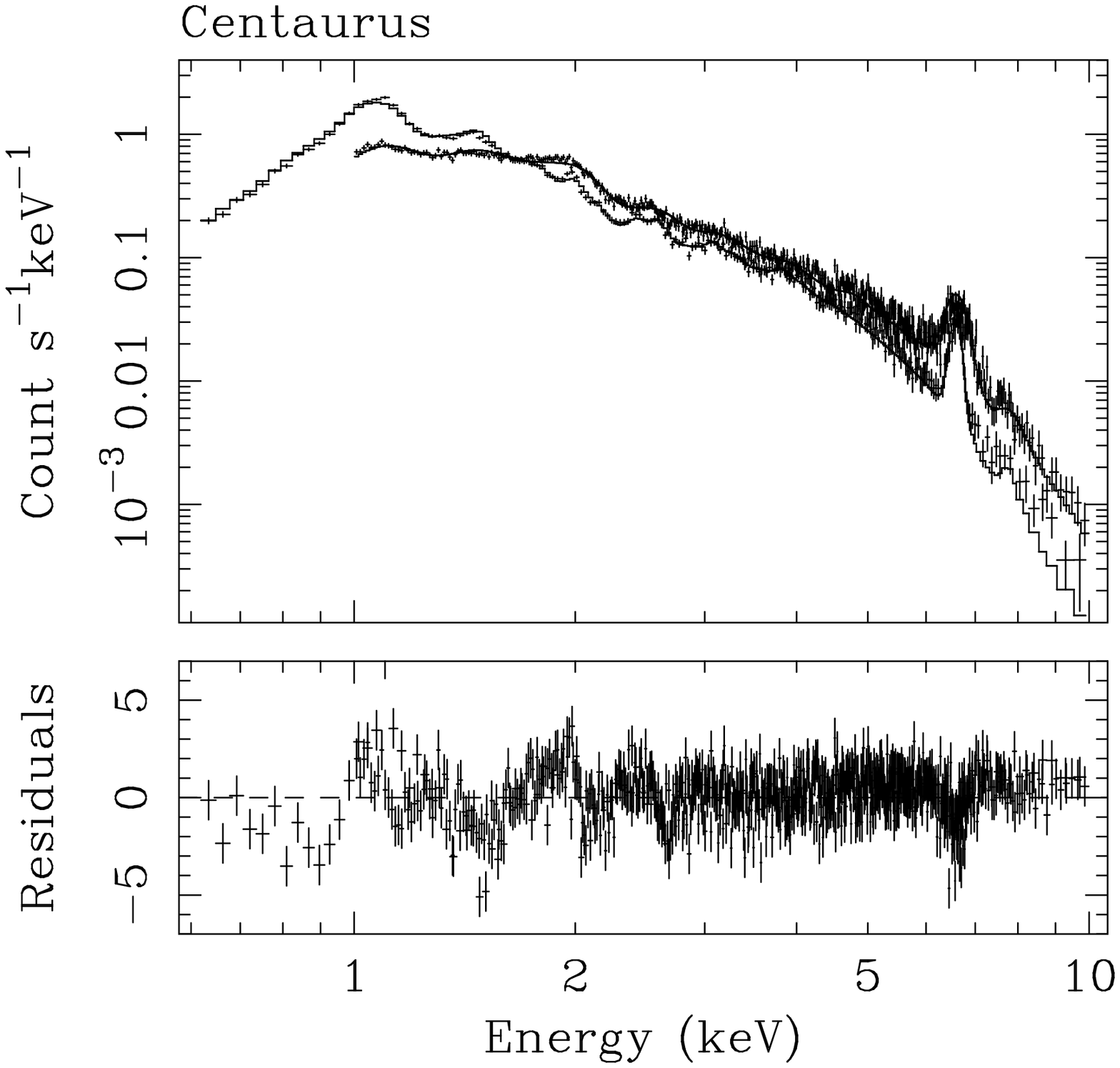,width=0.65\textwidth,angle=270}
\hspace{-2.5cm}\psfig{figure=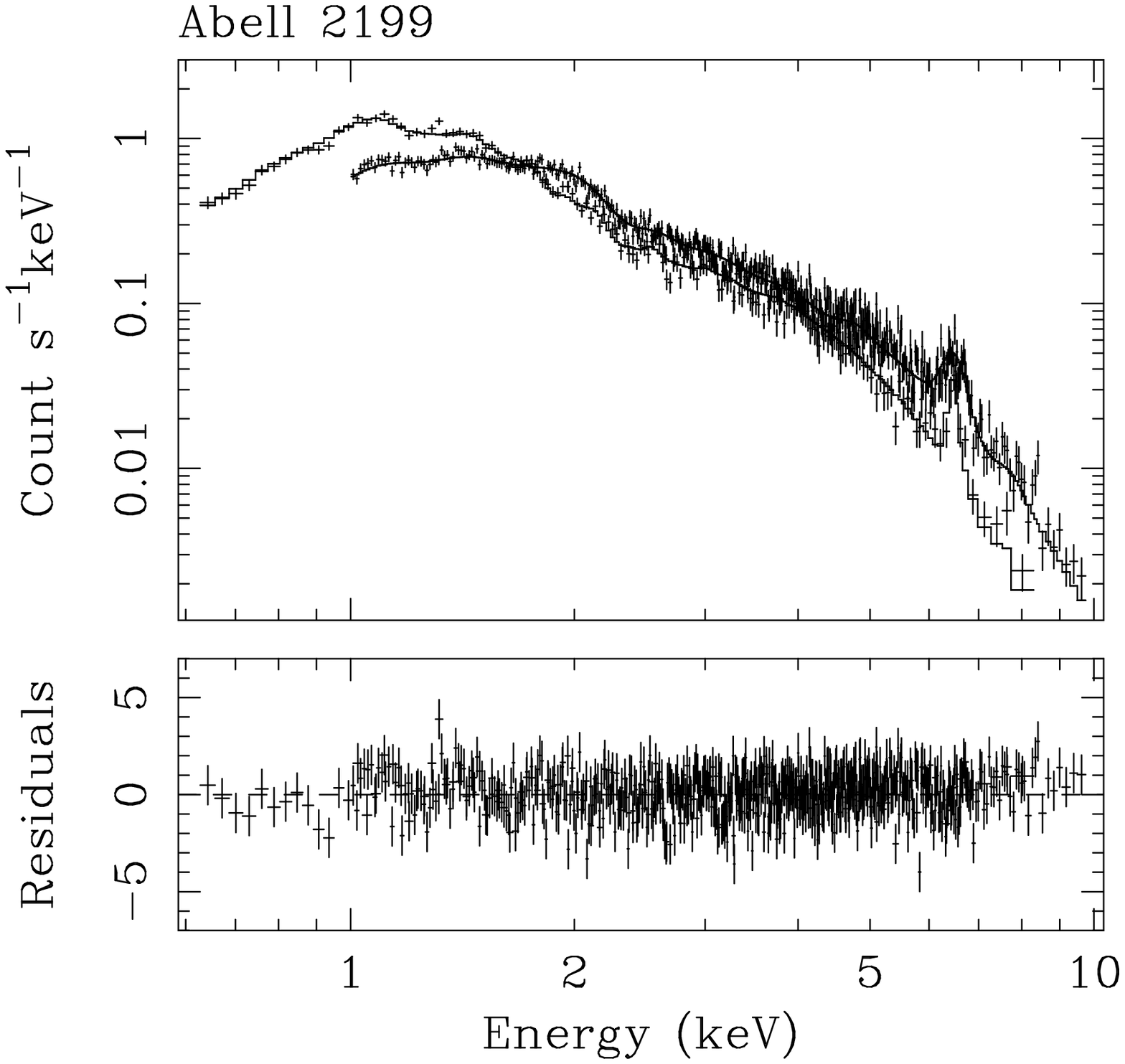,width=0.65\textwidth,angle=270}
}
\caption{(Upper panels) The ASCA data and best-fitting models 
and (lower panels) residuals to the fits (in units of $\chi$) 
for (clockwise from top left) Abell 426, the Virgo Cluster, Abell 2199 and 
the Centaurus Cluster. For clarity, only the S1 and G2 data and 
the results determined with spectral model C are shown.} 
\end{figure*}

\clearpage

\begin{figure*}
\hbox{
\hspace{1.5cm}\psfig{figure=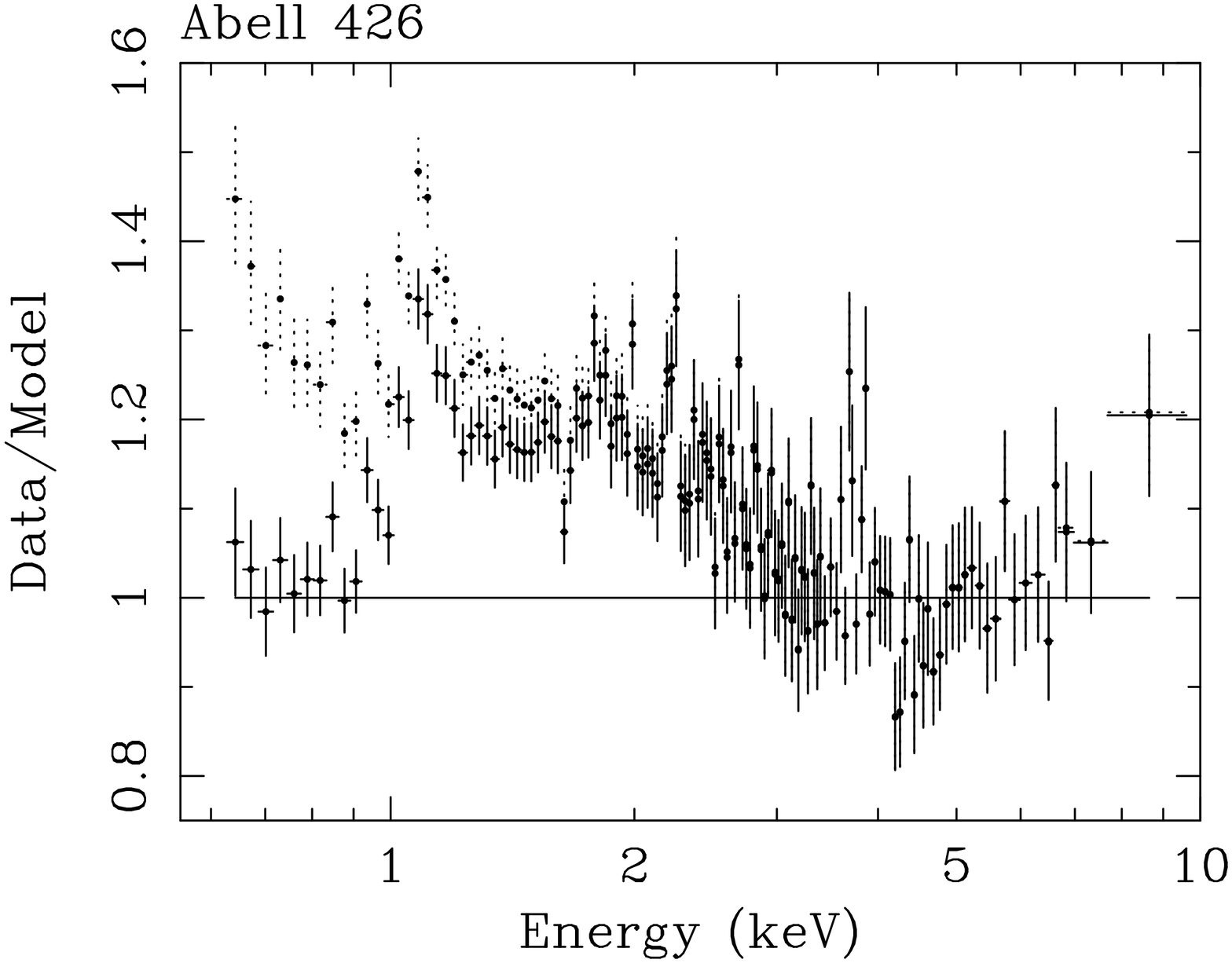,width=0.4\textwidth,angle=270}
\hspace{1cm}\psfig{figure=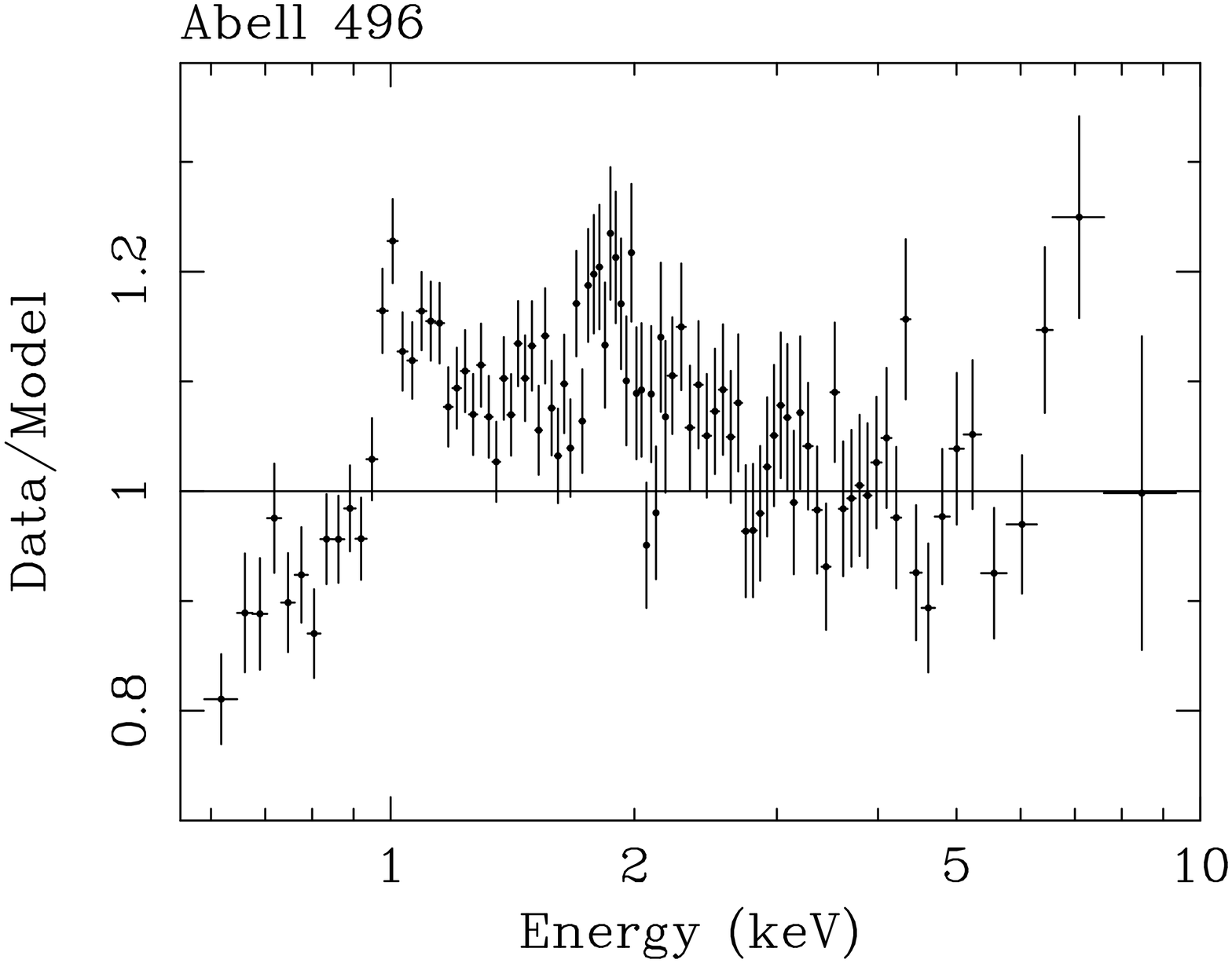,width=0.4\textwidth,angle=270}
}
\vspace{0.2cm}
 \hbox{
\hspace{1.5cm}\psfig{figure=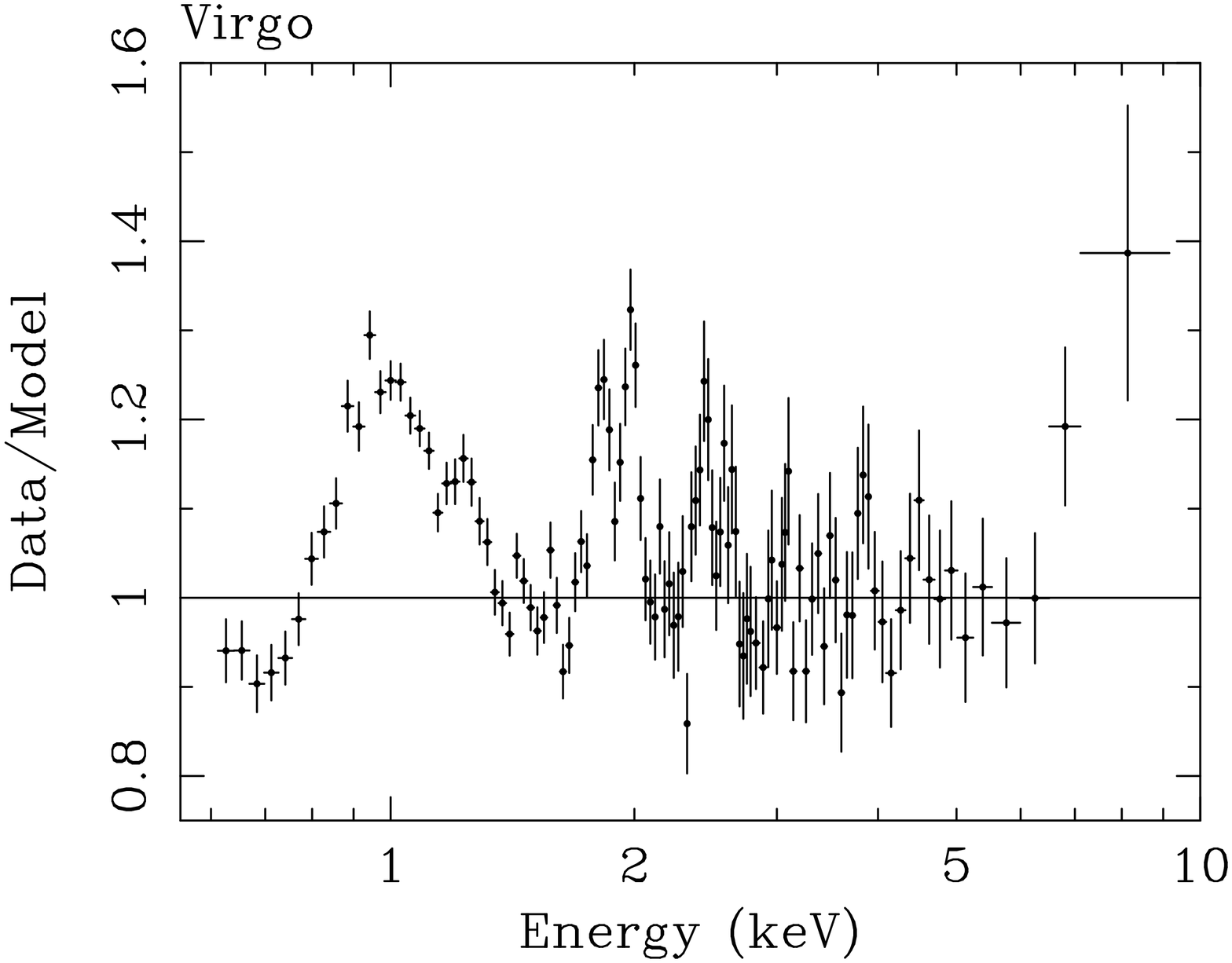,width=0.4\textwidth,angle=270}
\hspace{1cm}\psfig{figure=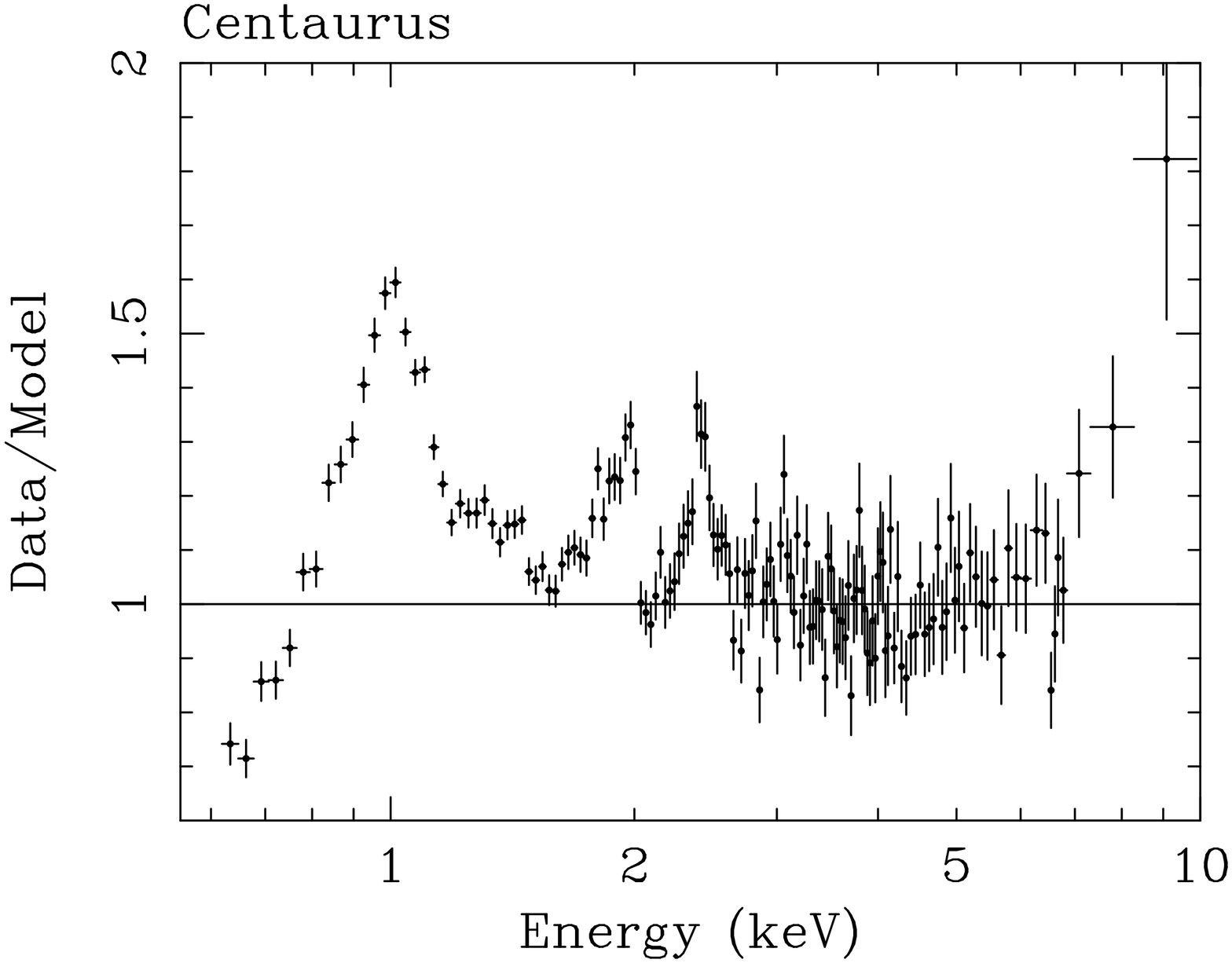,width=0.4\textwidth,angle=270}
}
\vspace{0.2cm}
 \hbox{
\hspace{1.5cm}\psfig{figure=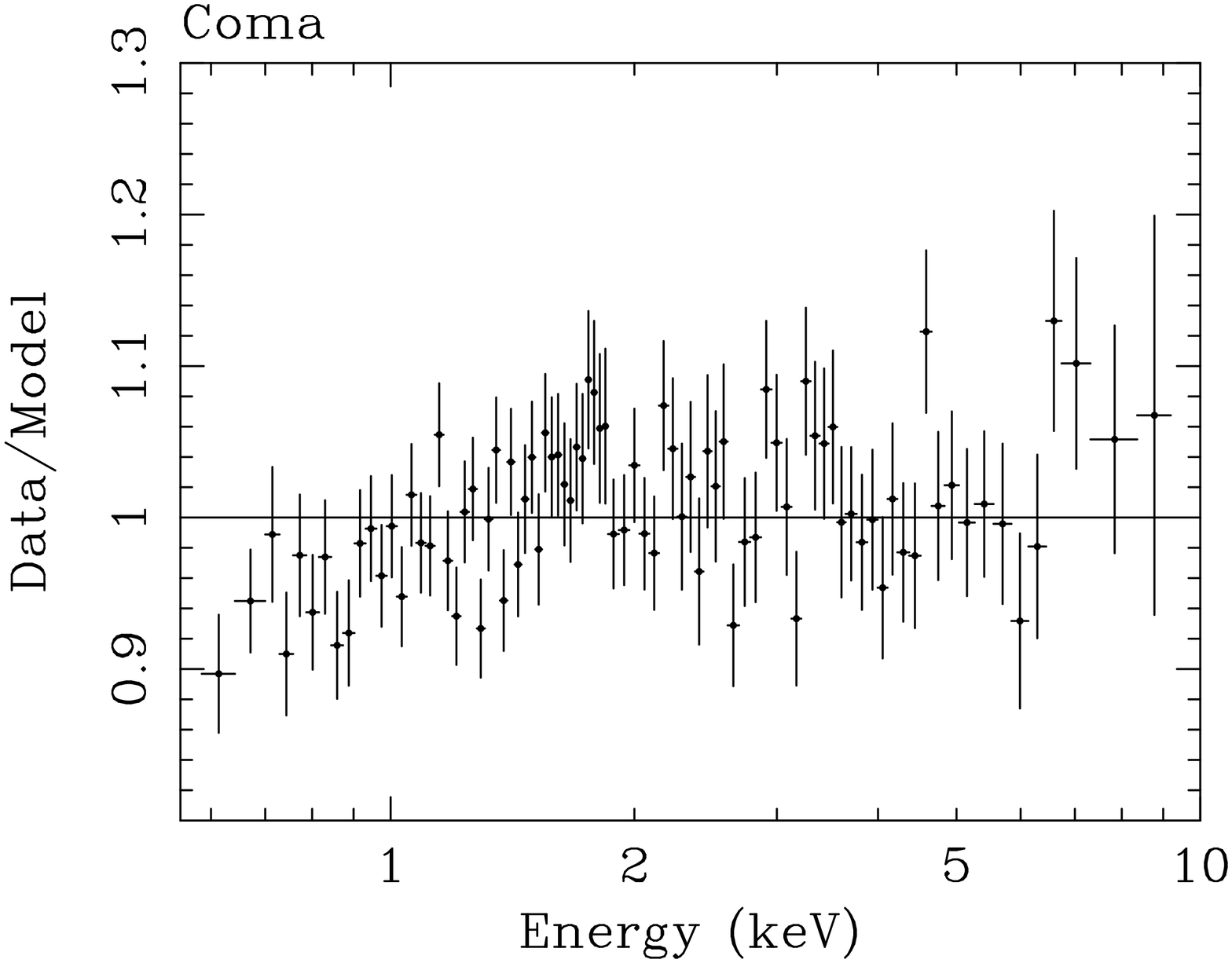,width=0.4\textwidth,angle=270}
\hspace{1cm}\psfig{figure=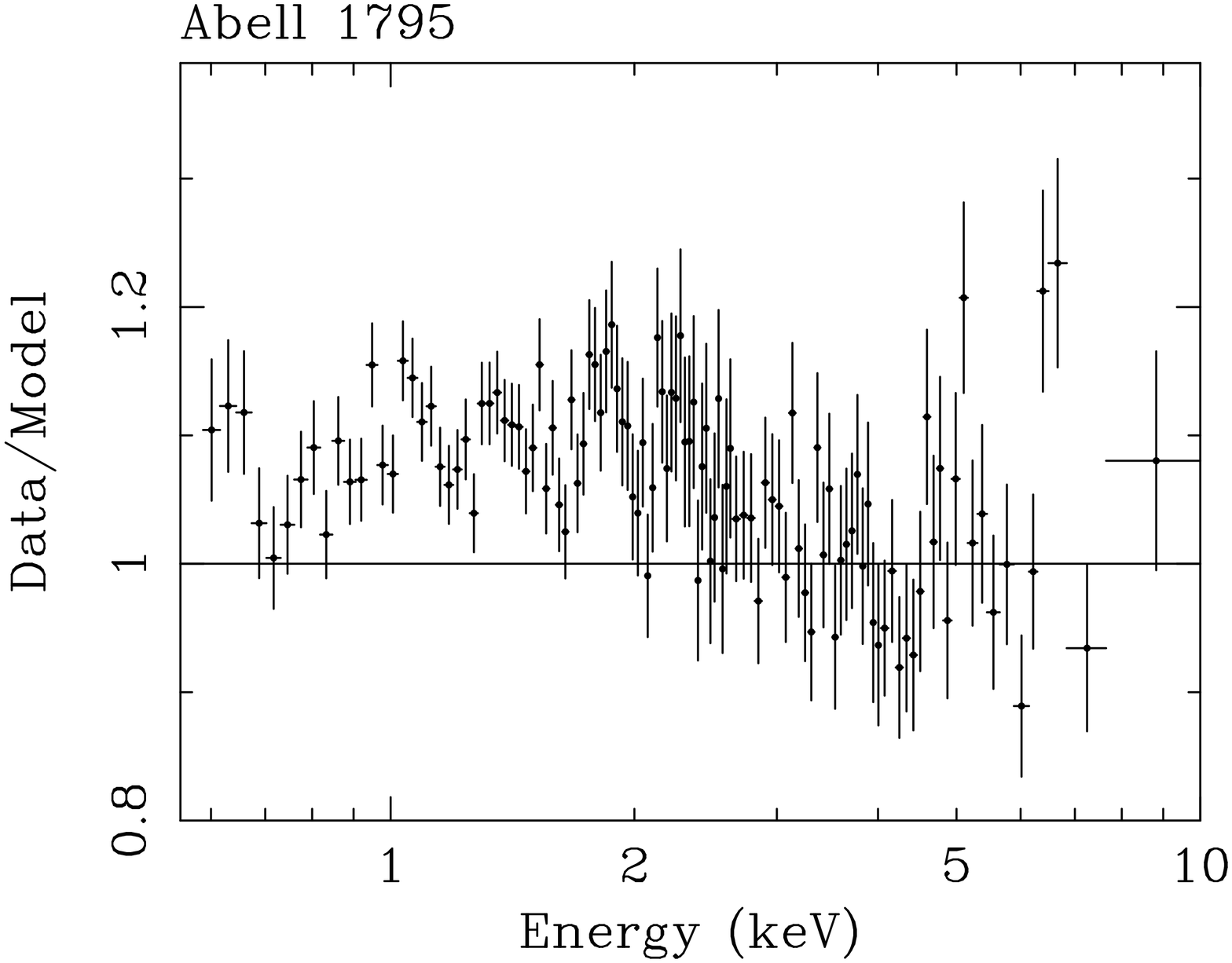,width=0.4\textwidth,angle=270}
}
\vspace{0.2cm}
 \hbox{
\hspace{1.5cm}\psfig{figure=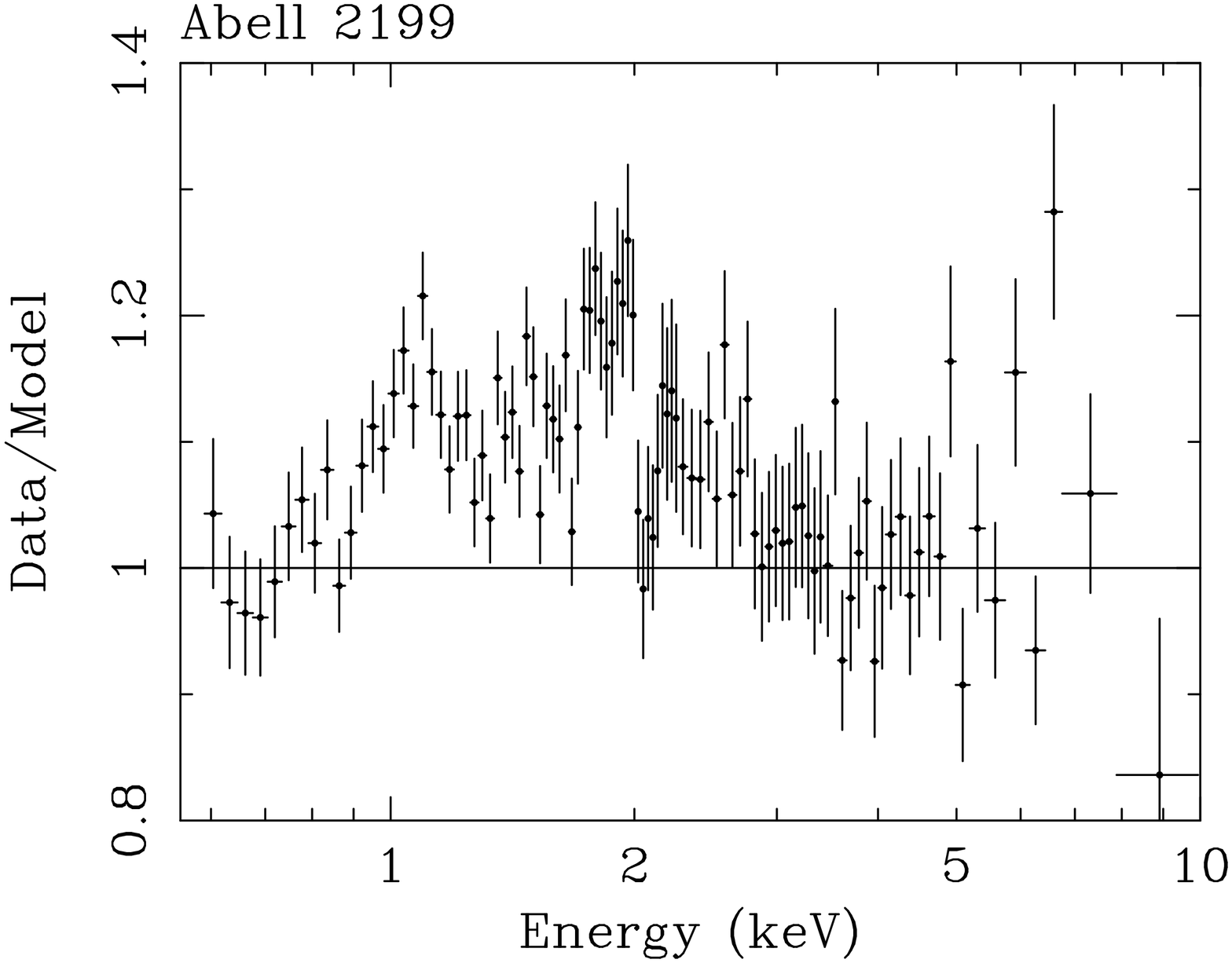,width=0.4\textwidth,angle=270}
\hspace{1cm}\psfig{figure=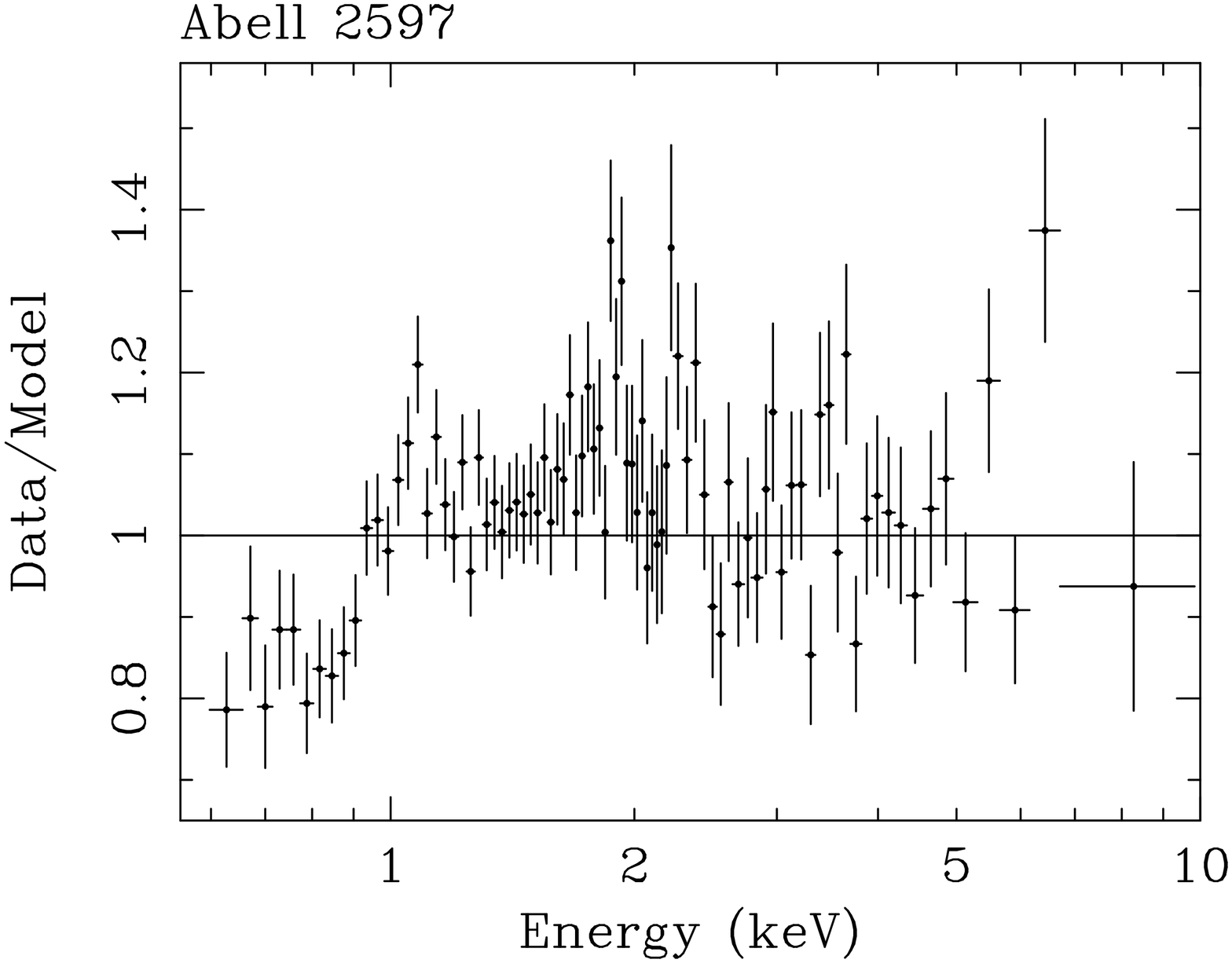,width=0.4\textwidth,angle=270}
}
 \caption{The residuals from fits 
to the SIS spectra in the $3.0-10.0$ keV energy range using 
a single-temperature emission model with Galactic absorption (model A).
The best-fitting models have then been extrapolated to cover the full 
$0.6-10.0$ keV band of the SIS instruments. 
For the cooling-flow clusters, note the clear excess in the residuals 
at energies between $\sim 0.8$ and 3.0 keV (the region dominated 
dominated by the Fe-L, Mg, Si and S line complexes) indicating 
the presence of gas cooler than the ambient cluster
temperatures. The deficits in the residuals at energies below 0.8~keV provide 
evidence for excess absorption. For Abell
426, the solid points show the results for an assumed  Galactic column
density of $1.0 \times 10^{21}$ \apc~(Section 3.3). The dotted points 
show the results using the nominal Dickey \& Lockman (1990) value 
of $1.49 \times 10^{21}$\apc. Note also the
positive residuals at high energies in the data for Abell 426, the Virgo
Clusters and the Centaurus Cluster (see Section 4).}
\end{figure*}

\clearpage

\begin{figure*}
\hbox{
\hspace{-0.2cm}\psfig{figure=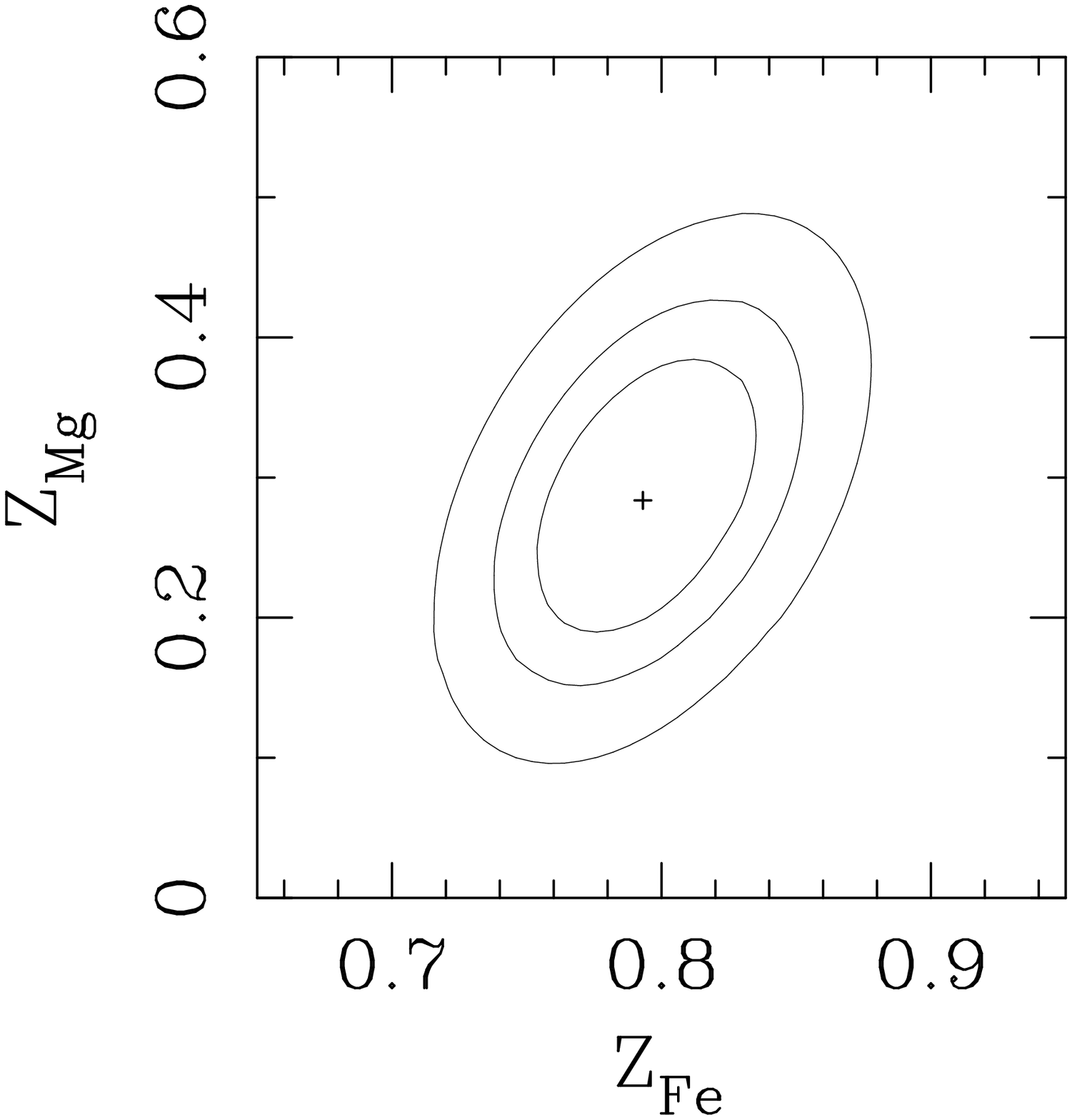,width=0.45\textwidth,angle=270}
\hspace{-2.0cm}\psfig{figure=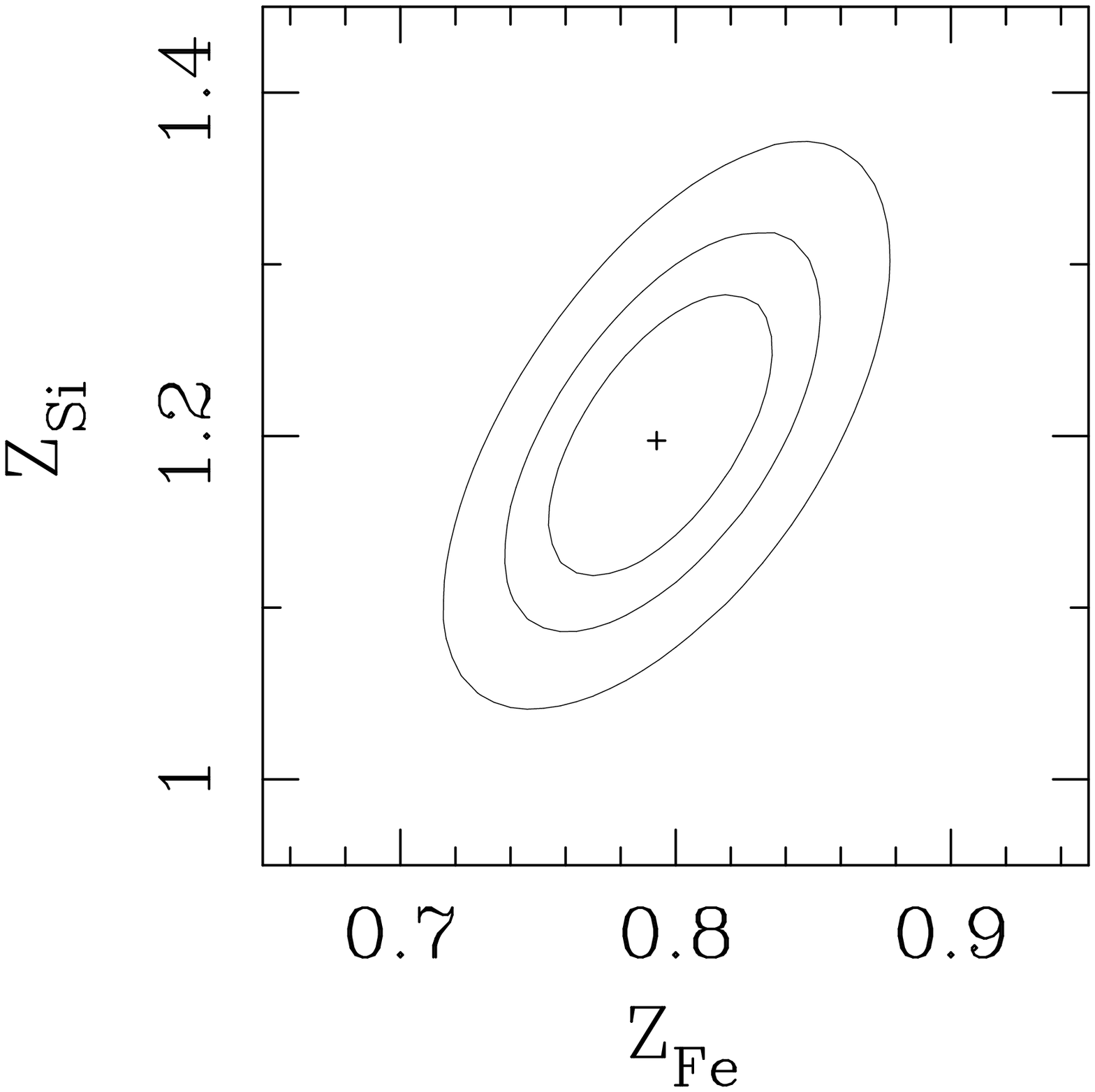,width=0.45\textwidth,angle=270}
\hspace{-2.0cm}\psfig{figure=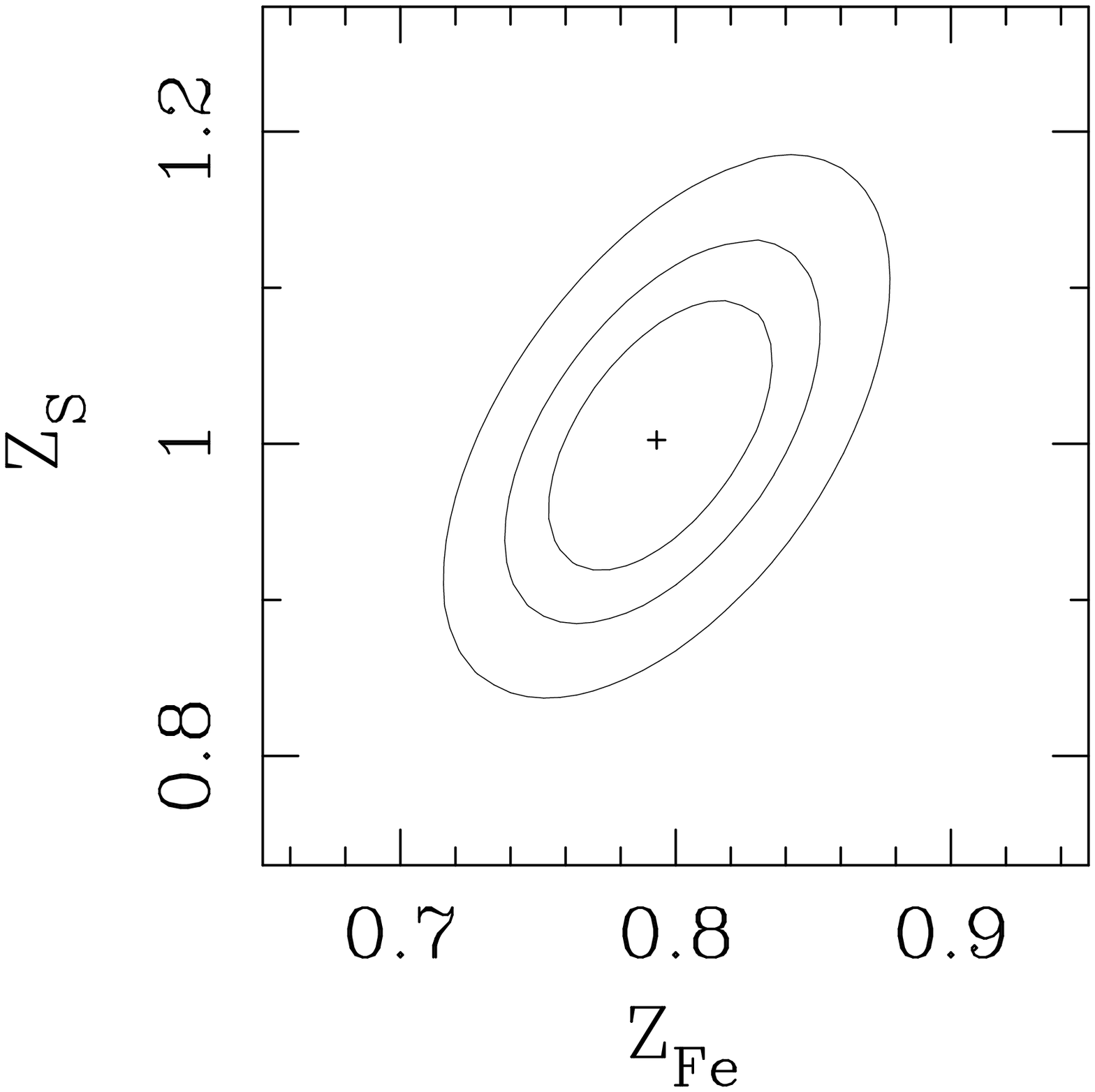,width=0.45\textwidth,angle=270}
}
\caption{Joint confidence contours on the abundances of (left) Mg and Fe,
(centre) Si and Fe, and (right) S and Fe determined from the ASCA SIS spectra 
for the Virgo Cluster. The  solar photospheric abundance scale of Anders 
\& Grevesse (1989) is assumed. Contours mark the regions of 68, 90 and 99 per
cent confidence ($\Delta \chi^2=2.30,4.61$ and 9.21, respectively).}
\vskip 2cm
\end{figure*}

\begin{figure*}
\hbox{
\hspace{0cm}\psfig{figure=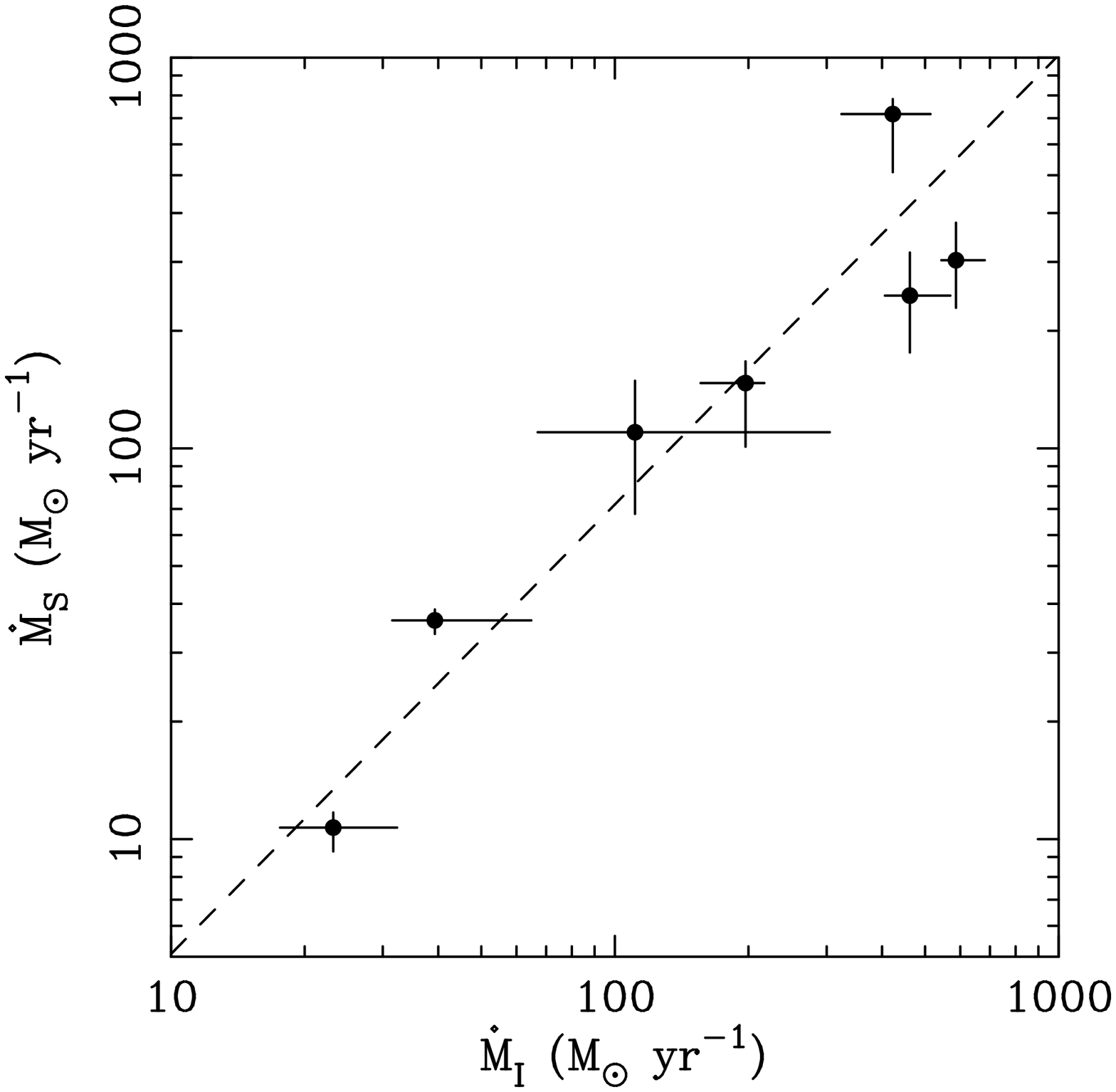,width=0.65\textwidth,angle=270}
\hspace{-2.5cm}\psfig{figure=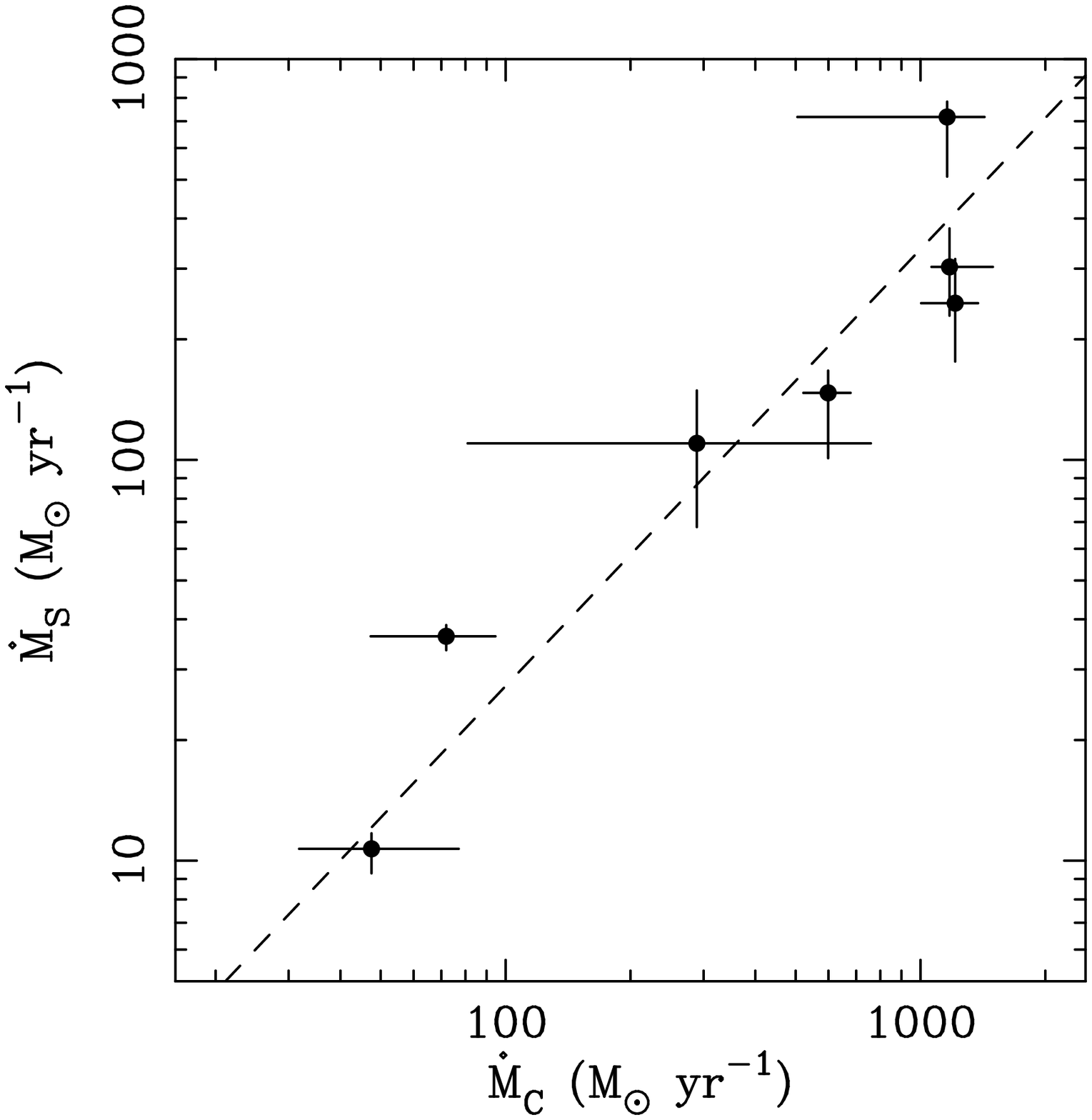,width=0.65\textwidth,angle=270}
}
\caption{(a) The mass deposition rates determined from the ASCA spectra 
(${\dot M_{\rm S}}$) versus the values measured from the deprojection analysis 
of the ROSAT images (${\dot M_{\rm I}}$). No correction for the 
effects of intrinsic absorption on the deprojection results has been made. 
The dashed line is the best-fitting power-law model which has a slope, 
$Q=1.15\pm 0.33$ and a normalization, $P= 0.36\pm 0.65$. 
(b) as for (a) but 
with the deprojection results corrected for the effects of 
intrinsic absorption by a uniform screen of cold gas with the best-fit 
parameters determined from the spectral analysis (${\dot M_{\rm C}}$). 
The best-fitting 
power-law model has a slope, $Q=1.09\pm 0.32$ and a normalization, 
$P= 0.18\pm 0.37$. (Errors on $P$ and $Q$ are $1\sigma$
uncertainties determined by bootstrap re-sampling). }
\end{figure*}

\clearpage 

\begin{figure*}
\centerline{\hspace{2cm}\psfig{figure=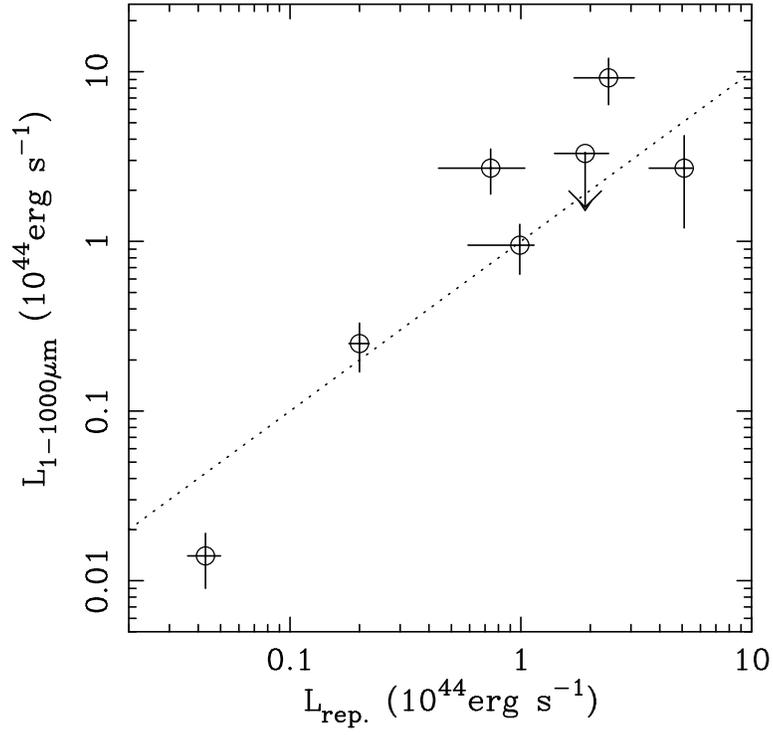,width=0.85\textwidth
,angle=270}}
\caption{The $1-1000\mu$m luminosities determined from the IRAS
60 and 100$\mu$m fluxes (using equation 2) plotted as a function of the 
predicted reprocessed X-ray luminosities from the cooling flows.
The dotted curve is the line of equality between the values.} 
\end{figure*}

\clearpage

\begin{figure*}
\vspace{0.5cm}
\hbox{
\hspace{1.66cm}\psfig{figure=a426_break_final.ps,width=0.352\textwidth,angle=270}
\hspace{1.82cm}\psfig{figure=a496_break_final.ps,width=0.352\textwidth,angle=270}
}
\vspace{-0.1cm}
\hbox{
\hspace{1.5cm}\psfig{figure=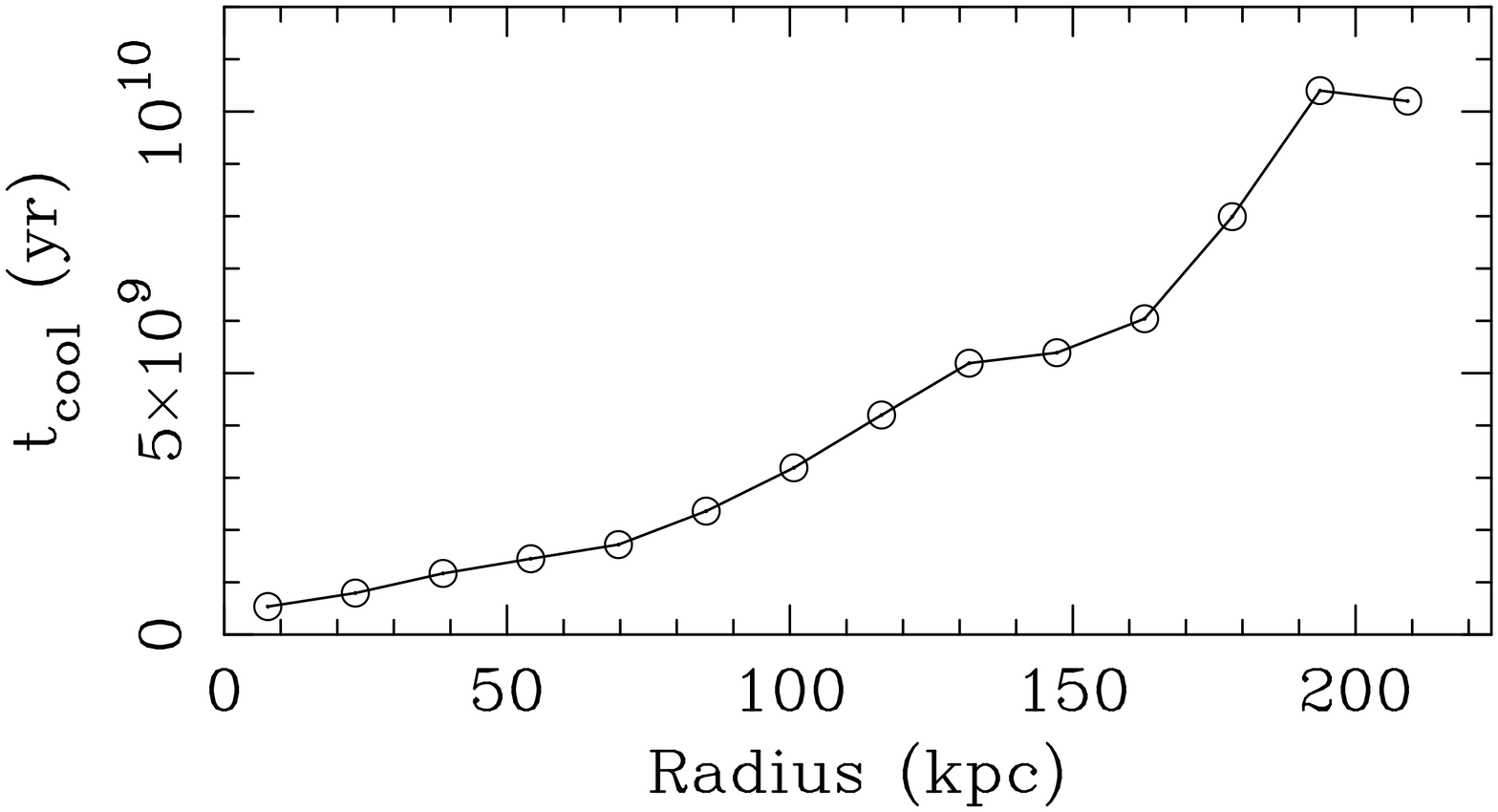,width=0.4\textwidth,angle=270}
\hspace{1.0cm}\psfig{figure=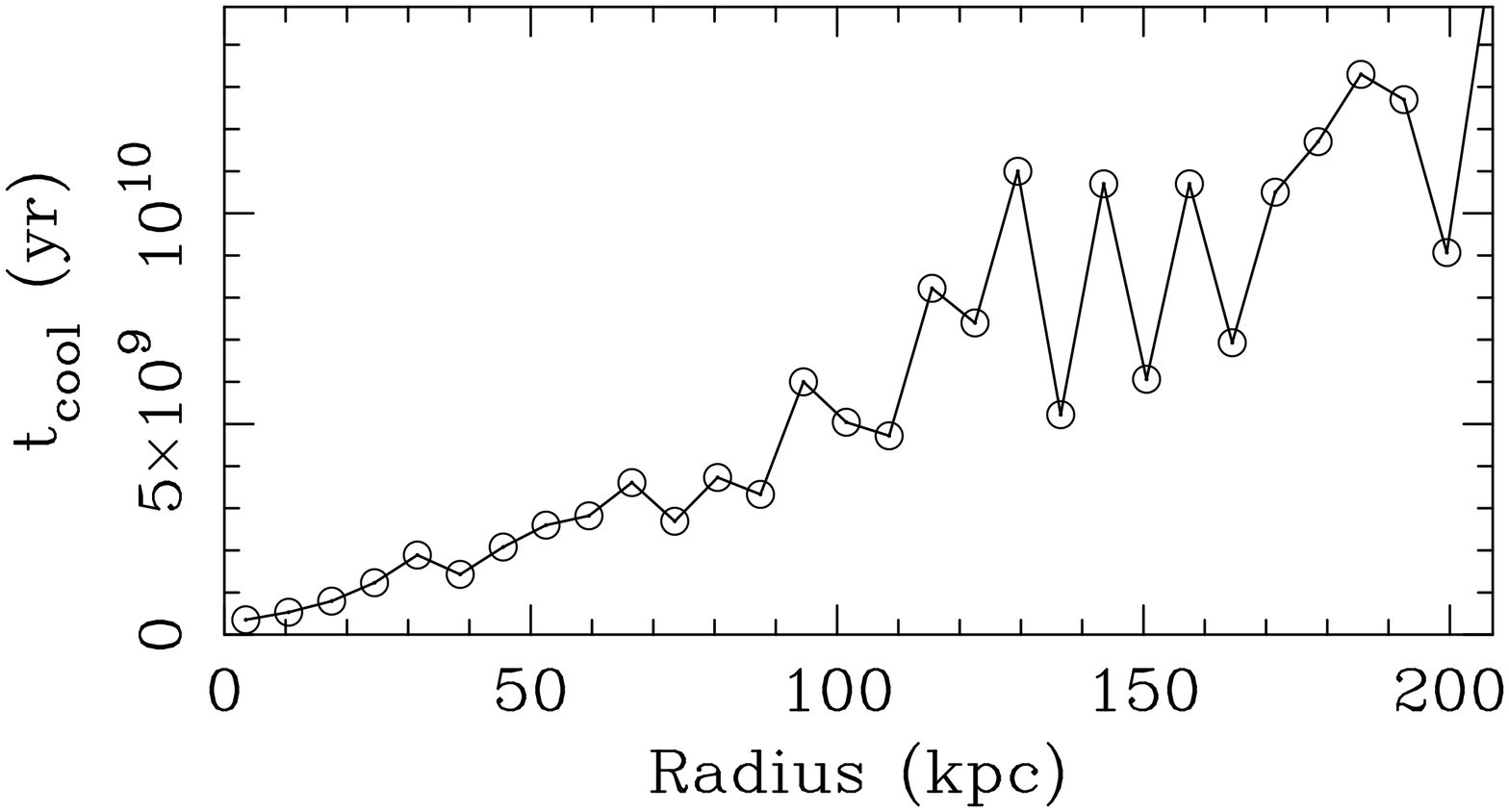,width=0.4\textwidth,angle=270}
}
\vspace{-0.2cm}
 \hbox{
\hspace{1.66cm}\psfig{figure=a1795_break_final.ps,width=0.352\textwidth,angle=270}
\hspace{1.82cm}\psfig{figure=a2199_break_final.ps,width=0.352\textwidth,angle=270}
}
\vspace{-0.1cm}
\hbox{
\hspace{1.5cm}\psfig{figure=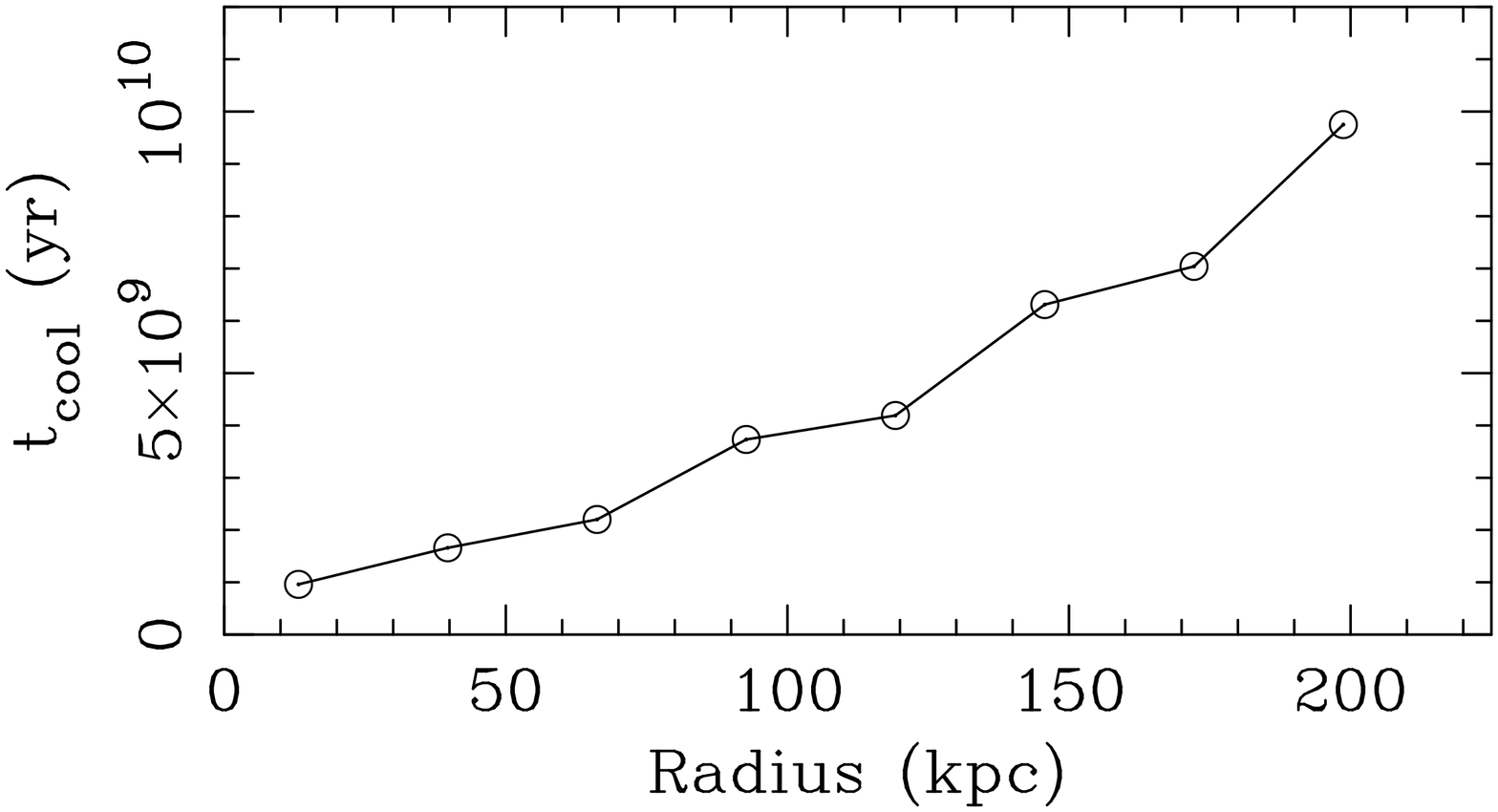,width=0.4\textwidth,angle=270}
\hspace{1.0cm}\psfig{figure=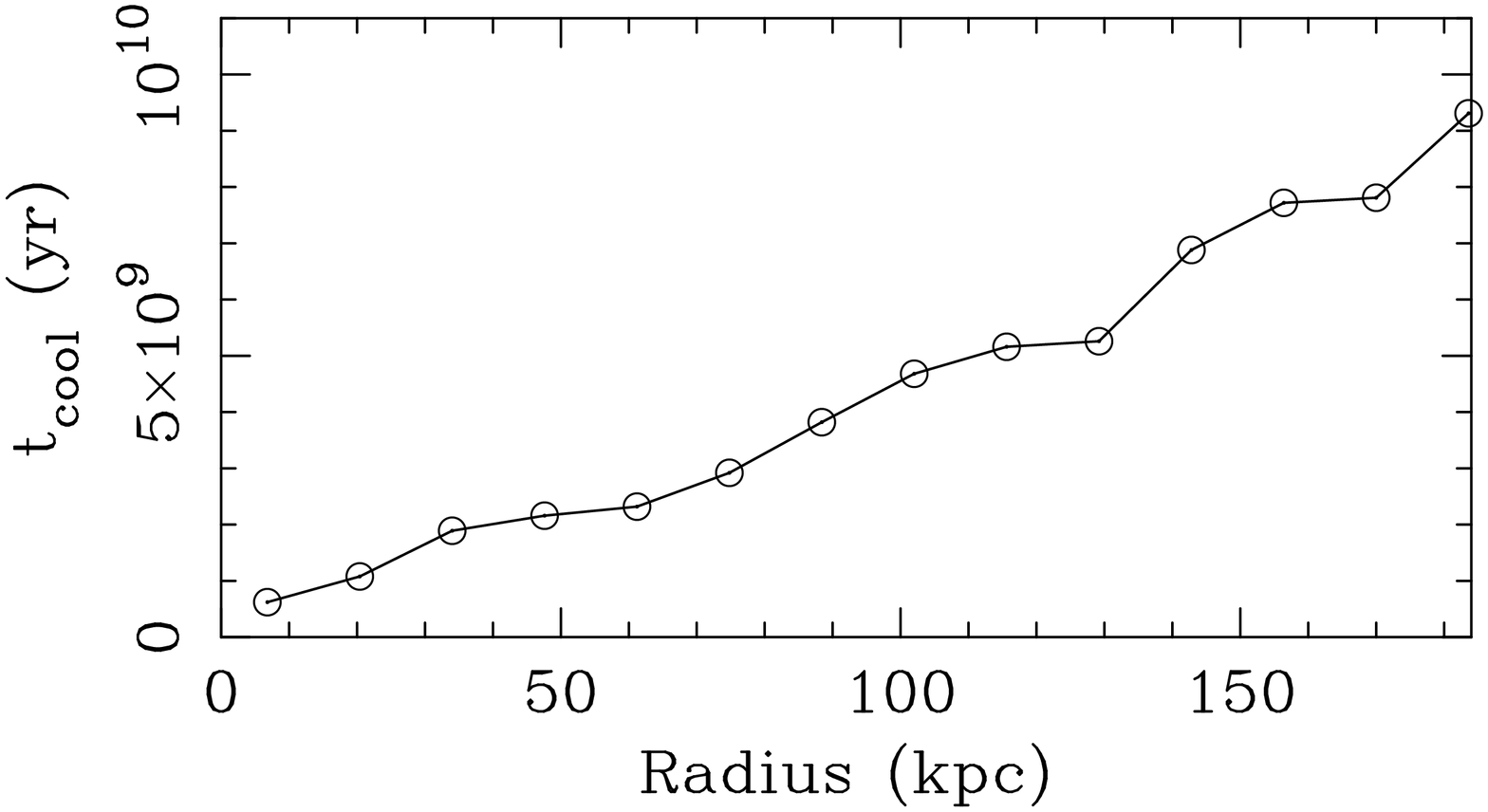,width=0.4\textwidth,angle=270}
}
\caption{(Upper panels) The results from the fits with simple 
broken power-law models 
to the mass deposition profiles determined from the 
deprojection study (no correction for intrinsic absorption has been made). 
Only those data (marked with circles) from radii interior to the 
90 percentile upper limit to the cooling radius were included in the
fits. The vertical dashed lines mark the cooling radii in the clusters.  
The dotted lines mark the 90 percentile lower limits on the cooling radii. 
The lower panels show the mean cooling time of the cluster gas as a function of
radius.} 
\end{figure*}

\clearpage

\begin{figure*}
\hbox{
\hspace{0cm}\psfig{figure=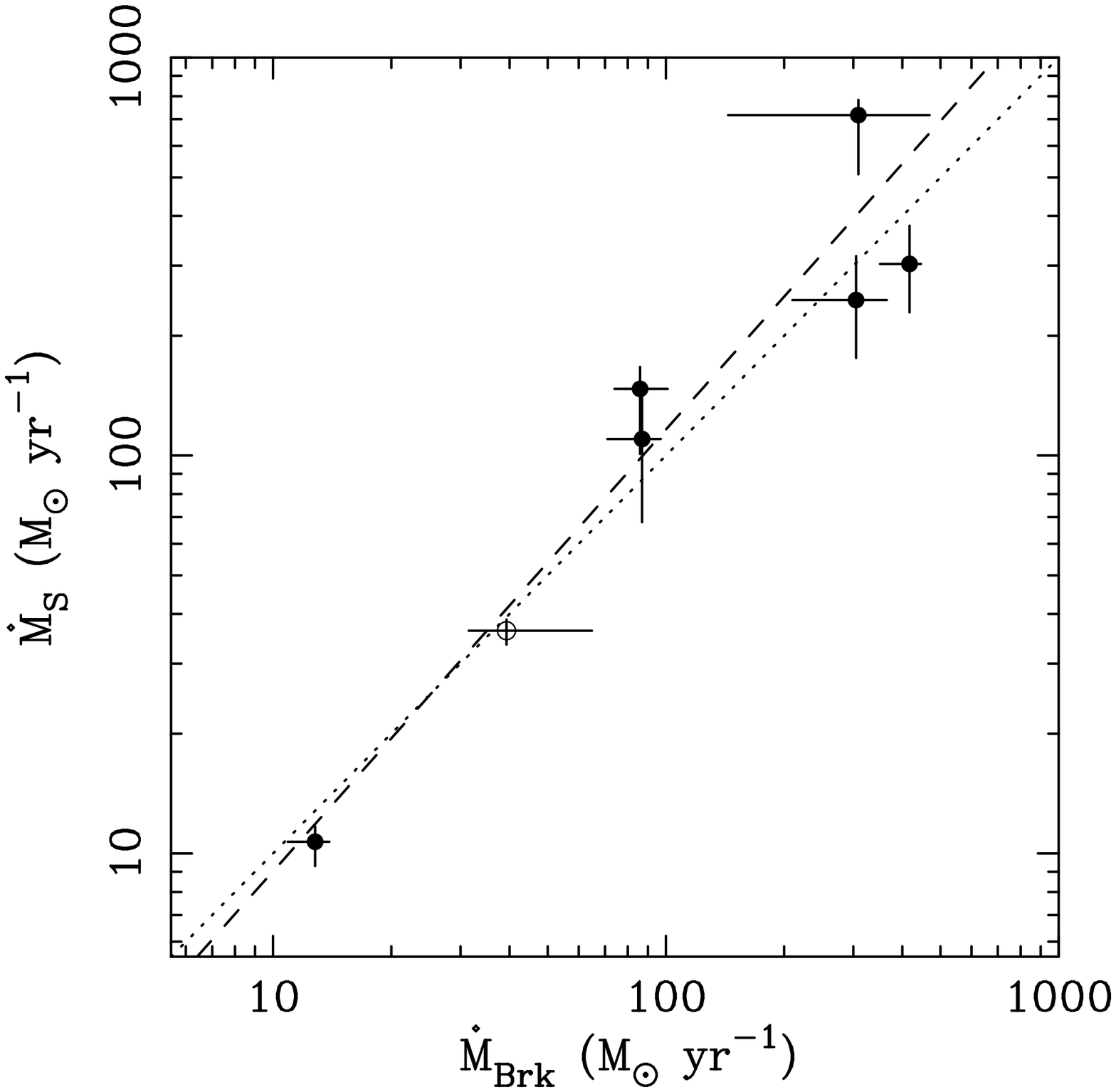,width=0.65\textwidth,angle=270}
\hspace{-2.5cm}\psfig{figure=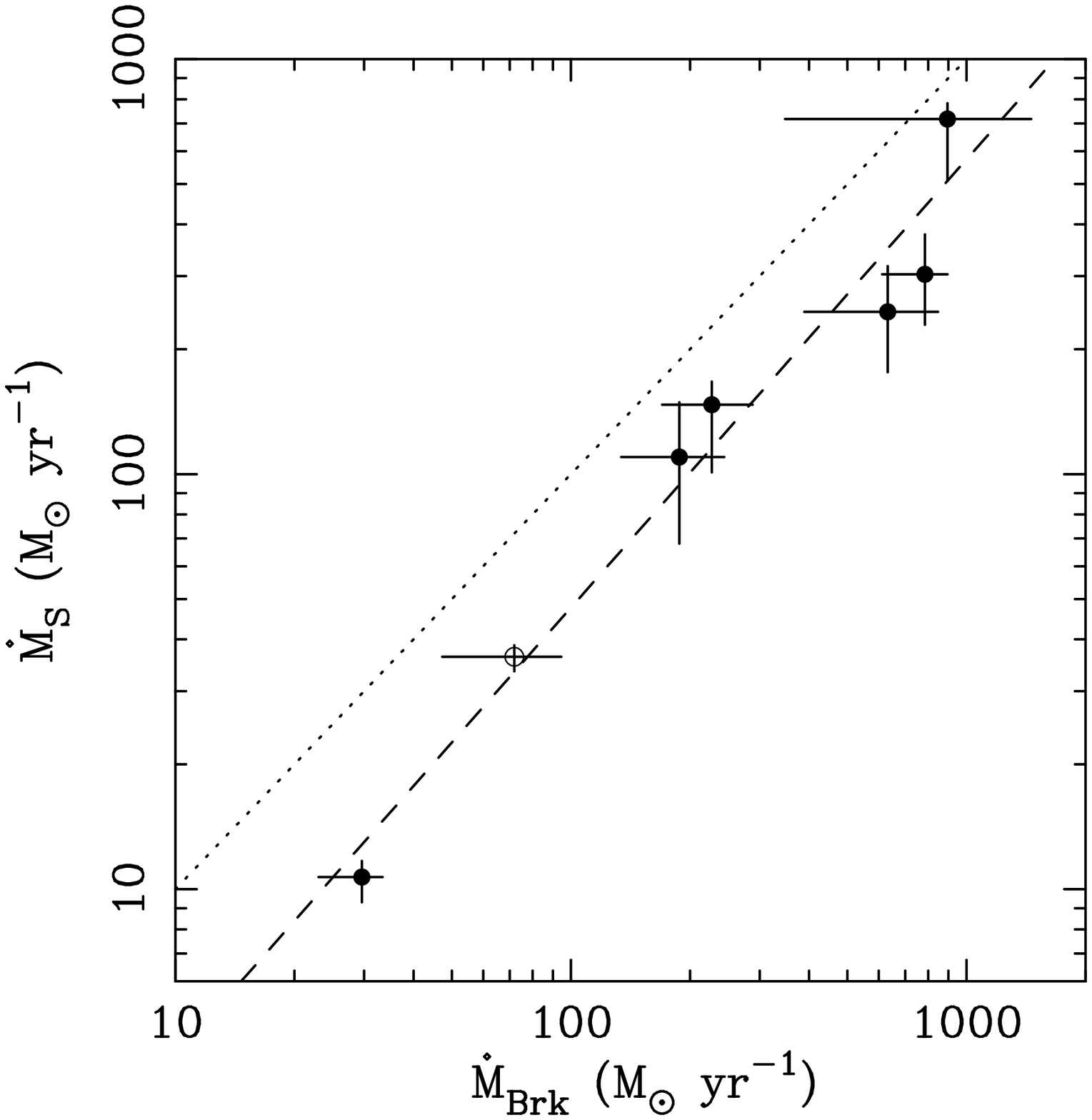,width=0.65\textwidth,angle=270}
}
\caption{The mass deposition rates determined from the 
ASCA spectra (${\dot M_{\rm S}}$) versus the values inferred from the
breaks in the mass deposition profiles (${\dot M_{\rm Brk}}$) determined 
from the deprojection study. The left and right panels show the results 
obtained without and with corrections for intrinsic absorption due to a 
uniform screen of cold gas, respectively. The dashed curves show the best 
fitting power-law models
(see text for details). The dotted lines are the lines of equality 
($y=x$) between the values. The Centaurus Cluster (plotted as an open 
circle) does not exhibit an obvious break in its mass deposition profile 
within the cooling radius and so we assume ${\dot M_{\rm Brk}} = 
{\dot M_{\rm C}}$ or ${\dot M_{\rm Brk}} = {\dot M_{\rm I}}$ from Table
10, as appropriate. } 
\end{figure*}

\clearpage

\begin{figure*}
\centerline{\hspace{-2cm}\psfig{figure=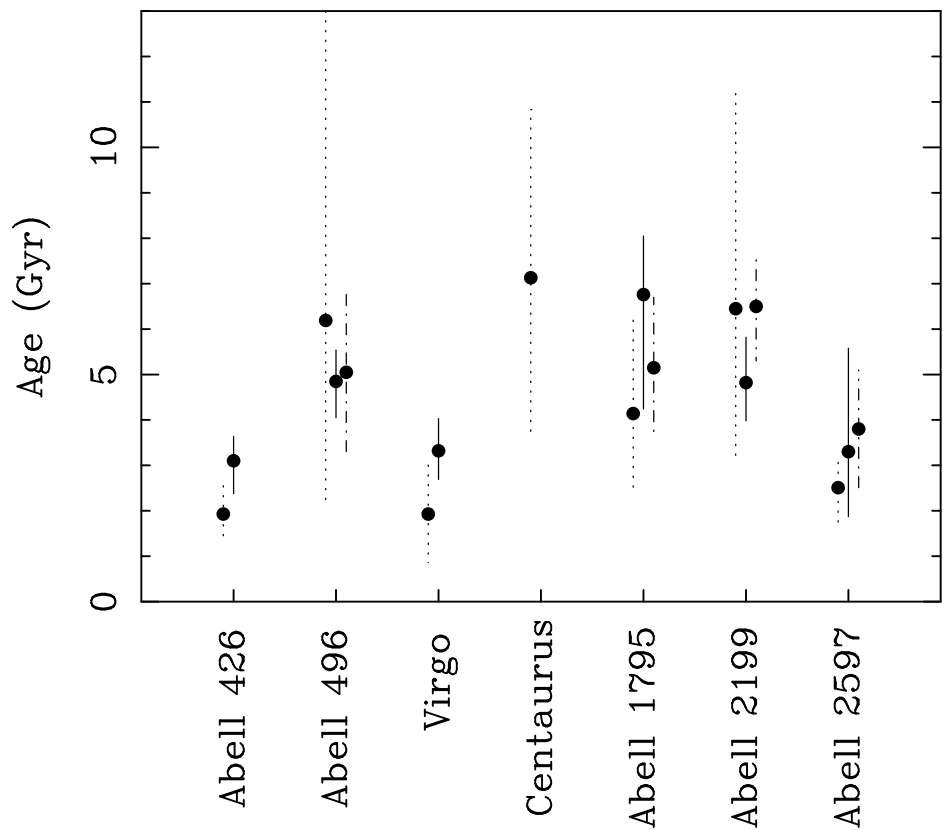,width=1.0\textwidth
,angle=270}}
\caption{A comparison of the results on the ages of the cooling flows 
from the three methods described in Section 8. The dot-dashed
lines are the results from the X-ray colour deprojection study of 
Allen \& Fabian (1997; method 1). The dotted lines show the ages 
inferred from the comparison of the mass deposition rates 
determined from the spectral and image deprojection methods (method 2). 
For method 2, the average of the absorption-corrected and uncorrected
results is shown, except for Abell 2597, for which the uncorrected 
result was unbounded and therefore, only the absorption-corrected result
used. (The extent of the dotted curves mark the extrema obtained from the absorption-corrected 
and uncorrected analyses). The solid lines are the measurements based on the identifications 
of breaks in the mass deposition profiles (method 3).} 
\end{figure*}

\clearpage 

\begin{figure*}
\vskip 1cm
\centerline{\hspace{0cm}\psfig{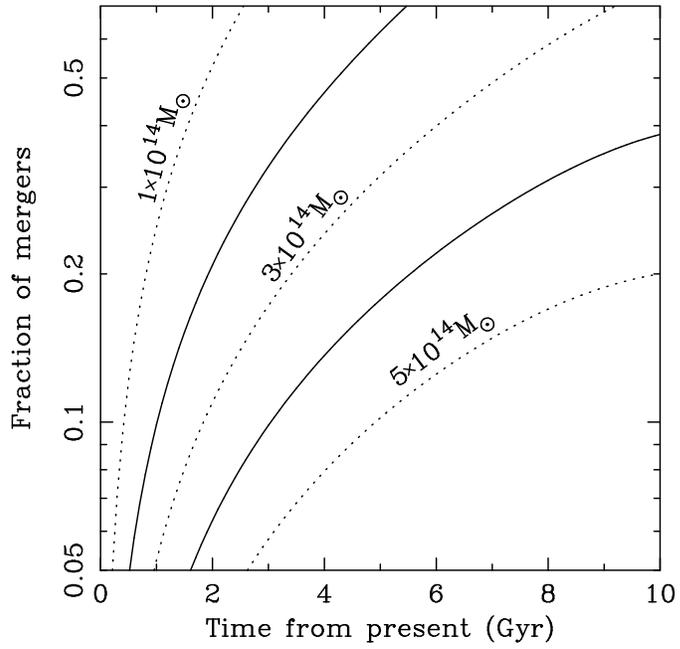}}
\caption{The fraction of current $10^{15}\Msun$ clusters which have
experienced a merger with a subcluster of mass $1-5 \times 10^{14}\Msun$ 
(top to bottom) within a given time from the present ($\Omega=1.0, \Lambda = 
0, b=1$ and a cosmic fluctuation index, $n=-1.5$ are assumed.)} 
\end{figure*}


\begin{thebibliography}{}
\bibitem{} Allen S.W., 1995, MNRAS, 276, 947
\bibitem{} Allen S.W., 1998, MNRAS, 296, 392
\bibitem{} Allen S.W., 2000, MNRAS, 315, 269
\bibitem{} Allen S.W., Fabian A.C., 1994, MNRAS, 269, 409
\bibitem{} Allen S.W., Fabian A.C., 1997, MNRAS, 286, 583
\bibitem{} Allen S.W., Fabian A.C., 1998, MNRAS, 297, L63
\bibitem{} Allen S.W., Di Matteo T., Fabian A.C., 2000, MNRAS, 311, 493
%\bibitem{} Allen S.W., Fabian A.C., Johnstone R.M., Nulsen P.E.J., Edge A.C., 1992, MNRAS, 254, 51 
%\bibitem{} Allen S.W., Fabian A.C., Johnstone R.M., White D.A., Daines S.J., Edge A.C., Stewart G.C., 1993, MNRAS, 262, 901
%\bibitem{} Allen S.W., Fabian A.C., Edge A.C., B\"ohringer H., White D.A., 1995, MNRAS, 275, 741
%\bibitem{} Allen S.W., Fabian A.C., Edge A.C., Bautz M.W., Furuzawa A., Tawara Y., 1996, MNRAS, 283, 263
\bibitem{} Akritas M.G., Bershady M.A., 1996, ApJ, 470, 706
\bibitem{} Anders E., Grevesse N., 1989, Geochemica et Cosmochimica Acta 53, 197
\bibitem{} Arnaud, K.A., 1988, in Fabian A.C., ed., Cooling flows in clusters and galaxies, Kluwer, Dordrecht, p. 31 
\bibitem{} Arnaud, K.A., 1996, in Astronomical Data Analysis Software and Systems V, eds. Jacoby G. and Barnes J., ASP Conf. Series volume 101, p17
\bibitem{} Arnaud K.A., Mushotzky R.F., 1998, ApJ, 501, 119
\bibitem{} Bailey M.E., 1980, MNRAS, 191, 195
\bibitem{} Balucinska-Church M., McCammon D., 1992, ApJ, 400, 699
\bibitem{} Binney J., Tremaine S., 1987, Galactic Dynamics, Princeton Univ. Press, Princeton
%\bibitem{} Briel U., Henry J.P., 1996, ApJ, 472, 131
\bibitem{} Buote D.A., Canizares C.R., Fabian A.C., 1999, MNRAS, 310, 483
%\bibitem{} Burns J.O. \etal, 1999, in preparation
\bibitem{} Cardiel N., Gorgas J., Aragon-Salamanca A., 1995, MNRAS, 277, 502
\bibitem{} Cardiel N., Gorgas J., Aragon-Salamanca A., 1998, MNRAS, 298, 977
\bibitem{} Cavaliere A., Menci N., Tozzi P., 1999, MNRAS, 308, 599
\bibitem{} Cox C.V., Bregman J.N., Schombert J.M., 1995, ApJS, 99, 405
\bibitem{} Crawford C.S., Allen S.W., Ebeling H., Edge A.C., Fabian A.C., 1999, MNRAS, 306, 857
\bibitem{} Daines S.J., Fabian A.C., Thomas, P.A., 1994, MNRAS, 268, 1060
\bibitem{} David L.P., Harnden F.R., Kearns K.E., Zombeck M.V., 1996, The ROSAT HRI Calibration Report, $ftp://legacy.gsfc.nasa.gov/rosat/doc/hri/hri\_report$
\bibitem{} Dickey J.M., Lockman F.J., 1990, ARA\&A, 28, 215
\bibitem{} Di Matteo T., Quataert E., Allen S.W., Narayan R., Fabian A.C., 1999, MNRAS, 311, 507 
\bibitem{} Dotani T. \etal, 1996, ASCA News, 4, 3
\bibitem{} Dupke R., White R., 2000a, ApJ, 528, 139 
\bibitem{} Dupke R., White R., 2000b, ApJ, 537, 123
%\bibitem{} Dwek E., Rephaeli Y., Mather J.C., 1990, ApJ, 350, 104
\bibitem{} Edge A.C., Stewart G.C., Fabian A.C., 1992, MNRAS, 258, 177
%\bibitem{} Eke V.R., Cole S., Frenk C.S., 1996, MNRAS, 282, 263
\bibitem{} Ezawa H., Fukazawa Y., Makishima K., Ohashi T., Takahara F., Xu H., Yamasaki N.Y., 1997, ApJ, 490, L33
\bibitem{} Fabian A.C., 1994, A\&AR, 32, 277
\bibitem{} Fabian A.C., Crawford C.S., 1990, MNRAS, 247, 439 
\bibitem{} Fabian A.C., Johnstone R.M., Daines S.J., 1994a, MNRAS, 271, 737
\bibitem{} Fabian A.C., Hu E.M., Cowie L.L., Grindlay J., 1981, ApJ, 248, 47
\bibitem{} Fabian A.C., Arnaud K.A., Bautz M.W., Tawara Y., 1994b, ApJ, 436, L63
\bibitem{} Ferland G.J., Fabian A.C., Johnstone R.M., 2000, MNRAS, submitted 
\bibitem{} Finoguenov A., Ponman T.J., 1999, MNRAS, 305, 325
\bibitem{} Fusco-Femiano R., Dal Fiume, D., Feretti L., Giovannini G., Grandi P., Matt G., Molendi S., Santangelo A., 1999, ApJ, 513, L21
\bibitem{} Fukazawa Y. \etal 1994, PASJ, 46, L55
\bibitem{} Gibson B.K., Loewenstein M., Mushotzky R.F., 1997, MNRAS, 290, 623
\bibitem{} Hatsukade I., Kawarabata K., Takenada K., Ishizaka J., 1997, in Makino F., Mitsuda K., eds., X-ray Imaging and Spectroscopy of Cosmic Hot Plasmas, Universal Academy Press, Tokyo, p. 105
\bibitem{} Helou G., Khan I.R., Malek L., Boehmer L., 1988, ApJS, 68, 151
\bibitem{} Henkel C., Wiklind T., 1997, SSRv, 81, 1
\bibitem{} Honda H., Hirayama M., Watanabe M., Kunieda H., Tawara Y., Yamashita K., Ohashi T., Hughes J.P., Henry J.P., 1996, ApJ, 473, L71
\bibitem{} Hwang U., Mushotzky R.F., Burns J.O., Fukazawa Y., White R.A., 1999, ApJ, 516, 604
\bibitem{} Ikebe Y., Makishima K., Fukazawa Y., Tamura T., Xu H., Ohashi T., Matsushita K., 1999, ApJ, 525, 58 
%\bibitem{} Irwin J.A., Sarazin C.L., 1995, ApJ, 355, 497
\bibitem{} Ishimaru Y., Arimoto N., 1997, PASJ, 49, 1
\bibitem{} Johnstone R.M., Fabian A.C., Nulsen P.E.J., 1987, MNRAS, 224, 75
\bibitem{} Johnstone R.M., Fabian A.C., Edge A.C., Thomas P.A., 1992, MNRAS, 255, 431
\bibitem{} Johnstone R.M., Fabian A.C., Taylor G.B., 1998, MNRAS, 298, 854
%\bibitem{} Jones C., Forman W., 1984, ApJ, 276, 38
\bibitem{} Kaastra J.S., Mewe R., 1993, Legacy, 3, HEASARC, NASA
%\bibitem{} Kitayama T., Suto Y., 1996, ApJ, 469, 480
%\bibitem{} Kim K.T., Kronberg P.P., Dewdney P.E., Landecker T.L., 1990, ApJ, 355, 29
\bibitem{} Lacey C., Cole S., 1993, MNRAS, 262, 627 
\bibitem{} Liedhal D.A., Osterheld A.L., Goldstein W.H., 1995, ApJ, 438, L115 
%\bibitem{} Magorrian J. \etal, 1998, AJ, 115, 2285
\bibitem{} Makishima K., 1997, in Makino F., Mitsuda K., eds., X-ray Imaging and Spectroscopy of Cosmic Hot Plasmas, Universal Academy Press, Tokyo, p. 137
\bibitem{} Markevitch M., Forman W.R., Sarazin C.L., Vikhlinin A., 1998, ApJ, 503, 77
\bibitem{} Markevitch M., Vikhlinin A., Forman W.R., Sarazin C.L., 1999, ApJ, 527, 545 
\bibitem{} Matsumoto H., Koyama K., Awaki H., Tomida H., Tsuru T., Mushotzky R., Hatsukade I., 1996, PASJ, 48, 201
\bibitem{} Matsuzawa H., Matsuoka M., Ikebe Y., Mihara T., Yamashita K., 1996, PASJ, 48, 565
\bibitem{} McGlynn T.A., Fabian A.C., 1984, MNRAS, 208, 709
\bibitem{} McNamara B.R., O'Connell R.W., 1989, AJ, 98, 2018
\bibitem{} Mushotzky R.F., Lowenstein M., Arnaud K.A., Tamura T., Fukazawa Y., Matsushita K., Kikuchi K., Hatsukade I., 1996, 466, 686
\bibitem{} Nagataki S., Sato K., 1998, ApJ, 504, 629
\bibitem{} Nandra K., George I.M., Mushotzky R.F., Turner T.J., Yaqoob T., 1997, ApJ, 477, 602
\bibitem{} Navarro J.F., Frenk C.S., White S.D.M., 1995, MNRAS, 275, 720
\bibitem{} Nulsen P.E.J., 1986, MNRAS, 221, 377
\bibitem{} Nulsen P.E.J., 1998, MNRAS, 297, 1109
\bibitem{} Nulsen P.E.J., B\"ohringer H., 1995, MNRAS, 274, 1093 
\bibitem{} Nulsen P.E.J., Stewart G.C., Fabian A.C., 1984, MNRAS, 208, 185
\bibitem{} O'Dea C.P., Baum S.A., 1996, in Cold gas at high redshift, eds.  Bremer M.N., van der Werf P.P., R\"ottgering H.J.A., Carilli C.L., Kluwer, Dordrecht, p.199
\bibitem{} O'Dea C.P., Baum S.A., Maloney P.R., Tacconi L.J., Sparks W.B., 1994, ApJ, 422, 467
\bibitem{} Ohashi T., Honda H., Ezawa H., Kikuchi K., 1997, in Makino F., Mitsuda K., eds., X-ray Imaging and Spectroscopy of Cosmic Hot Plasmas, Universal Academy Press, Tokyo, p. 49
%\bibitem{} Oukbir J., Bartlett J.G., Blanchard A., 1997, A\&A, 320, 365
\bibitem{} Owen F.N., Eilek J.A., 1998, ApJ, 493, 730
%\bibitem{} Owen F.N. \etal 1999, preprint
\bibitem{} Orr A., Yaqoob T., Parmar A.N., Piro L., White N.E., Grandi P., A\&A, 337, 685
\bibitem{} Pedlar A., Ghataure H.S., Davies R.D., Harrison B.A., Perley R., Crane P.C., Unger S.W., 1990, MNRAS, 246, 477
\bibitem{} Peres C.B., Fabian A.C., Edge A.C., Allen S.W., Johnstone R.M., White D.A., 1998, MNRAS, 298, 416 
\bibitem{} Raymond J.C., Smith B.W., 1977, ApJS, 35, 419
\bibitem{} Reisenegger A., Miralda-Escud\'e J., Waxman E., 1996, ApJ, 457, 11L
\bibitem{} Rephaeli Y., Gruber D., Blanco P., 1999, ApJ, 511, 21L
\bibitem{} Roettiger K., Burns J.O., Loken C., 1996, ApJ, 473, 651
\bibitem{} Schindler S., 1996, A\&A, 305, 756
\bibitem{} Stark A.A., Gammie C.F., Wilson R.W., Bally J., Linke R.A., Heiles C. \& Hurwitz M., 1992. ApJS, 79, 77
\bibitem{} Tamura T. \etal, 1996, PASJ, 48, 671
\bibitem{} Tanaka Y., Inoue H., Holt S.S., 1994, PASJ, 46, L37
\bibitem{} Taylor G.B., Barton E.J., Ge J., 1994, AJ, 107, 1942 
\bibitem{} Taylor G.B., Allen S.W., Fabian A.C., 1999, in Diffuse Thermal and Relativistic Plasma in Galaxy Clusters, eds. B\"ohringer H., Feretti L., Schuecker P., MPE Report 271
\bibitem{} Thomas P.A., Fabian A.C., Nulsen P.E.J, 1987, MNRAS, 228, 973
\bibitem{} Tucker W., David L.P., ApJ, 1997, 484, 602
\bibitem{} Voit G.M., Donahue M., 1995, ApJ, 452, 164
\bibitem{} Watanabe M., Yamashita K., Kunieda H., Tawara Y., 1997, in Makino F., Mitsuda K., eds., X-ray Imaging and Spectroscopy of Cosmic Hot Plasmas,  Universal Academy Press, Tokyo, p. 131
%\bibitem{} White D.A., 1999, MNRAS, in press
\bibitem{} White D.A., Jones C., Forman W., 1997, MNRAS, 292, 419
\bibitem{} White D.A., Fabian A.C., Johnstone R.M., Mushotzky R.F., Arnaud K.A., 1991, MNRAS, 252, 72
\bibitem{} Wise M.W., O'Connell R.W., Bregman J.N., Roberts M.S., 1993, ApJ, 405, 94
\bibitem{} Wise M.W., Sarazin C.L., 2000, ApJ, submitted
\bibitem{} Xu H., Makishima K., Fukazawa Y., Ikebe Y., Kikuchi K., Ohashi T., Tamura T., 1998, ApJ, 500, 738


\end{thebibliography}
\end{document}